\documentclass[12pt]{article}
\RequirePackage[OT1]{fontenc}

\usepackage{amsmath,amsfonts,amsthm,amssymb,amsbsy}
\usepackage[toc,title,page]{appendix}
\usepackage{array,epsfig,fancyhdr,rotating}
\usepackage{color, dsfont}
\usepackage{enumerate}
\usepackage{float}
\usepackage{graphicx,psfrag,epsf}
\usepackage[]{hyperref}  
\usepackage{multirow}
\usepackage[sectionbib]{natbib}
\usepackage{sectsty, secdot}
\usepackage{subcaption}
\usepackage{url} 

\newtheorem{algorithm}{Algorithm}
\newtheorem{assumption}{Assumption}

\newtheorem{theorem}{Theorem}[section]
\newtheorem{corollary}[theorem]{Corollary}

\newtheorem{example}{Example}
\newtheorem{lemma}[theorem]{Lemma}

\newtheorem{remark}{Remark}

\def\AIC{\textsc{aic}}
\def\BIC{\textsc{bic}}

\begin{document}

%%%%%%%%%%%%%%%%%%%%%%%%%%%%%%%%%%%%%%%%%%%%%%%%%%%%%%%%%%%%%%%%%%%%%%%%%%%%%%%%%%%%%%%%%%%%%%%%%%%%%%%%%%%%%%%%%%%%%%%%%%%%
%%%%%%%%%%%%%%%%%%%%%%%%%%%%%%%%%%%%%%%%%%%%%%%%%%%%%%%%%%%%%%%%%%%%%%%%%%%%%%%%%%%%%%%%%%%%%%%%%%%%%%%%%%%%%%%%%%%%%%%%%%%%

\title{\Large\bf Multinomial Link Models}
\author{Tianmeng Wang$^1$, Liping Tong$^2$, and Jie Yang$^1$\\ 
	$^1$University of Illinois at Chicago and $^2$Advocate Aurora Health
}

\maketitle

%%%%%%%%%%%%%%%%%%%%%%%%%%%%%%%%%%%%%%%%%%%%%%%%%%%%%%%%%%%%%%%%%%%%%%%%%%%%%%%%%%%%%%%%%%%%%%%%%%%%%%%%%%%%%%%%%%%%%%%%%%%%

\begin{abstract}
We propose a new family of regression models for analyzing categorical responses, called multinomial link models. It consists of four classes, namely, mixed-link models that generalize existing multinomial logistic models and their extensions, two-group models that can incorporate the observations with NA or unknown responses, dichotomous conditional link models that handle longitudinal binary responses, and po-npo mixture models that are more flexible than partial proportional odds models. By characterizing the feasible parameter space, deriving necessary and sufficient conditions, and developing validated algorithms to guarantee the finding of feasible maximum likelihood estimates, we solve the infeasibility issue of existing statistical software when estimating parameters for cumulative link models. We also provide explicit formulae and detailed algorithms for computing the Fisher information matrix and selecting the best models among the new family. The applications to real datasets show that the new models can fit the data significantly better, correct misleading conclusions due to missing responses, and make more informative statistical inference.
\end{abstract}


\section{Introduction}\label{sec:intro}

We consider experiments or observational studies with categorical responses, which naturally arise in many different scientific disciplines \citep{agresti2018introduction}. When responses are binary, generalized linear models have been commonly used \citep{pmcc1989, dobson2018} to analyze the data. When responses have three or more categories, multinomial logistic models have been widely used in the literature \citep{pmcc1995, atkinson1999, bu2020}, which cover four kinds of logit models, namely baseline-category, cumulative,  adjacent-categories, and continuation-ratio logit models.

Following the notations of \cite{bu2020}, there are $d$ covariates and $m\geq 2$ distinct covariate settings ${\mathbf x}_i = (x_{i1}, \ldots, x_{id})^T$, $i=1, \ldots, m$. At the $i$th setting, $n_i>0$ categorical responses are collected i.i.d.~from a discrete distribution with $J$ categories, which are summarized into a multinomial response
${\mathbf Y}_i=(Y_{i1},\cdots,Y_{iJ})^T \sim {\rm Multinomial}(n_i; \pi_{i1},\cdots,\pi_{iJ})$,
where $\pi_{ij}$ is the probability that the response falls into the $j$th category at the $i$th setting. 
Throughout this paper, we assume $\pi_{ij} \in (0,1)$ for all $i=1, \ldots, m$ and $j=1, \ldots, J$. 
The four logit models with partial proportional odds (ppo, see \cite{lall2002, bu2020}) can be written as follows:
\begin{eqnarray}
  \log\left(\frac{\pi_{ij}}{\pi_{iJ}}\right) &=& {\mathbf h}_j^T({\mathbf x}_i)\boldsymbol\beta_j+{\mathbf h}_c^T({\mathbf x}_i)\boldsymbol\zeta~, \>\mbox{baseline-category};\label{eq:model_baseline}\\
    \log\left(\frac{\pi_{i1}+\cdots + \pi_{ij}}{\pi_{i,j+1}+\cdots + \pi_{iJ}}\right) &=& {\mathbf h}_j^T({\mathbf x}_i)\boldsymbol\beta_j+{\mathbf h}_c^T({\mathbf x}_i)\boldsymbol\zeta~, \>\mbox{cumulative}; \label{eq:cumulative_model}\\ 
  \log\left(\frac{\pi_{ij}}{\pi_{i,j+1}}\right) &=& {\mathbf h}_j^T({\mathbf x}_i)\boldsymbol\beta_j+{\mathbf h}_c^T({\mathbf x}_i)\boldsymbol\zeta~, \>\mbox{adjacent-categories}; \label{eq:adjacent_model}\\
	\log\left(\frac{\pi_{ij}}{\pi_{i,j+1} + \cdots + \pi_{iJ}}\right) &=& {\mathbf h}_j^T({\mathbf x}_i)\boldsymbol\beta_j+{\mathbf h}_c^T({\mathbf x}_i)\boldsymbol\zeta~, \>\mbox{continuation-ratio}, \label{eq:continuation_model}
\end{eqnarray}
where $i=1, \ldots, m$, $j=1, \ldots, J-1$, ${\mathbf h}_j^T(\cdot) = (h_{j1}(\cdot), \ldots, h_{jp_j}(\cdot))$ are known predictor functions associated with the parameters $\boldsymbol\beta_j = (\beta_{j1}, \ldots, \beta_{jp_j})^T$ for the $j$th response category, and ${\mathbf h}_c^T(\cdot) = (h_{1}(\cdot), \ldots, h_{p_c}(\cdot))$ are known predictor functions associated with the parameters $\boldsymbol\zeta = (\zeta_{1}, \ldots, \zeta_{p_c})^T$ that are common for all categories.  As special cases, ${\mathbf h}_j^T({\mathbf x}_i) \equiv 1$ leads to proportional odds (po) models assuming the same parameters for different categories \citep{pmcc1980}, and ${\mathbf h}_c^T({\mathbf x}_i) \equiv 0$ leads to nonproportional odds (npo) models allowing all parameters to change across categories \citep{agresti2013}. The corresponding expressions for po and  npo models can be found in the Supplementary Material (Sections~S.7 and S.8) of \cite{bu2020}. 

In the literature, the baseline-category logit model~\eqref{eq:model_baseline} is also known as the (multiclass) logistic regression model \citep{hastie2009elements}, which is commonly used for nominal responses that do not have a natural ordering \citep{agresti2013}. Models~\eqref{eq:cumulative_model}, \eqref{eq:adjacent_model}, and \eqref{eq:continuation_model} are typically used for ordinal or hierarchical responses with either a natural ordering or a hierarchical structure. According to \cite{wang2023identifying}, however, even for nominal responses, one can use the Akaike information criterion ($\AIC$, \cite{AIC1973information,hastie2009elements}) or Bayesian information criterion ($\BIC$, \cite{hastie2009elements}) to choose a working order of the response categories, treat the responses as ordinal ones, and apply models~\eqref{eq:cumulative_model}, \eqref{eq:adjacent_model}, or \eqref{eq:continuation_model}, which may significantly improve the prediction accuracy.

The four logit models \eqref{eq:model_baseline}, \eqref{eq:cumulative_model}, \eqref{eq:adjacent_model}, \eqref{eq:continuation_model} can be rewritten into a unified form \citep{pmcc1995, atkinson1999, bu2020}
\begin{equation}\label{eq:logitunifiedmodel}
\bar{\mathbf C}^T\log(\bar{\mathbf L}{\bar{\boldsymbol\pi}}_i)=
\bar{\mathbf X}_i{\boldsymbol\theta}, \qquad     i=1,\cdots,m,
\end{equation}
where 
$\bar{\mathbf C}^T$ is a $J\times(2J-1)$ constant matrix, 
$\bar{\mathbf L}$ is a $(2J-1)\times J$ constant matrix depending on \eqref{eq:model_baseline}, \eqref{eq:cumulative_model}, \eqref{eq:adjacent_model}, or \eqref{eq:continuation_model}, $\bar{\boldsymbol\pi}_i = (\pi_{i1}, \ldots, \pi_{iJ})^T$, $\bar{\mathbf X}_i$ is a $J\times p$ matrix depending on ${\mathbf h}_j^T({\mathbf x}_i)$, $j=1, \ldots, J-1$ and ${\mathbf h}_c^T({\mathbf x}_i)$, $p=p_1 + \cdots + p_{J-1} + p_c$, and $\boldsymbol{\theta} = (\boldsymbol{\beta}_1^T, \ldots, \boldsymbol{\beta}_{J-1}^T, \boldsymbol{\zeta}^T)^T$. 

Along another line in the literature, cumulative logit models \eqref{eq:cumulative_model} have been extended to cumulative link models or ordinal regression models \citep{pmcc1980, agresti2013, ytm2016}. In our notations, the cumulative link models can be written as
\begin{equation}\label{eq:clm}
g\left(\pi_{i1}+\cdots + \pi_{ij}\right) = {\mathbf h}_j^T({\mathbf x}_i)\boldsymbol\beta_j+{\mathbf h}_c^T({\mathbf x}_i)\boldsymbol\zeta\ ,
\end{equation}
where the link function $g(\rho)$ can be logit, probit, log-log, complementary log-log, and Cauchit (see Table~\ref{tab:link_functions}). The cumulative link model \eqref{eq:clm} with logit link is the same as the cumulative logit model \eqref{eq:cumulative_model}.

Baseline-category logit model \eqref{eq:model_baseline} has been extended with probit link, known as multinomial probit models \citep{aitchison1970, agresti2013, greene2018econometric}. In our notations, 
\begin{equation}\label{eq:bclm}
g\left(\frac{\pi_{ij}}{\pi_{ij}+\pi_{iJ}}\right) = {\mathbf h}_j^T({\mathbf x}_i)\boldsymbol\beta_j+{\mathbf h}_c^T({\mathbf x}_i)\boldsymbol\zeta\ ,
\end{equation}
where the link function $g(\rho)$ can be logit or probit. Model~\eqref{eq:bclm} with logit link is the same as the baseline-category logit model \eqref{eq:model_baseline}. Examples can be found in \cite{agresti2010, agresti2013}.

Continuation-ratio logit model \eqref{eq:continuation_model} has been extended with complementary log-log link by \cite{oconnell2006} and other links by \cite{ai2023locally}. In our notations, 
\begin{equation}\label{eq:crlm}
g\left(\frac{\pi_{ij}}{\pi_{ij}+\cdots + \pi_{iJ}}\right) = {\mathbf h}_j^T({\mathbf x}_i)\boldsymbol\beta_j+{\mathbf h}_c^T({\mathbf x}_i)\boldsymbol\zeta\ ,
\end{equation}
where the link function $g(\rho)$ can be logit, probit, log-log, complementary log-log, and Cauchit. Model~\eqref{eq:crlm} with logit link is the same as the continuation-ratio logit model \eqref{eq:continuation_model}.

Given so many multinomial models have been proposed or extended for categorical data, however, there are many challenges arising from real data analysis (Section~\ref{sec:applications}). In this paper, we propose four new classes of multinomial regression models, namely, {\it (i) mixed-link models} allowing different link functions across categories (Sections~\ref{sec:mixed_link_model} and \ref{sec:MLM_example_extrauma}), which cover all the multinomial models that we reviewed above; {\it (ii) two-group models} incorporating observations with NA or unknown responses (Sections~\ref{sec:two_group_link_model} and \ref{sec:metabolic_syndrome}); {\it (iii) dichotomous conditional link models} dealing with longitudinal binary responses (Sections~\ref{sec:multinomial_conditional_link} and \ref{sec:six_cities_cond_link_model}); and {\it (iv) po-npo mixture models} admitting more flexible structures than ppo models (Sections~\ref{sec:po_npo_mixture_model} and \ref{sec:police_po-npo_mixture_model}).
We unify all the four classes of models into a new family of multinomial regression models, called {\it multinomial link models} (see Section~\ref{sec:mlm}), and provide detailed theoretical results on their feasible parameter spaces (Section~\ref{sec:feasible_parameter}) and information matrices (Section~\ref{sec:fisher_information}), as well as justified algorithms for finding feasible parameter estimates (Section~\ref{sec:formulae_and_algorithms}) and selecting the most appropriate models given a dataset (Section~\ref{sec:model_selection}). Our theoretical results and algorithms solve the infeasibility issue commonly existing in current statistical software on fitting cumulative link models (Section~\ref{sec:MLM_existing_issue}). The family of multinomial link models is much broader than existing multinomial regression models. It provides potential users a comprehensive toolbox for categorical data analysis. Its unified structure allows the users to select models and variables more conveniently, correct misleading statements due to missing data, and draw more reliable conclusions (Section~\ref{sec:applications}).

\section{Multinomial Link Models}
\label{sec:mlm}

Inspired by the unified form \eqref{eq:logitunifiedmodel} of multinomial logistic models, in this section, we propose a new family of multinomial regression model, called the multinomial link model, which covers four classes of new models, namely the mixed-link models allowing separate link functions for different categories (Section~\ref{sec:mixed_link_model}), two-group models incorporating NA or unknown responses (Section~\ref{sec:two_group_link_model}), dichotomous conditional link models dealing with longitudinal binary responses (Section~\ref{sec:multinomial_conditional_link}), and po-npo mixture models allowing more flexible model structures than ppo models (Section~\ref{sec:po_npo_mixture_model}). 

\subsection{Multinomial link models in a unified form}
\label{sec:mlm_matrix}

In matrix form, a multinomial link model can be written as
\begin{equation}\label{eq:mlm_in_matrix}
{\mathbf g}\left(\frac{{\mathbf L}{\boldsymbol\pi}_i}{{\mathbf R}\boldsymbol\pi_i + \pi_{iJ} {\mathbf b}}\right)={\boldsymbol\eta}_i={\mathbf X}_i{\boldsymbol\theta}\ ,
\end{equation}
where ${\mathbf g} = (g_1, \ldots, g_{J-1})^T$ is a vector of $J-1$ link functions, ${\mathbf L}$ and ${\mathbf R}$ are $(J-1)\times (J-1)$ constant matrices, ${\mathbf b} \in \mathbb{R}^{J-1}$ is a constant vector, ${\boldsymbol\pi}_i = (\pi_{i1}, \ldots, \pi_{i,J-1})^T \in \mathbb{R}^{J-1}$, $\pi_{iJ} = 1 -  \sum_{j=1}^{J-1} \pi_{ij}$~, ${\boldsymbol\eta}_i = (\eta_{i1}, \ldots, \eta_{i,J-1})^T \in \mathbb{R}^{J-1}$,  ${\mathbf X}_i = ({\mathbf f}_1({\mathbf x}_i), \ldots, {\mathbf f}_{J-1}({\mathbf x}_i))^T \in \mathbb{R}^{(J-1)\times p}$ with ${\mathbf f}_j({\mathbf x}_i) = (f_{j1}({\mathbf x}_i), \ldots, f_{jp}({\mathbf x}_i))^T$,
and the regression parameter vector $\boldsymbol\theta=(\theta_1, \ldots, \theta_p)^T$ consists of $p$ unknown parameters in total. Note that the vector ${\mathbf g}$ of link functions applies to the ratio of two vectors  component-wise, which can be denoted as ${\mathbf g}\left( ({\mathbf L}{\boldsymbol\pi}_i) \oslash ({\mathbf R} {\boldsymbol\pi}_i + \pi_{iJ} {\mathbf b})\right)$ with the notation of element-wise division ``$\oslash$'' (also known as Hadamard division). That is, if we denote ${\mathbf L}  = ({\mathbf L}_1, \ldots, {\mathbf L}_{J-1})^T$, ${\mathbf R} = ({\mathbf R}_1, \ldots, {\mathbf R}_{J-1})^T$ and ${\mathbf b} = (b_1, \ldots, b_{J-1})^T$, then the multinomial link model~\eqref{eq:mlm_in_matrix} can be written in its equation form
\begin{equation}\label{eq:mlm_j}
g_j\left(\frac{{\mathbf L}^T_j \boldsymbol\pi_i}{{\mathbf R}^T_j \boldsymbol\pi_i + \pi_{iJ} b_j}\right) = \eta_{ij} = {\mathbf f}_j^T({\mathbf x}_i)\boldsymbol\theta, \quad j=1, \ldots, J-1.
\end{equation}
To simplify the notation,
we define
\begin{equation}\label{eq:rho_ij_general}
\rho_{ij} = \frac{{\mathbf L}^T_j \boldsymbol\pi_i}{{\mathbf R}^T_j \boldsymbol\pi_i + \pi_{iJ} b_j}, \quad j=1, \ldots, J-1,
\end{equation}
and $\boldsymbol\rho_i = (\rho_{i1}, \ldots, \rho_{i, J-1})^T = \left({\mathbf L}{\boldsymbol\pi}_i\right)\oslash\left({\mathbf R}\boldsymbol\pi_i + \pi_{iJ} {\mathbf b}\right)$. The notations of $\boldsymbol\pi_i$, $\boldsymbol\eta_i$, ${\mathbf X}_i$ here are different from those in \cite{bu2020}. For readers' convenience, we list all notations used in the main text in Appendix~\ref{subsec:list}. Special classes of the multinomial link models \eqref{eq:mlm_in_matrix} or \eqref{eq:mlm_j} with explicit ${\mathbf L}$, ${\mathbf R}$, and ${\mathbf b}$ can be found in Appendices~\ref{subsec:m_link_m}, \ref{subsec:more_two_group}, and \ref{sec:more_on_dichotomous_condi_link_model}. 

One major benefit by taking the unified form \eqref{eq:mlm_in_matrix} or \eqref{eq:mlm_j} is that the corresponding theoretical results (Section~\ref{sec:parameter_fisher_information}), algorithms (Section~\ref{sec:formulae_and_algorithms}), and model selection techniques can be applied to all models covered by the same unified form (see Section~\ref{sec:feasible_parameter} for necessary and sufficient conditions for ${\mathbf L}, {\mathbf R}, {\mathbf b}$, such that the multinomial link model \eqref{eq:mlm_in_matrix} or \eqref{eq:mlm_j} is well defined).

\subsection{Link functions and mixed-link models}\label{sec:mixed_link_model}

For multinomial link models~\eqref{eq:mlm_in_matrix} or \eqref{eq:mlm_j}, we assume that {\it (i)} the link functions $g_1, \ldots, g_{J-1}$ are well defined from $\rho\in (0,1)$ to $\eta \in (-\infty, \infty)$, which are part of the model assumptions. In this paper, we also require that {\it (ii)} $g_1^{-1}, \ldots, g_{J-1}^{-1}$ exist and are differentiable from $\eta \in (-\infty, \infty)$ to $\rho \in (0,1)$; and {\it (iii)} $(g_j^{-1})'(\eta) > 0$ for all $\eta \in (-\infty, \infty)$ and $j=1, \ldots, J-1$. If a link function $g$ under consideration is decreasing, one may replace it with $-g$ to satisfy our requirements. Such a replacement leads to a model that is mathematically equivalent to the previous one.

In the statistical literature, many link functions have been proposed. For examples, logit, probit, log-log, and complementary log-log links were used by \cite{pmcc1989} for binary responses; Cauchit link can be tracked back to \cite{morgan1992note} for distributions with many extreme values; $t$ link was suggested first by \cite{albert1993bayesian} and also connected to robit regression \citep{liu2004robit} and Gosset link family \citep{koenker2009parametric}; Pregibon link family \citep{pregibon1980goodness, koenker2009parametric, smith2020exploration} was introduced as a two-parameter generalization of the logit link. We skip Pregibon link in this paper since its image usually does not cover the whole real line. 

In Table~\ref{tab:link_functions} we list possible link functions considered for multinomial link models. It should be noted that the $t$ link family $g(\rho) = F_\nu^{-1}(\rho)$ incorporates logit, which can be approximated by $F_7$ according to \cite{liu2004robit}, and probit as a limit when $\nu$ goes to $\infty$. Here $F_\nu$ and $f_\nu$ are the cumulative distribution function (cdf) and probability density function (pdf) of $t$-distribution with the number $\nu$ of degrees of freedom, respectively.

\begin{table}[ht]
\caption{Possible Link Functions for Multinomial Link Models}
\label{tab:link_functions}
\footnotesize
\begin{center}
\begin{tabular}{|l|c|c|c|}\hline
Name & $\eta = g(\rho)$ & $\rho = g^{-1}(\eta)$ & $(g^{-1})'(\eta)$ \\ \hline
logit & $\log \left(\frac{\rho}{1-\rho}\right)$ & $\frac{e^\eta}{1+e^\eta}$ & $\frac{e^\eta}{(1+e^\eta)^2} = \rho (1-\rho)$ \\
probit & $\Phi^{-1}(\rho)$ & $\Phi (\eta)$ & $\phi (\eta) = \frac{1}{\sqrt{2\pi}} e^{-\frac{\eta^2}{2}}$ \\
log-log & $-\log(-\log(\rho))$ & $ \exp\{-e^{-\eta}\}$ & $\exp\{-e^{-\eta}-\eta\} = \rho e^{-\eta}$ \\
complementary log-log & $\log(-\log(1-\rho))$ & $1-\exp\{-e^\eta\}$ & $\exp\{-e^\eta + \eta\}  = (1-\rho) e^\eta$ \\
Cauchit & $\tan(\pi(\rho-1/2))$ & $\frac{1}{2} + \frac{1}{\pi}\cdot {\rm arctan}(\eta)$ & $\frac{1}{\pi (1+\eta^2)}$ \\
t/robit/Gosset, $\nu>0$ & $F_\nu^{-1}(\rho)$  & $F_\nu(\eta)$ & $f_\nu(\eta) = \frac{\Gamma\left(\frac{\nu+1}{2}\right)}{\sqrt{\nu \pi}\ \Gamma\left(\frac{\nu}{2}\right)} \left(1 + \frac{\eta^2}{\nu}\right)^{-\frac{\nu+1}{2}}$   \\
\hline
\end{tabular}
\end{center}
\normalsize
\end{table}

In this section, we introduce a special class of the multinomial link models~\eqref{eq:mlm_j}, which allows separate links for different categories. 
We show later in  Section~\ref{sec:MLM_example_extrauma} that a multinomial regression model with mixed links can fit some real data significantly better.

\begin{example}\label{ex:mixed_link_ppo} {\bf Mixed-link models with ppo}\quad {\rm
Inspired by the extended models~\eqref{eq:clm}, \eqref{eq:bclm}, \eqref{eq:crlm}, we extend models~\eqref{eq:model_baseline}, \eqref{eq:cumulative_model}, \eqref{eq:adjacent_model}, and \eqref{eq:continuation_model} by allowing mixed links to the following {\it mixed-link model} with ppo: 
\begin{equation}\label{eq:mixed_link_ppo}
g_j\left(\rho_{ij}\right) = {\mathbf h}_j^T({\mathbf x}_i)\boldsymbol\beta_j+{\mathbf h}_c^T({\mathbf x}_i)\boldsymbol\zeta\ ,
\end{equation}
where $i=1, \ldots, m$; $j=1, \ldots, J-1$; $g_1, \ldots, g_{J-1}$ are given link functions; and
\begin{equation}\label{eq:rho_ij}
\rho_{ij} = \left\{
\begin{array}{cl}
\frac{\pi_{ij}}{\pi_{ij} + \pi_{iJ}} & \mbox{, for baseline-category mixed-link models;}\\
\pi_{i1} + \cdots + \pi_{ij} & \mbox{, for cumulative mixed-link models;}\\
\frac{\pi_{ij}}{\pi_{ij} + \pi_{i,j+1}} & \mbox{, for adjacent-categories mixed-link models;}\\
\frac{\pi_{ij}}{\pi_{ij} + \cdots + \pi_{iJ}} & \mbox{, for continuation-ratio mixed-link models.}
\end{array}
\right.
\end{equation}
The mixed-link model~\eqref{eq:mixed_link_ppo}+\eqref{eq:rho_ij} covers all models reviewed in Section~\ref{sec:intro}. It is a special class of the multinomial link model~ \eqref{eq:mlm_j}
(see Appendix~\ref{subsec:m_link_m}).
\hfill{$\Box$}
}\end{example}

\subsection{NA category and two-group models}\label{sec:two_group_link_model}

In practice, it is fairly common to encounter observations with NA or unknown responses. If the missing mechanism is not at random, the analysis after removing those observations can be misleading \citep{bland2015introduction}. According to \cite{wang2023identifying}, one may treat NA as a special category and use $\AIC$ or $\BIC$ to choose a working order of the response categories including NA. However, for some real applications, a multinomial model with a working order for all response categories may not fit the data well (see  Section~\ref{sec:metabolic_syndrome}).

In this section, we introduce a special class of the multinomial link model \eqref{eq:mlm_in_matrix} or \eqref{eq:mlm_j}, which allows response categories consisting of two overlapped groups, called a {\it two-group model}. One group of $k + 1 \geq 2$ categories are controlled by a baseline-category mixed-link model (see Example~\ref{ex:mixed_link_ppo}), while the other group of $J-k \geq 3$ categories are controlled by a cumulative, adjacent-categories, or continuation-ratio mixed-link model (see \eqref{eq:rho_ij}). The two groups share a common category so that all categories are connected via the shared category. A special case of two-group models is that the two groups share the same baseline category $J$ (see Appendix~\ref{subsec:more_two_group} and Example~\ref{ex:two_group_J}). A general two-group model is described as follows.

\begin{example}\label{ex:two_group_s} {\bf Two-group models with ppo}\quad {\rm
In this model, we assume that $J\geq 4$ and the response categories consist of two groups. The first group $\{1, \ldots, k, s\}$ is controlled by a baseline-category mixed-link model with the baseline category $s$, where $1\leq k \leq J-3$ and $k+1 \leq s \leq J$, while the other group $\{k+1, \ldots, J\}$ is controlled by a cumulative, adjacent-categories, or continuation-ratio mixed-link model with $J$ as the baseline category. The two groups share the category $s$ to connect all the $J$ categories. 
The two-group model with ppo is defined by equation~\eqref{eq:mixed_link_ppo} plus
\begin{equation}\label{eq:rho_ij_two_group_s}
\rho_{ij} = \left\{\begin{array}{cl}
\frac{\pi_{ij}}{\pi_{ij} + \pi_{is}} & \mbox{, for }j=1, \ldots, k;\\
\frac{\pi_{i,k+1} + \cdots + \pi_{ij}}{\pi_{i,k+1} + \cdots + \pi_{iJ}} & \mbox{, for baseline-cumulative and }j=k+1, \ldots, J-1;\\
\frac{\pi_{ij}}{\pi_{ij}+\pi_{i,j+1}} & \mbox{, for baseline-adjacent and }j=k+1, \ldots, J-1;\\
\frac{\pi_{ij}}{\pi_{ij} + \cdots + \pi_{iJ}} & \mbox{, for baseline-continuation and }j=k+1, \ldots, J-1.
\end{array}\right.
\end{equation}

The two-group model~\eqref{eq:mixed_link_ppo}+\eqref{eq:rho_ij_two_group_s} consists of three subclasses, namely baseline-cumulative, baseline-adjacent, and baseline-continuation mixed-link  models with ppo, which are all special cases of the multinomial link model~\eqref{eq:mlm_in_matrix} or \eqref{eq:mlm_j}
(see Appendix~\ref{subsec:more_two_group}).
\hfill{$\Box$}
}\end{example}

Other possible structures, such as cumulative-continuation (two-group), three-group or multi-group models, can be defined similarly, which are still covered by the general multinomial link model \eqref{eq:mlm_in_matrix} or \eqref{eq:mlm_j}.
 
\subsection{Longitudinal responses and dichotomous conditional link model}\label{sec:multinomial_conditional_link}

In practice, categorical responses may be collected from the same subject on a regular basis, especially in clinical trials, leading to longitudinal categorical responses, which are much more difficult to model than a single categorical response.
Inspired by \cite{evans2013marginal}, in this section, we propose a dichotomous conditional link model for longitudinal binary responses, or more generally, binary responses with a sequential order.

Suppose for each of the experimental settings or covariate vectors $\{{\mathbf x}_i, i=1, \ldots, m\}$, a sequence of categorical responses $Z_{it} \in {\cal I}_t$ are collected at time $t=1, \ldots, T$. For binary responses, for example, ${\cal I}_t = \{0, 1\}$ for each $t$. We denote ${\cal T} = {\cal I}_1\times \cdots \times {\cal I}_T$ as the collection of all possible longitudinal outcomes.

\begin{example}\label{ex:binary_cond_link_model}{\bf Dichotomous (or binary) conditional link model}\quad {\rm
In this case, ${\cal I}_1 = \cdots = {\cal I}_T = \{0,1\}$, and $J=2^T$. We let $\sigma : {\cal T} \rightarrow {\cal J} = \{1, \ldots, J\}$ be defined as $\sigma(z_1, \ldots, z_T) = \sum_{t=1}^T z_t 2^{t-1}$ if $z_t=1$ for at least one $t$; and $2^T$ if $z_1 = \cdots = z_T=0$. Then {\it (i)} for $j = \sigma(1, 0, \ldots, 0) = 1$, 
\[
g_1(P(Z_{i1}=1)) = g_1\left(\frac{{\mathbf L}_{1}^T {\boldsymbol{\pi}}_{i}}{{\mathbf R}_{1}^T {\boldsymbol{\pi}}_{i} + \pi_{iJ} b_1}\right) = \eta_{i1} = {\mathbf f}_1^T({\mathbf x}_i) \boldsymbol{\theta}
\]
with ${\mathbf L}_1^T = (1,0,1,0,\ldots,1)$, ${\mathbf R}_1^T = (1, 1, \ldots, 1)$, and $b_1=1$; and {\it (ii)} given $t \in \{2, \ldots, T\}$, $z_1, \ldots, z_{t-1}$ $\in \{0,1\}$, for $j=\sigma(z_1, \ldots, z_{t-1},1, 0, \ldots, 0) = \sum_{l=1}^{t-1} z_l 2^{l-1} + 2^{t-1}$,
\[
g_j(P(Z_{it}=1\mid Z_{i1}=z_1, \ldots, Z_{i,t-1}=z_{t-1})) = g_j\left(\frac{{\mathbf L}_{j}^T {\boldsymbol{\pi}}_{i}}{{\mathbf R}_{j}^T {\boldsymbol{\pi}}_{i} + \pi_{iJ} b_j}\right) = \eta_{ij} = {\mathbf f}_j^T({\mathbf x}_i) \boldsymbol{\theta}
\]
with ${\mathbf L}_j^T = (L_{j1}, \ldots, L_{j,J-1}), {\mathbf R}_j^T = (R_{j1}, \ldots, R_{j,J-1}) \in \mathbb{R}^{J-1}$ and $b_j\in \mathbb{R}$, where $L_{jl} = 1$ if $l\equiv \sum_{r=1}^{t-1}z_r 2^{r-1} + 2^{t-1}\mod 2^t$, and $0$ otherwise; $R_{jl} = 1$ if $l\equiv \sum_{r=1}^{t-1}z_r 2^{r-1}\mod 2^{t-1}$, and $0$ otherwise; and $b_j = 1$ if $z_1 = \cdots = z_{t-1} =0$, and $0$ otherwise.
\hfill{$\Box$}}\end{example}

The dichotomous conditional link model is a special class of the multinomial link model~\eqref{eq:mlm_in_matrix} or \eqref{eq:mlm_j}. 
It is different in nature from the multivariate logistic models or log-linear regression models \citep{pmcc1995} proposed for similar purposes (see Appendix~\ref{sec:more_on_multi_condi_link_model}). 
The dichotomous conditional link model described here can lead to more informative inference in practice (see Section~\ref{sec:six_cities_cond_link_model} for a real data example, and  Appendix~\ref{sec:more_on_dichotomous_condi_link_model} for more technical details).

\subsection{Partially equal coefficients and po-npo mixture models}\label{sec:po_npo_mixture_model}

If we check the right hand sides of models \eqref{eq:model_baseline}, \eqref{eq:cumulative_model}, \eqref{eq:adjacent_model}, \eqref{eq:continuation_model}, \eqref{eq:clm}, \eqref{eq:bclm}, \eqref{eq:crlm}, ppo \citep{lall2002,bu2020} is the most flexible structure for model matrices in the literature, which allows that the parameters of some predictors are the same across different categories (i.e., the po component ${\mathbf h}_c^T({\mathbf x}_i)\boldsymbol\zeta$), while the parameters of some other predictors are different across categories (i.e., the npo component ${\mathbf h}_j^T({\mathbf x}_i)\boldsymbol\beta_j$). 

For some applications (see Section~\ref{sec:police_po-npo_mixture_model} for a real data example), however, it can be significantly better if we allow some (but not all) categories share the same  coefficients for some predictors. For example, the first and second categories share the same coefficients for $x_{i1}$ and $x_{i2}$ (i.e., follow a po model), while the third and fourth categories have their own coefficients for $x_{i1}$ and $x_{i2}$ (i.e., follow a npo model). 
The corresponding model matrix is
\begin{equation*}
{\mathbf X}_i= \begin{pmatrix}
1 & & & & & & & & x_{i1} & x_{i2}\\
& 1 & & & & & & & x_{i1} & x_{i2}\\
& & 1 & x_{i1} & x_{i2} & & & & &\\
& & & & & 1 & x_{i1} & x_{i2} & &\\
\end{pmatrix}
\end{equation*}
with parameters $\boldsymbol\theta=(\beta_{11},\beta_{21},\beta_{31},\beta_{32},\beta_{33},\beta_{41},\beta_{42},\beta_{43},\zeta_{1},\zeta_{2})^T$. It is not a ppo model.

In this section, we introduce a special class of the multinomial link models~\eqref{eq:mlm_in_matrix} or \eqref{eq:mlm_j}, called {\it po-npo mixture models}, which allows the regression coefficients/parameters for a certain predictor to be partially equal, that is, equal across some, but not all, categories.

\begin{example}\label{ex:po_ppo_mixture}{\bf Po-npo mixture model}\quad {\rm
We assume that the model matrix of model~\eqref{eq:mlm_in_matrix} takes the form of
\begin{equation}\label{eq:po_npo_Xi}
{\mathbf X}_i= \begin{pmatrix}
{\mathbf h}_1^T({\mathbf x}_i) &   & & {\mathbf h}_{c1}^T({\mathbf x}_i)\\
&   \ddots &  & \vdots \\
&   & {\mathbf h}_{J-1}^T({\mathbf x}_i) & {\mathbf h}_{c,J-1}^T({\mathbf x}_i)
\end{pmatrix}\>\in \mathbb{R}^{(J-1)\times p}\ ,
\end{equation}
where ${\mathbf h}_{cj} ({\mathbf x}_i) = (h_{cj1}({\mathbf x}_i), \ldots, h_{cjp_c}({\mathbf x}_i))^T$ are known functions to determine the $p_c$ predictors associated with the $j$th category. If we write $\boldsymbol{\theta} = (\boldsymbol{\beta}_1^T, \ldots, \boldsymbol{\beta}_{J-1}^T, \boldsymbol{\zeta}^T)^T$, the po-npo mixture model can be written as
\begin{equation}\label{eq:po_npo_model}
g_j\left(\rho_{ij}\right) = {\mathbf h}_j^T({\mathbf x}_i)\boldsymbol\beta_j+{\mathbf h}_{cj}^T({\mathbf x}_i)\boldsymbol\zeta\ ,
\end{equation}
where $\rho_{ij}$ is given by \eqref{eq:rho_ij_general}.

One special case with ${\mathbf h}_{c1}({\mathbf x}_i) =\cdots = {\mathbf h}_{c,J-1}({\mathbf x}_i) \equiv {\mathbf h}_c ({\mathbf x}_i)$ leads to the classical ppo model (see \eqref{eqn:Xi_ppo_J-1} in Appendix~\ref{subsec:m_link_m}).
Another special case with $\rho_{ij}$ given by \eqref{eq:rho_ij} leads to a generalization of Example~\ref{ex:mixed_link_ppo}, called mixed-link models with po-npo mixture (see Appendix~\ref{sec:summary_notation} for other special classes of multinomial link models).
\hfill{$\Box$}
}\end{example}

\section{Feasible Parameter Space}\label{sec:feasible_parameter}

In this section, we discuss the necessary and sufficient conditions such that the multinomial link model~\eqref{eq:mlm_in_matrix} or \eqref{eq:mlm_j} is well defined and the parameters $\boldsymbol{\theta}$ are feasible.

\subsection{Feasibility of parameters}\label{sec:feasibility_parameter}

It is known that the parameter estimates $\hat{\boldsymbol{\theta}}$ found by R or SAS for cumulative logit models \eqref{eq:cumulative_model} might be infeasible. That is, some categorical probability $\pi_{ij}(\hat{\boldsymbol{\theta}}) \notin (0,1)$. For example, \cite{huang2025constrained} reported in their Example~8 that $44$ out of $1,000$ fitted parameters by SAS PROC LOGISTIC command for cumulative logit models lead to $\pi_{ij}(\hat{\boldsymbol{\theta}})<0$  for some $i=1, \ldots, m$ and $j=1, \ldots, J$  (see Section~\ref{sec:MLM_existing_issue} for a more comprehensive simulation study).

In this section, we provide explicit formulae for $\pi_{ij}$'s as functions of parameters $\boldsymbol{\theta}$ and ${\mathbf x}_i$'s under a general multinomial link model \eqref{eq:mlm_in_matrix} or \eqref{eq:mlm_j}, and characterize the space of feasible parameters for searching parameter estimates.
 
Given the parameters $\boldsymbol{\theta} \in \mathbb{R}^p$ and a setting ${\mathbf x}_i \in \mathbb{R}^d$, $\eta_{ij} = {\mathbf f}_j^T({\mathbf x}_i) \boldsymbol{\theta} \in (-\infty, \infty)$ according to \eqref{eq:mlm_j}, and $\rho_{ij} = g_j^{-1} (\eta_{ij}) \in (0,1)$  as defined in \eqref{eq:rho_ij_general}, $j=1, \ldots, J-1$. To generate multinomial responses under the multinomial link model, we require $\pi_{ij} \in (0, 1)$, $j=1, \ldots, J$. 
To solve $\boldsymbol\pi_i = (\pi_{i1}, \ldots, \pi_{i,J-1})^T$ and $\pi_{iJ}$ from $\boldsymbol\rho_i = (\rho_{i1}, \ldots, \rho_{i,J-1})^T$, we denote ${\mathbf D}_i = {\rm diag}(\boldsymbol\rho_i^{-1}) {\mathbf L} - {\mathbf R} \in \mathbb{R}^{(J-1)\times (J-1)}$, where ${\rm diag}(\boldsymbol\rho_i^{-1}) = {\rm diag}\{\rho_{i1}^{-1}, \ldots, \rho_{i,J-1}^{-1}\} \in \mathbb{R}^{(J-1)\times (J-1)}$. The explicit formulae are provided as follows.

\begin{lemma}\label{lem:pi_from_rho}
Suppose $\rho_{ij} \in (0, 1), j=1, \ldots, J-1$, ${\mathbf D}_i^{-1}$ exists, and all the $J-1$ coordinates of ${\mathbf D}_i^{-1} {\mathbf b}$ are positive. Then model~\eqref{eq:mlm_in_matrix} implies a unique $\boldsymbol\pi_i$ as a function of $\boldsymbol\rho_i$~:
\begin{equation}\label{eq:pi_from_rho}
\boldsymbol\pi_i = \frac{{\mathbf D}_i^{-1} {\mathbf b}}{1 + {\mathbf 1}^T_{J-1} {\mathbf D}_i^{-1} {\mathbf b}}
\end{equation}
as well as $\pi_{iJ} = (1 + {\mathbf 1}^T_{J-1} {\mathbf D}_i^{-1} {\mathbf b})^{-1}$, such that $\boldsymbol\pi_{ij} \in (0,1)$ for all $j=1, \ldots, J$, where ${\mathbf 1}_{J-1}$ is a vector consisting of $J-1$ ones.
\end{lemma}

The proof of Lemma~\ref{lem:pi_from_rho}, as well as other proofs, is relegated to Appendix~\ref{sec:proofs}.

According to Lemma~\ref{lem:pi_from_rho}, it is sufficient for $\pi_{ij} \in (0, 1)$ to let ${\mathbf D}_i^{-1}$ exist and all the $J-1$ coordinates of ${\mathbf D}_i^{-1} {\mathbf b}$ to be positive. 
Given the data with the observed set of distinct settings $\{{\mathbf x}_1, \ldots, {\mathbf x}_m\}$, we define the {\it feasible parameter space} of model~\eqref{eq:mlm_in_matrix} or \eqref{eq:mlm_j} as 
\begin{equation}\label{eq:Theta_in_general}
\boldsymbol{\Theta} = \left\{ \boldsymbol\theta \in \mathbb{R}^p \mid {\mathbf D}_i^{-1} \mbox{ exists, all the } J-1 \mbox{ coordinates of }{\mathbf D}_i^{-1} {\mathbf b} \mbox{ are positive}, i=1, \ldots, m \right\}\ .
\end{equation}
Here $\boldsymbol\Theta$ is for a general multinomial link model, which is typically either $\mathbb{R}^p$ itself or an open subset of $\mathbb{R}^p$. For many specific classes of multinomial link models, we can obtain simplified or more detailed conditions for $\boldsymbol{\Theta}$ (see Section~\ref{sec:model_regularity}).

For typical applications in practice, to find the parameter estimates numerically, we often specify a bounded subset of $\boldsymbol\Theta$, which is expected to contain the true $\boldsymbol\theta$ as an interior point, as the working parameter space to achieve desired theoretical properties \citep{ferguson1996course}.

\subsection{Model regularities}
\label{sec:model_regularity}

In this section, we explore regularity conditions for ${\mathbf L}$, ${\mathbf R}$, and ${\mathbf b}$, such that the multinomial link model~\eqref{eq:mlm_in_matrix} or \eqref{eq:mlm_j} is well defined. We break the relevant conditions into five assumptions in this section.

To simply the notation, we let $\preceq$ denote the element-wise $\leq$~. That is, $0 \preceq (a_1, \ldots, a_n)^T$ if $0\leq a_i$ for each $i$, and $(c_1, \ldots, c_n)^T \preceq (a_1, \ldots, a_n)^T$ if $c_i\leq a_i$ for each $i$. We also let $\boldsymbol{\Pi}_0 = \{(\pi_1, \ldots, \pi_{J-1})^T \in \mathbb{R}^{J-1} \mid \pi_j > 0, j=1, \ldots, J-1; \sum_{j=1}^{J-1} \pi_j < 1\}$ denote the collection of $\boldsymbol{\pi}_i$ under our consideration.

\begin{assumption}\label{as:A1}
For $j=1, \ldots, J-1$, {\it (i)} $0 \preceq {\mathbf L}_j$ and  ${\mathbf 1}_{J-1}^T {\mathbf L}_j > 0$; {\it (ii)} ${\mathbf L}_j \preceq {\mathbf R}_j$~; {\it (iii)} $b_j\geq 0$; and {\it (iv)} ${\mathbf 1}_{J-1}^T ({\mathbf R}_j - {\mathbf L}_j) > 0$ if $b_j=0$. 
\end{assumption}

\begin{lemma}\label{thm:rho_ij_in_0_1}
As defined in \eqref{eq:rho_ij_general}, $\rho_{ij}\in (0,1)$ and ${\mathbf L}_j^T \boldsymbol{\pi}_i > 0$ for all $\boldsymbol{\pi}_i \in \boldsymbol{\Pi}_0$ and $j=1, \ldots, J-1$, if and only if Assumption~\ref{as:A1} is satisfied.
\end{lemma}

Technically speaking, if ${\mathbf L}$, ${\mathbf R}$ and ${\mathbf b}$ lead to a well-defined model~\eqref{eq:mlm_in_matrix}, so do $-{\mathbf L}$, $-{\mathbf R}$ and $-{\mathbf b}$. To skip this trivially equivalent case, we add ${\mathbf L}_j^T \boldsymbol{\pi}_i > 0$ in Lemma~\ref{thm:rho_ij_in_0_1}. To ensure the uniqueness of $\boldsymbol{\pi}_i$ as a function of $\boldsymbol{\rho}_i$~, we need the following additional assumption:

\begin{assumption}\label{as:1_b>0}
${\mathbf 1}_{J-1}^T {\mathbf b} > 0$.    
\end{assumption}

\begin{lemma}\label{thm:unique_pi}
Suppose ${\mathbf L}$, ${\mathbf R}$ and ${\mathbf b}$ satisfy Assumption~\ref{as:A1}. Given $\rho_{ij}\in (0, 1)$, $j=1, \ldots, J-1$, if there is at most one $\boldsymbol{\pi}_i \in \boldsymbol{\Pi}_0$ satisfying \eqref{eq:rho_ij_general}, then ${\mathbf b}$ must satisfy Assumption~\ref{as:1_b>0}.    
\end{lemma}

According to Lemmas~\ref{thm:rho_ij_in_0_1} and \ref{thm:unique_pi}, Assumptions~\ref{as:A1} and \ref{as:1_b>0} are necessary conditions of ${\mathbf L}$, ${\mathbf R}$ and ${\mathbf b}$ for the multinomial link model~\eqref{eq:mlm_in_matrix} or \eqref{eq:mlm_j} to be well defined.

Now we explore the connection between ${\mathbf L}$, ${\mathbf R}$, ${\mathbf b}$ and the feasible parameter space $\boldsymbol{\Theta}$. According to the proof of Lemma~\ref{lem:pi_from_rho}, equations~\eqref{eq:rho_ij_general} imply $({\mathbf D}_i + {\mathbf b}{\mathbf 1}_{J-1}^T)\boldsymbol{\pi}_i = {\mathbf b}$. Based on the Sherman-Morrison-Woodbury formula \citep{golub2013}, $({\mathbf D}_i + {\mathbf b} {\mathbf 1}_{J-1}^T)^{-1}$ exists if ${\mathbf D}_i^{-1}$ exists and $1 + {\mathbf 1}_{J-1}^T {\mathbf D}_i^{-1} {\mathbf b} \neq 0$. On the other hand, $\pi_{iJ} = (1 + {\mathbf 1}_{J-1}^T {\mathbf D}_i^{-1} {\mathbf b})^{-1} > 0$ according to Lemma~\ref{lem:pi_from_rho}. We may add the following assumption to ensure the existence of $\boldsymbol{\pi}_i$ (not necessarily in $\boldsymbol{\Pi}_0$ yet):

\begin{assumption}\label{as:Di_inverse}
Given any $\rho_{ij}\in (0,1)$, $j=1, \ldots, J-1$, we always have ${\mathbf D}_i^{-1}$ exists and $1 + {\mathbf 1}_{J-1}^T {\mathbf D}_i^{-1} {\mathbf b} > 0$.    
\end{assumption}

With Assumption~\ref{as:Di_inverse}, we can solve $\boldsymbol{\pi}_i$ uniquely via \eqref{eq:pi_from_rho}. To ensure that $\boldsymbol{\pi}_i \in \boldsymbol{\Pi}_0$~, we still need the following assumption:

\begin{assumption}\label{as:Di_b>0}
Given any $\rho_{ij}\in (0,1)$, $j=1, \ldots, J-1$, all the $J-1$ coordinates of ${\mathbf D}_i^{-1} {\mathbf b}$ are strictly positive.    
\end{assumption}

\begin{theorem}\label{thm:feasibility_with_assumptions}
Consider the multinomial link model~\eqref{eq:mlm_in_matrix} or \eqref{eq:mlm_j}. If ${\mathbf L}, {\mathbf R}, {\mathbf b}$ satisfy Assumptions~\ref{as:A1}, \ref{as:1_b>0}, and \ref{as:Di_inverse}, then ${\mathbf D}_i^{-1}$ always exists and the feasible parameter space $\boldsymbol{\Theta} = \{ \boldsymbol{\theta} \in \mathbb{R}^p \mid \mbox{all the } J-1 \mbox{ coordinates of }{\mathbf D}_i^{-1}{\mathbf b}\mbox{ are positive, }i=1,\ldots,m\}$. If furthermore ${\mathbf L}, {\mathbf R}, {\mathbf b}$ satisfy Assumption~\ref{as:Di_b>0}, then $\boldsymbol{\Theta} = \mathbb{R}^p$. In both cases, $\boldsymbol{\pi}_i$ can be solved uniquely via \eqref{eq:pi_from_rho}.    
\end{theorem}

With the aid of Theorem~\ref{thm:feasibility_with_assumptions}, we can justify the feasibility of any parameters for many models proposed in this paper. 

\begin{theorem}\label{thm:models_assumption_1234}
For baseline-category mixed-link models, adjacent-categories mixed-link models, continuation-ratio mixed-link models, baseline-adjacent (two-group) mixed-link models, baseline-continuation (two-group) mixed-link models, and dichotomous conditional link models, Assumptions~\ref{as:A1}, \ref{as:1_b>0}, \ref{as:Di_inverse}, and \ref{as:Di_b>0} are all satisfied, and thus $\boldsymbol{\Theta} = \mathbb{R}^p$.     
\end{theorem}

Whenever a cumulative component gets involved in a multinomial link model, the feasibility of parameters becomes an issue and must be examined in practice. The following theorem provides simplified conditions for cumulative-related models proposed in this paper.

\begin{theorem}\label{thm:cumulative_models_assumption_1234}
For multinomial link models involving cumulative components, we have the following results:
\begin{itemize}
\item[(i)] For cumulative mixed-link models, Assumptions~\ref{as:A1}, \ref{as:1_b>0}, and \ref{as:Di_inverse} are satisfied, and $\boldsymbol\Theta = \{ \boldsymbol\theta \in \mathbb{R}^p \mid \rho_{i1} < \cdots < \rho_{i,J-1}, i=1, \ldots, m\}$, where $\rho_{ij} = g_j^{-1}({\mathbf f}_j^T({\mathbf x}_i) \boldsymbol\theta)$ in general. 
\item[(ii)] For baseline-cumulative (two-group) mixed-link models with $s=J$, Assumptions~\ref{as:A1}, \ref{as:1_b>0}, and \ref{as:Di_inverse} are satisfied, and $\boldsymbol\Theta = \{ \boldsymbol\theta \in \mathbb{R}^p \mid \rho_{i, k+1} < \cdots < \rho_{i,J-1}, i=1, \ldots, m\}$. 
\item[(iii)] For baseline-cumulative (two-group) mixed-link models with $s\neq J$, Assumptions~\ref{as:A1} and \ref{as:1_b>0} are satisfied, ${\mathbf D}_i^{-1}$ exists, and $\boldsymbol\Theta = \{ \boldsymbol\theta \in \mathbb{R}^p \mid \rho_{i, k+1} < \cdots < \rho_{i,J-1}, i=1, \ldots, m\}$. 
\end{itemize}
\end{theorem}

Since Assumptions~\ref{as:A1}, \ref{as:1_b>0}, \ref{as:Di_inverse}, and \ref{as:Di_b>0} are all about ${\mathbf L}, {\mathbf R}, {\mathbf b}$, and no restrictions on linear predictors are posted, the conclusions in  Theorems~\ref{thm:models_assumption_1234} and \ref{thm:cumulative_models_assumption_1234} are applicable for both ppo models and po-npo mixture models (see Example~\ref{ex:po_ppo_mixture}).

On the other hand, from Theorem~\ref{thm:cumulative_models_assumption_1234} we can see that some multinomial link models do not satisfy Assumptions~\ref{as:Di_inverse} or \ref{as:Di_b>0}, we relax them into the following assumption with the notation:
\[
{\mathbf P}_0 = \left\{(\rho_1, \ldots, \rho_{J-1}) \in \mathbb{R}^{J-1} \mid \rho_j = \frac{{\mathbf L}_j^T \boldsymbol{\pi}}{{\mathbf R}_j^T \boldsymbol{\pi} + (1-{\mathbf 1}_{J-1}^T \boldsymbol{\pi}) b_j}, j=1, \ldots, J-1; \boldsymbol{\pi}\in \boldsymbol{\Pi}_0\right\}
\]
for given ${\mathbf L}, {\mathbf R}$ and ${\mathbf b}$.

\begin{assumption}\label{as:P_0_to_Pi_0_assumption}
Given any $\boldsymbol{\rho}_i \in {\mathbf P}_0$~, ${\mathbf D}_i$ is invertible and all the $J-1$ coordinates of ${\mathbf D}_i^{-1}{\mathbf b}$ are strictly positive.
\end{assumption}

If a multinomial link model satisfies Assumptions~\ref{as:A1}, \ref{as:1_b>0}, \ref{as:Di_inverse} and \ref{as:Di_b>0}, it must satisfy Assumption~\ref{as:P_0_to_Pi_0_assumption}, which is designed for cumulative-related multinomial link models.

\begin{lemma}\label{lem:cumulative_assumption_5}
Both cumulaive mixed-link models and baseline-cumulative (two-group) mixed-link models satisfy Assumption~\ref{as:P_0_to_Pi_0_assumption}.    
\end{lemma}    

Once the multinomial link model satisfies Assumptions~\ref{as:A1}, \ref{as:1_b>0}, and \ref{as:P_0_to_Pi_0_assumption}, we are able to develop  algorithms that guarantee to find feasible parameter estimates (see algorithms in Section~\ref{sec:formulae_and_algorithms}). For general multinomial link models, one can always use \eqref{eq:Theta_in_general} to validate the feasibility of $\boldsymbol{\theta}$.

\section{Information Matrix and Model Selection}\label{sec:parameter_fisher_information}

\subsection{Fisher information matrix}
\label{sec:fisher_information}

There are many reasons that we need to calculate the Fisher information matrix ${\mathbf F}(\boldsymbol{\theta})$, for examples, when finding the maximum likelihood estimate (MLE) $\hat{\boldsymbol\theta}$ of $\boldsymbol{\theta}$ using the Fisher scoring method (see Section~\ref{sec:fisher_scoring}), constructing confidence intervals of $\boldsymbol\theta$ (see Section~\ref{sec:confidence_test}), or finding optimal designs of experiment \citep{atkinson2007,bu2020}. Inspired by Theorem~2.1 in \cite{bu2020} for multinomial logistic models~\eqref{eq:logitunifiedmodel}, in this section, we provide explicit formulae for calculating ${\mathbf F}(\boldsymbol{\theta})$, $\boldsymbol{\theta}\in \boldsymbol{\Theta}$ for a general multinomial link model \eqref{eq:mlm_in_matrix} or \eqref{eq:mlm_j}.

Suppose for distinct ${\mathbf x}_i$~, $i=1,\cdots,m$, we have independent multinomial responses
${\mathbf Y}_i=(Y_{i1},\cdots,Y_{iJ})^T \sim {\rm Multinomial}(n_i; \pi_{i1},\cdots,\pi_{iJ})$,
where $n_i=\sum_{j=1}^J Y_{ij}$~. 
Then the log-likelihood for the multinomial model is
\begin{eqnarray*}
	l(\boldsymbol\theta) &=&\log \left(\prod_{i=1}^m \frac{{n_i}!}{{Y_{i1}}!\cdots{Y_{iJ}}!} \pi_{i1}^{Y_{i1}}\cdots\pi_{iJ}^{Y_{iJ}}\right)\\
    &=& \sum_{i=1}^m {\mathbf Y}_i^T\log \bar{\boldsymbol\pi}_i + \sum_{i=1}^m \log (n_i!) - \sum_{i=1}^m\sum_{j=1}^J \log (Y_{ij}!)\ ,
\end{eqnarray*}
where $\bar{\boldsymbol\pi}_i = (\pi_{i1}, \ldots, \pi_{iJ})^T = (\boldsymbol\pi_i^T, \pi_{iJ})^T$, $\log \bar{\boldsymbol\pi}_i=(\log \pi_{i1}, \cdots,\log \pi_{iJ})^T$.

Using matrix differentiation formulae (see, e.g., Chapter~17 in \cite{seber2008}), we obtain the score vector $\partial l/\partial \boldsymbol\theta^T$ and the Fisher information matrix for general model~\eqref{eq:mlm_in_matrix} as follows (see Appendix~\ref{subsec:supp_fisher_mat} for more details).

\begin{theorem}\label{thm:fisher_information}
	Consider the multinomial link model \eqref{eq:mlm_in_matrix} with distinct settings ${\mathbf x}_1, \ldots, {\mathbf x}_m$ and independent response observations. Suppose $\boldsymbol{\theta} \in \boldsymbol{\Theta}$ as defined in \eqref{eq:Theta_in_general}. Then the score vector
\begin{equation}\label{eq:partial_l_partial_theta}
	\frac{\partial l}{\partial \boldsymbol\theta^T} = \sum_{i=1}^{m}{\mathbf Y}_i^T{\rm diag}(\bar{\boldsymbol\pi}_i)^{-1}\frac{\partial \bar{\boldsymbol\pi}_i}{\partial \boldsymbol\theta^T}
\end{equation}
satisfying $E(\partial l/\partial \boldsymbol\theta^T)=0$, and the Fisher information matrix
\begin{equation}\label{eq:Fisher_F_i}
		{\mathbf F}=\sum_{i=1}^{m}n_i{\mathbf F}_i\ ,
	\end{equation}
where
\begin{eqnarray}
		{\mathbf F}_i &=& \left(\frac{\partial \bar{\boldsymbol\pi}_i}{\partial \boldsymbol\theta^T}\right)^T {\rm diag}(\bar{\boldsymbol\pi}_i)^{-1}\frac{\partial \bar{\boldsymbol\pi}_i}{\partial \boldsymbol\theta^T}\ ,\label{eq:F_i}\\
\frac{\partial \bar{\boldsymbol\pi}_i}{\partial \boldsymbol\theta^T}
	&=& {\mathbf E}_i {\mathbf D}_i^{-1} \cdot {\rm diag}\left({\mathbf L}\boldsymbol\pi_i\right) \cdot {\rm diag} \left(\boldsymbol\rho_i^{-2} \right) \cdot {\rm diag}\left(\left({\mathbf g}^{-1}\right)'\left(\boldsymbol\eta_i\right)\right) \cdot {\mathbf X}_i\ ,\label{eq:p_pi_p_theta}\\
{\mathbf E}_i &=& 
\left[\begin{array}{c}
	{\mathbf I}_{J-1}\\ {\mathbf 0}_{J-1}^T
\end{array}\right] -
\bar{\boldsymbol{\pi}}_i {\mathbf 1_{J-1}^T}\quad \in \mathbb{R}^{J\times (J-1)}\ ,\label{eq:E_i}
\end{eqnarray}
${\rm diag}(({\mathbf g}^{-1})'(\boldsymbol\eta_i)) = {\rm diag}\{(g_1^{-1})'\left(\eta_{i1}\right), \ldots, (g_{J-1}^{-1})'\left(\eta_{i, J-1}\right)\} \in \mathbb{R}^{(J-1)\times (J-1)}$, ${\mathbf I}_{J-1}$ is the identity matrix of order $J-1$, ${\mathbf 0}_{J-1}$ is a vector of zeros in $\mathbb{R}^{J-1}$, and ${\mathbf 1}_{J-1}$ is a vector of ones in $\mathbb{R}^{J-1}$. 
\hfill{$\Box$}
\end{theorem}
Theorem~\ref{thm:fisher_information} covers the conclusion of Theorem~2.1 in \cite{bu2020} as a special case.

\subsection{Positive definiteness of Fisher information matrix}\label{sec:positive_definite}

In this section, we explore when the Fisher information matrix ${\mathbf F}$ is positive definite, which is critical not only for the existence of ${\mathbf F}^{-1}$, but also for the existence of unbiased estimates of 
a feasible parameter $\boldsymbol\theta$ with finite variance \citep{stoica2001} and relevant optimal design problems \citep{bu2020}.

To investigate the rank of ${\mathbf F}$, we denote a $J\times (J-1)$ matrix
\begin{equation}\label{eq:C_i}
{\mathbf C}_i = \left({\mathbf c}_{i1}, \ldots, {\mathbf c}_{i,J-1}\right) = 	{\mathbf E}_i {\mathbf D}_i^{-1} \cdot {\rm diag}\left({\mathbf L}\boldsymbol\pi_i\right) \cdot {\rm diag} \left(\boldsymbol\rho_i^{-2} \right) \cdot {\rm diag}\left(\left({\mathbf g}^{-1}\right)'\left(\boldsymbol\eta_i\right)\right) \ ,
\end{equation}
where ${\mathbf c}_{i1}, \ldots, {\mathbf c}_{i,J-1} \in \mathbb{R}^J$ are column vectors. Then $\partial \bar{\boldsymbol\pi}_i/\partial \boldsymbol\theta^T = {\mathbf C}_i {\mathbf X}_i$ according to \eqref{eq:p_pi_p_theta}. We further denote a $(J-1)\times (J-1)$ matrix ${\mathbf U}_i = (u_{st}(\boldsymbol\pi_i))_{s,t=1,\ldots, J-1} = {\mathbf C}_i^T {\rm diag} (\bar{\boldsymbol\pi}_i)^{-1} {\mathbf C}_i$~, whose $(s,t)$th entry $u_{st}(\boldsymbol\pi_i) = {\mathbf c}_{is}^T {\rm diag} (\bar{\boldsymbol\pi}_i)^{-1} {\mathbf c}_{it}$~. Then ${\mathbf F}_i = {\mathbf X}_i^T {\mathbf U}_i {\mathbf X}_i$ according to Theorem~\ref{thm:fisher_information}.

\begin{lemma}\label{lem:Fi}
Suppose $\boldsymbol{\theta} \in \boldsymbol{\Theta}$.
Then ${\rm rank}({\mathbf U}_i) = J-1$ and ${\rm rank}({\mathbf F}_i) = {\rm rank}({\mathbf X}_i)$.
\hfill{$\Box$}
\end{lemma}

We further define an $m(J-1)\times m(J-1)$ matrix ${\mathbf U} = ({\mathbf U}_{st})_{s,t=1,\ldots, J-1}$ with ${\mathbf U}_{st}= {\rm diag}\{n_1 u_{st}({\boldsymbol\pi}_1), \ldots, n_m u_{st}({\boldsymbol\pi}_m)\}$.
Recall that the model matrix for a general multinomial link model \eqref{eq:mlm_in_matrix} is
\begin{equation}\label{eq:general_X_i}
{\mathbf X}_i = ({\mathbf f}_1({\mathbf x}_i), \ldots, {\mathbf f}_{J-1} ({\mathbf x}_i))^T = (f_{jl}({\mathbf x}_i))_{j=1, \ldots, J-1; l=1, \ldots, p}\ .
\end{equation}

To explore the positive definiteness of ${\mathbf F}$, we define a $p\times m(J-1)$ matrix
\begin{equation}\label{eq:H_general}
{\mathbf H} = ({\mathbf f}_1({\mathbf x}_1), \ldots, {\mathbf f}_1({\mathbf x}_m), \ldots, {\mathbf f}_{J-1}({\mathbf x}_1), \ldots, {\mathbf f}_{J-1}({\mathbf x}_m)) = 
\left[\begin{array}{ccc}
{\mathbf F}_{11}^T & \cdots & {\mathbf F}_{J-1,1}^T\\
\vdots & \cdots & \vdots\\
{\mathbf F}_{1p}^T & \cdots & {\mathbf F}_{J-1,p}^T
\end{array}\right]  \ ,  
\end{equation}
where ${\mathbf F}_{jl} = (f_{jl}({\mathbf x}_1), \ldots, f_{jl}({\mathbf x}_m))^T \in \mathbb{R}^m$.

\begin{example}\label{ex:ppo_model_H} {\bf General ppo model}\quad {\rm
For ppo models including Examples~\ref{ex:mixed_link_ppo} and \ref{ex:two_group_s},
\[
{\mathbf H}=
\left(\begin{array}{ccc}
{\mathbf H}_{1} & &\\  & \ddots & \\& &{\mathbf H}_{J-1} \\{\mathbf H}_{c} & \cdots & {\mathbf H}_{c}
\end{array}\right) \>\in \> \mathbb{R}^{p\times m(J-1)}\ ,
\]
where ${\mathbf H}_{j}= ({\mathbf h}_j({\mathbf x}_1), \cdots, {\mathbf h}_j({\mathbf x}_m)) \in \mathbb{R}^{p_j \times m}$, and ${\mathbf H}_{c} = ({\mathbf h}_c({\mathbf x}_1), \cdots, {\mathbf h}_c({\mathbf x}_m)) \in \mathbb{R}^{p_c \times m}$.
}\hfill{$\Box$}
\end{example}

\medskip
\noindent
{\bf Example~\ref{ex:po_ppo_mixture}}. {\it (continued)}\quad For po-npo mixture models, 
\[
{\mathbf H}=
\left(\begin{array}{ccc}
{\mathbf H}_{1} & &\\  & \ddots & \\& &{\mathbf H}_{J-1} \\{\mathbf H}_{c1} & \cdots & {\mathbf H}_{c,J-1}
\end{array}\right) \>\in \> \mathbb{R}^{p\times m(J-1)}\ ,
\]
with ${\mathbf H}_{j}= ({\mathbf h}_j({\mathbf x}_1), \ldots, {\mathbf h}_j({\mathbf x}_m)) \in \mathbb{R}^{p_j \times m}$ and ${\mathbf H}_{cj} = ({\mathbf h}_{cj}({\mathbf x}_1), \ldots, {\mathbf h}_{cj}({\mathbf x}_m)) \in \mathbb{R}^{p_c \times m}$. 
\hfill{$\Box$}

\begin{theorem}\label{thm:F=HUH^T}
For the multinomial link model~\eqref{eq:mlm_in_matrix} with $n_i>0$ independent observations at distinct setting ${\mathbf x}_i$, $i=1, \ldots, m$, its Fisher information matrix ${\mathbf F} = {\mathbf H} {\mathbf U} {\mathbf H}^T$. Since $(g_j^{-1})'(\eta_{ij}) \neq 0$ for all $i=1, \ldots, m$ and $j=1, \ldots, J-1$, then ${\mathbf F}$ at a feasible parameter vector $\boldsymbol{\theta}$ is positive definite if and only if ${\mathbf H}$ is of full row rank.
\end{theorem}

According to Theorem~\ref{thm:F=HUH^T}, the positive definiteness of ${\mathbf F}$ at a feasible $\boldsymbol{\theta}$ depends only on the predictor functions and the distinct settings ${\mathbf x}_1, \ldots, {\mathbf x}_m$~. From an experimental design point of view, one needs to collect observations from a large enough set $\{{\mathbf x}_1, \ldots, {\mathbf x}_m\}$ of distinct experiments settings. From a data analysis point of view, given the data with the set $\{{\mathbf x}_1, \ldots, {\mathbf x}_m\}$ of distinct settings, there is an upper bound of model complexity, beyond which not all parameters are estimable with a finite variance.

\subsection{Confidence intervals and hypothesis tests for parameters}\label{sec:confidence_test}

In this paper, we use maximum likelihood for estimating $\boldsymbol\theta$. That is, we look for $\hat{\boldsymbol\theta}$ which maximizes the likelihood function $L(\boldsymbol\theta)$ or the log-likelihood function $l(\boldsymbol\theta) = \log L(\boldsymbol\theta)$, known as the maximum likelihood estimate (MLE, see Section~1.3.1 in \cite{agresti2013} for justifications on adopting MLE).

Under regularity conditions (see, e.g., Section~5f in \cite{rao1973linear} or Chapter~18 in \cite{ferguson1996course}), asymptotically  $\hat{\boldsymbol\theta}$ is unbiased, normally distributed with
${\rm Cov}(\hat{\boldsymbol\theta}) = {\mathbf F}(\boldsymbol{\theta})^{-1}$.
Denoting ${\mathbf F}(\hat{\boldsymbol\theta})^{-1} = \left(\hat\sigma_{ij}\right)_{i,j=1,\ldots, p}$~, we construct approximate confidence intervals
$
\theta_i \in (\hat{\theta}_i - z_{\alpha/2} \sqrt{\hat{\sigma}_{ii}}, \ \hat{\theta}_i + z_{\alpha/2} \sqrt{\hat{\sigma}_{ii}})
$
with $\boldsymbol\theta = (\theta_1, \ldots, \theta_p)^T$ and $\hat{\boldsymbol\theta} = (\hat{\theta}_1, \ldots, \hat{\theta}_p)^T$, where $z_{\alpha/2}$ is the upper $(\alpha/2)$th quantile of $N(0,1)$ (see Section~\ref{sec:MLM_example_fly} for a real data example). 

To test $H_0: \boldsymbol\theta = \boldsymbol\theta_0$~, we may use Wald statistic \citep{wald1943tests, agresti2013}, which asymptotically is
$
W = (\hat{\boldsymbol\theta} - \boldsymbol\theta_0)^T {\mathbf F}(\hat{\boldsymbol\theta}) (\hat{\boldsymbol\theta} - \boldsymbol\theta_0) \stackrel{H_0}{\sim} \chi^2_p
$~.

Suppose $\boldsymbol\theta = (\boldsymbol\theta_1^T, \boldsymbol\theta_2^T)^T$ with $\boldsymbol\theta_1 \in \mathbb{R}^{r}$ and $\boldsymbol\theta_2 \in \mathbb{R}^{p-r}$. To test $H_0: \boldsymbol\theta_1 = {\mathbf 0}_r$, we may use the likelihood-ratio test \citep{wilks1935likelihood, wilks1938large, agresti2013} with the test statistic
\[
\Lambda = -2 \log \frac{\max_{\boldsymbol\theta_2} L(\boldsymbol\theta_1 = {\mathbf 0}_r, \boldsymbol\theta_2)}{\max_{\boldsymbol\theta} L(\boldsymbol\theta)} \stackrel{H_0}{\sim} \chi^2_r
\]
asymptotically. It may be used before removing more than one predictors simultaneously for variable selection purposes.

\subsection{Model selection}\label{sec:model_selection}

Given data $({\mathbf x}_i, {\mathbf y}_i), i=1,\cdots,m$, where ${\mathbf y}_i=(y_{i1},\cdots,y_{iJ})^T$ satisfying 
$n_i=\sum_{j=1}^J y_{ij}$~. Suppose the MLE $\hat{\boldsymbol\theta} \in \mathbb{R}^p$ has been obtained. Following Lemma~\ref{lem:pi_from_rho}, we obtain $\hat\pi_{ij} = \pi_{ij}(\hat{\boldsymbol\theta})$, $i=1, \ldots, m$, $j=1, \ldots, J$. Then the maximized log-likelihood
\[
	l(\hat{\boldsymbol\theta}) = \sum_{i=1}^m \log (n_i!) + \sum_{i=1}^m {\mathbf y}_i^T\log \hat{\bar{\boldsymbol\pi}}_i  - \sum_{i=1}^m\sum_{j=1}^J \log (y_{ij}!)\ ,
\]
where $\log \hat{\bar{\boldsymbol\pi}}_i=(\log \hat\pi_{i1}, \cdots,\log \hat\pi_{iJ})^T$. We may use $\AIC$ or $\BIC$ to choose the most appropriate model (see, e.g., \cite{hastie2009elements}, for a good review). More specifically, $\AIC = -2l(\hat{\boldsymbol\theta}) + 2 p$, and $\BIC = -2l(\hat{\boldsymbol\theta}) + (\log n)p$, where $n = \sum_{i=1}^m n_i$~. A smaller $\AIC$ or $\BIC$ value indicates a better model \citep{burnham2004aic}.

\section{Algorithms and Comparison Study}
\label{sec:formulae_and_algorithms}

To facilitate the readers, we provide a summary of notations for specifying a multinomial link model in Appendix~\ref{sec:summary_notation}.
In this section, we provide detailed formulae and algorithms for finding a feasible MLE of $\boldsymbol{\theta}$ for a general multinomial link model \eqref{eq:mlm_in_matrix} or \eqref{eq:mlm_j}, given a dataset in its summarized form $\{({\mathbf x}_i, {\mathbf y}_i) \mid i=1, \ldots, m\}$, where ${\mathbf x}_i \in \mathbb{R}^d, i=1, \ldots, m$ are distinct settings, ${\mathbf y}_i = (y_{i1}, \ldots, y_{iJ})^T, i=1, \ldots, m$ are vectors of nonnegative integers with $\sum_{j=1}^J y_{ij} = n_i >0$, $i=1, \ldots, m$.
We also provide an algorithm for calculating the gradient $\partial l/\partial \boldsymbol\theta^T$ and Fisher information matrix ${\mathbf F}$ in Appendix~\ref{subsec:calculate_gradient_fisher}, and a backward selection algorithm for finding the most appropriate po-npo mixture model in Appendix~\ref{sec:po_npo_mixture_algorithm}.

\subsection{Fisher scoring method for estimating parameters}\label{sec:fisher_scoring}

For numerically finding the MLE $\hat{\boldsymbol\theta}$, we adopt the Fisher scoring method described, for examples, in  
\cite{osborne1992fisher} or Chapter~14 in \cite{lange2010numerical}. That is, if we have $\boldsymbol\theta^{(t)}$ at the $t$th iteration, we obtain
\[
\boldsymbol\theta^{(t+1)} = \boldsymbol\theta^{(t)} + \delta \left({\mathbf F}(\boldsymbol\theta^{(t)})\right)^{-1} \frac{\partial l}{\partial \boldsymbol\theta^T}(\boldsymbol\theta^{(t)})
\]
at the $(t+1)$th iteration, where $\delta \in (0,1]$ is a  step length that is chosen to let $l({\boldsymbol\theta}^{(t+1)}) > l({\boldsymbol\theta}^{(t)})$ and $\pi_{ij}(\boldsymbol\theta^{(t+1)}) \in (0,1)$ for all $i=1, \ldots, m$ and $j=1, \ldots, J$, ${\mathbf F}(\boldsymbol\theta^{(t)})$ is the Fisher information matrix at $\boldsymbol\theta = \boldsymbol\theta^{(t)}$, and $(\partial l/\partial \boldsymbol\theta^T)(\boldsymbol\theta^{(t)})$ is expression~\eqref{eq:partial_l_partial_theta} evaluated at $\boldsymbol\theta = \boldsymbol\theta^{(t)}$. Theoretical justifications and more discussions on the Fisher scoring method can be found in \cite{osborne1992fisher}, \cite{lange2010numerical}, and references therein.

\begin{algorithm} Fisher scoring algorithm for finding a feasible MLE $\hat{\boldsymbol\theta}$ for model~\eqref{eq:mlm_in_matrix} or \eqref{eq:mlm_j}
\label{algo:fisher_scoring}
\begin{itemize}
    \item[$0^\circ$] Input: Data ${\mathbf x}_i = (x_{i1}, \ldots, x_{id})^T, {\mathbf y}_i = (y_{i1}, \ldots, y_{iJ})^T$, $i=1, \ldots, m$; the tolerance level of relative error $\epsilon > 0$ (e.g., $\epsilon = 10^{-6}$); and the step length $\delta \in (0,1)$ of linear search (e.g., $\delta = 0.5$).
    \item[$1^\circ$] Obtain a feasible initial parameter estimate  ${\boldsymbol\theta}^{(0)} \in \boldsymbol\Theta \subseteq \mathbb{R}^p$ (see Algorithms~\ref{algo:initial_theta} and \ref{algo:initial_theta_supplement}). Set $t=0$.
    \item[$2^\circ$] Given ${\boldsymbol\theta}^{(t)}$, calculate the gradient $\partial l/\partial \boldsymbol\theta^T|_{\boldsymbol{\theta}=\boldsymbol{\theta}^{(t)}}$ and the Fisher information matrix ${\mathbf F}({\boldsymbol\theta}^{(t)})$ (see Algorithm~\ref{algo:fisher_matrix} in Appendix~\ref{subsec:calculate_gradient_fisher}). 
    \item[$3^\circ$] Set the initial power index $q=0$ for step length; calculate the maximum change 
    \[
     \Delta \boldsymbol\theta = {\mathbf F}(\boldsymbol\theta^{(t)})^{-1} \left.\frac{\partial l}{\partial \boldsymbol\theta^T}\right|_{\boldsymbol\theta = \boldsymbol\theta^{(t)}}
    \]
    and its Euclidean length $\|\Delta\boldsymbol\theta\|$.
    \item[$4^\circ$] Calculate a candidate for the next parameter estimate  $\boldsymbol\theta^* = {\boldsymbol\theta}^{(t)} + \delta^q\cdot \Delta \boldsymbol\theta$.
    \item[$5^\circ$] If $\delta^q \|\Delta\boldsymbol\theta\|/\max\{1, \|{\boldsymbol\theta}^{(t)}\|\} < \epsilon$, go to Step~$7^\circ$;\\
    else if $\boldsymbol\theta^* \notin \boldsymbol\Theta$, then replace $q$ with $q+1$ and go back to Step~$4^\circ$; \\
    else if $[l(\boldsymbol\theta^*)-l(\boldsymbol\theta^{(t)})]/\max\{1, |l(\boldsymbol\theta^{(t)})|\} < \epsilon$, then replace $q$ with $q+1$ and go back to Step~$4^\circ$.
    \item[$6^\circ$] Let $\boldsymbol\theta^{(t+1)} = \boldsymbol\theta^*$, replace $t$ with $t+1$, and go back to Step~$2^\circ$. 
    \item[$7^\circ$] Output $\hat{\boldsymbol\theta} = \boldsymbol\theta^{(t)}$ as the MLE of $\boldsymbol\theta$.
\end{itemize}
\end{algorithm}

In practice, it is sometimes tricky to calculate ${\mathbf F}(\boldsymbol\theta^{(t)})^{-1}$ numerically while keeping its positive definiteness, especially when some eigenvalue of ${\mathbf F}(\boldsymbol\theta^{(t)})$ is tiny, which should be continuously monitored to retain its numerical positive definiteness. Following a commonly used trust-region strategy (see, e.g., Section~4.4 in \cite{gill1981practical}), when the minimum eigenvalue of ${\mathbf F}(\boldsymbol\theta^{(t)})$ is less than a predetermined threshold $\lambda_0>0$, such as $10^{-6}$, we replace ${\mathbf F}(\boldsymbol\theta^{(t)})$ with ${\mathbf F}(\boldsymbol\theta^{(t)})+\lambda\mathbf{I}_p$, where $\lambda=\lambda_0-\min\{\textrm{eigenvalues of } {\mathbf F}(\boldsymbol\theta^{(t)})\}$, and $\mathbf{I}_p$ is an identity matrix of order $p$. Based on our experience, such a strategy works well even if the minimum eigenvalue of ${\mathbf F}(\boldsymbol\theta^{(t)})$ is numerically negative, which is possible in practice.

\subsection{Finding a feasible MLE}\label{sec:initial_parameter}

It can be verified that Algorithm~\ref{algo:fisher_scoring} is valid. First of all, according to, for example, Section~14.3 of \cite{lange2010numerical}, $l(\boldsymbol\theta^*) > l(\boldsymbol\theta^{(t)})$ for large enough $q$ or small enough $\delta^q$, if ${\mathbf F}(\boldsymbol\theta^{(t)})$ is positive definite. Secondly, since $\boldsymbol\Theta$ is open (see Section~\ref{sec:feasibility_parameter}), $\boldsymbol\theta^{(t)} \in \boldsymbol\Theta$ must be an interior point, then $\boldsymbol\theta^* \in \boldsymbol\Theta$ for large enough $q$.
That is, if $\boldsymbol{\theta}^{(t)}$ is feasible, then $\boldsymbol{\theta}^{(t+1)}$ is feasible as well. The remaining task is to find a feasible $\boldsymbol{\theta}^{(0)}$.

Step~1 of Algorithm~\ref{algo:fisher_scoring} is critical and nontrivial for multinomial link models when a cumulative component is involved (see Section~\ref{sec:model_regularity}). In this section we first provide Algorithm~\ref{algo:initial_theta} for finding a possible initial estimate $\boldsymbol\theta^{(0)}$ of $\boldsymbol\theta$, which is expected not far away from the MLE $\hat{\boldsymbol{\theta}}$. 
One needs to use \eqref{eq:Theta_in_general} to check whether the $\boldsymbol\theta^{(0)}$ obtained by Algorithm~\ref{algo:initial_theta} is feasible. If not, we provide Algorithm~\ref{algo:initial_theta_supplement} to pull $\boldsymbol\theta^{(0)}$ back into the feasible domain $\boldsymbol{\Theta}$.

\begin{algorithm}
Finding an initial estimate of $\boldsymbol\theta$ for model~\eqref{eq:mlm_in_matrix} or \eqref{eq:mlm_j}
\label{algo:initial_theta}
\begin{itemize}
\item[$0^\circ$] Input: ${\mathbf x}_i = (x_{i1}, \ldots, x_{id})^T, {\mathbf y}_i = (y_{i1}, \ldots, y_{iJ})^T$, $i=1, \ldots, m$.
    \item[$1^\circ$] Calculate $n_i = \sum_{j=1}^J y_{ij} > 0$ and $\pi_{ij}^{(0)} = (y_{ij} + 1)/(n_i + J)$, $j=1, \ldots, J$, and denote $\boldsymbol\pi_i^{(0)} = (\pi_{i1}^{(0)}, \ldots, \pi_{i,J-1}^{(0)})^T \in \mathbb{R}^{J-1}$, $i=1, \ldots, m$.
    \item[$2^\circ$] Calculate $\rho_{ij}^{(0)} = ({\mathbf L}_j^T{\boldsymbol\pi}_i^{(0)})/({\mathbf R}_j^T \boldsymbol\pi_i^{(0)} + \pi_{iJ}^{(0)} b_j), j=1, \ldots, J-1$ and denote $\boldsymbol\rho_i^{(0)} = (\rho_{i1}^{(0)}, 
    \ldots, \rho_{i,J-1}^{(0)})^T$, $i=1, \ldots, m$.
    \item[$3^\circ$] Calculate $\eta_{ij}^{(0)} = g_j(\rho_{ij}^{(0)})$, $j=1, \ldots, J-1$ and denote $\boldsymbol\eta_i^{(0)} = (\eta_{i1}^{(0)}, \ldots, \eta_{i,J-1}^{(0)})^T \in \mathbb{R}^{J-1}$, $i=1, \ldots, m$.
    \item[$4^\circ$] Calculate ${\mathbf X}_i$ according to \eqref{eq:general_X_i}, $i=1, \ldots, m$, and  $\boldsymbol\theta^{(0)} = (\mathbb{X}^T\mathbb{X})^{-}\mathbb{X}^T\mathbb{Y}$, where \[
    \mathbb{X} = 
    \left[\begin{array}{c}
    {\mathbf X}_1 \\ \vdots \\ {\mathbf X}_m
    \end{array}\right] \in \mathbb{R}^{m(J-1) \times p},\> \> \>
    \mathbb{Y} =
    \left[\begin{array}{c}
    \boldsymbol\eta_1^{(0)} \\ \vdots \\ \boldsymbol\eta_m^{(0)}
    \end{array}\right] \in \mathbb{R}^{m(J-1)}\ , 
    \]
    and $(\mathbb{X}^T\mathbb{X})^{-}$ is the Moore–Penrose inverse of $\mathbb{X}^T\mathbb{X}$ (see, e.g., Section~7.4 of \cite{seber2008}), which is the same as $(\mathbb{X}^T\mathbb{X})^{-1}$ when it exists.
    \item[$5^\circ$] Report $\boldsymbol\theta^{(0)} \in \mathbb{R}^p$ as a possible initial parameter estimate.
\end{itemize}
\end{algorithm}

Essentially, Algorithm~\ref{algo:initial_theta} finds an initial estimate $\boldsymbol{\theta}^{(0)}$ of $\boldsymbol{\theta}$, which approximately leads to $\hat{\pi}_{ij} = (y_{ij}+1)/(n_i+J)$. Such a $\boldsymbol{\theta}^{(0)}$ is computational convenient but may not be feasible.

For typically applications, model~\eqref{eq:mlm_j} has an intercept for each $j$, that is, $f_{jl_j}({\mathbf x}_i)\equiv 1$ for some $l_j \in \{1, \ldots, p\}$, which indicates $\theta_{l_j}$ to be the intercept of the $j$th category. Typically $l_1, \ldots, l_{J-1}$ are distinct (otherwise, two categories share the same intercept), that is, $f_{j_1l_j}({\mathbf x}_i)\equiv 0$ for all $j_1\neq j$. In that case, we recommend the following algorithm to find a feasible initial estimate of $\boldsymbol{\theta}$.

\begin{algorithm}
Finding a feasible initial estimate of $\boldsymbol\theta$ for model~\eqref{eq:mlm_j} with intercepts
\label{algo:initial_theta_supplement}
\begin{itemize}
\item[$0^\circ$] Input: ${\mathbf x}_i = (x_{i1}, \ldots, x_{id})^T, {\mathbf y}_i = (y_{i1}, \ldots, y_{iJ})^T$, $i=1, \ldots, m$, an infeasible $\boldsymbol\theta^{(0)} = (\theta_1^{(0)}, \ldots, \theta_p^{(0)})^T$ obtained by Algorithm~\ref{algo:initial_theta}, and the step length $\delta=0.5$.
    \item[$1^\circ$] Calculate  $\pi_{j}^{(0)} = (\sum_{l=1}^m y_{lj} + m)/(n + mJ) \in (0,1)$, $j=1, \ldots, J$, and let $\boldsymbol\pi^{(0)} = (\pi_1^{(0)}, \ldots, \pi_{J-1}^{(0)})^T$.
    \item[$2^\circ$] Calculate $\rho_{j}^{(0)} = ({\mathbf L}_j^T{\boldsymbol\pi}^{(0)})/({\mathbf R}_j^T \boldsymbol\pi^{(0)} + \pi_J^{(0)} b_j), j=1, \ldots, J-1$. 
    \item[$3^\circ$] Calculate $\eta_j^{(0)} = g_j(\rho_j^{(0)})$, $j=1, \ldots, J-1$.
    \item[$4^\circ$] Denote $\boldsymbol{\theta}^{(00)} = (\theta_1^{(00)}, \ldots, \theta_p^{(00)})^T$ with $\theta_l^{(00)} = \eta_j^{(0)}$ if $l = l_j$ for $j=1, \ldots, J-1$; and $0$ otherwise. (According to Theorem~\ref{thm:validation_algorithm_3}, $\boldsymbol{\theta}^{(00)} \in \boldsymbol\Theta$, which is an open set in $\mathbb{R}^p$, if the model satisfies Assumptions~\ref{as:A1}, \ref{as:1_b>0}, and \ref{as:P_0_to_Pi_0_assumption}.)
    \item[$5^\circ$] Let $\Delta\boldsymbol{\theta}=\boldsymbol{\theta}^{(0)}-\boldsymbol{\theta}^{(00)}$, and $\boldsymbol{\theta}^{(0,q)}=\boldsymbol{\theta}^{(00)}+\delta^q \Delta\boldsymbol{\theta}$, $q=0, 1, \ldots$.
    \item[$6^\circ$] Let $q^*$ be the smallest $q$ such that $\boldsymbol{\theta}^{(0,q)} \in \boldsymbol\Theta$.
    \item[$7^\circ$] Report $\boldsymbol{\theta}^{(0,q^*)}$ as a feasible initial estimate of $\boldsymbol{\theta}$.
\end{itemize}
\end{algorithm}

\begin{theorem}\label{thm:validation_algorithm_3}
Suppose a multinomial link model~\eqref{eq:mlm_in_matrix} or \eqref{eq:mlm_j} satisfies Assumptions~\ref{as:A1}, \ref{as:1_b>0}, and \ref{as:P_0_to_Pi_0_assumption}. Assume further it has a distinct intercept for each $j$, that is, $f_{jl_j}({\mathbf x}_i)\equiv 1$ for some $l_j \in \{1, \ldots, p\}$ and $f_{j_1l_j}({\mathbf x}_i)\equiv 0$ for all $j_1\neq j$. Then $\boldsymbol{\theta}^{(00)}$ and the initial estimate of $\boldsymbol{\theta}$ reported by Algorithm~\ref{algo:initial_theta_supplement} must be feasible.
\end{theorem}

According to Theorem~\ref{thm:validation_algorithm_3}, Algorithm~\ref{algo:initial_theta_supplement} is especially useful for cumulaive mixed-link models and baseline-cumulative (two-group) mixed-link
models. 

Based on our experience, the provided algorithms in this section work well for all examples that we explore in this paper.

\subsection{Comparison study}\label{sec:MLM_existing_issue}

As mentioned at the beginning of Section~\ref{sec:feasible_parameter}, an infeasibility issue has been discovered in existing statistical software when fitting cumulative logit models. In this section, we use a comprehensive simulation study to compare the performance of SAS, R package {\tt VGAM}, and our algorithms on fitting cumulative logit models. 

A trauma clinical trial with $n=802$ of trauma patients was studied by \cite{chuang1997}. There are five ordered response categories, namely, Death, Vegetative state, Major disability, Minor disability, and Good recovery, known as the Glasgow Outcome Scale (GOS) in the literature of critical care \citep{jennett1975}. An extended dataset (Table~V in \cite{chuang1997}) consists of $802$ observations with two covariates, trauma severity ($x_1 \in \{0,1\}$) and dose level ($x_2 \in \{1,2,3,4\}$). A main-effects cumulative logit model \eqref{eq:cumulative_model} with po was applied to this dataset \citep{chuang1997}, where the logit link was assumed for all categories. 

Following \cite{huang2025constrained}, we bootstrap the extended dataset (Table~V in \cite{chuang1997}) for 1,000 times. For each of the 1,000 bootstrapped datasets, we fit the main-effects cumulative logit model with po using SAS {\tt proc} {\tt logistic} procedure, R package \texttt{VGAM}, and our algorithms, respectively. 

When using SAS {\tt proc} {\tt logistic} (SAS studio version 3.81), warning messages are displayed for 44 out of the 1,000 datasets, saying that negative individual predicted probabilities were identified. Under such a scenario, SAS still outputs the results from the last iteration, but with some negative fitted category probabilities (see Appendix~\ref{sec:existing_issue_subsection}).

The \texttt{vglm} function in R package \texttt{VGAM} (version 1.1-11) has a similar issue. Among the 1,000 bootstrapped datasets, 4 of them encounter errors with NA probabilities produced, and 38 have negative fitted probabilities. When NA probabilities are generated, the \texttt{vglm} function simply stops running without outputting fitted parameters. As for the 38 cases with negative probabilities, the \texttt{vglm} function still outputs fitted parameters, but without calculated log-likelihood, which is needed for obtaining $\AIC$ and $\BIC$ values.

On the contrary, our algorithms work fine by providing strictly positive fitted probabilities for all the 1,000 bootstrapped datasets, which imply feasible parameter estimates for all cases. For the cases when SAS or R package {\tt VGAM} still works, the fitted models obtained by our algorithms essentially match the results based on SAS or R (more technical details are provided in Appendix~\ref{sec:existing_issue_subsection}).

\section{Applications}
\label{sec:applications}

In this section, we use real data examples to show that the proposed multinomial link models can be significantly better than existing models and draw more reliable or more informative conclusions.

\subsection{Trauma clinical trial with mixed-link models}\label{sec:MLM_example_extrauma}

In this section, we revisit the trauma clinical trial considered in Section~\ref{sec:MLM_existing_issue} to show that the mixed-link models (Example~\ref{ex:mixed_link_ppo}) can be significantly better than traditional logit models.

In this study, for the extended trauma dataset (Table~V in \cite{chuang1997}), we allow separate links for different categories and consider main-effects mixed-link models with po. For illustration purposes, we use logit, probit, loglog, and cloglog as our candidate set of link functions. The best model that we find for this dataset applies log-log, probit, log-log, and logit links to $j=1,2,3,4$, respectively. This mixed-link model achieves $\BIC$ $198.43$, while the original main-effects cumulative logit model with po has a $\BIC$ value $252.74$. 

\begin{figure}[ht]
\centering
\includegraphics[scale=0.45]{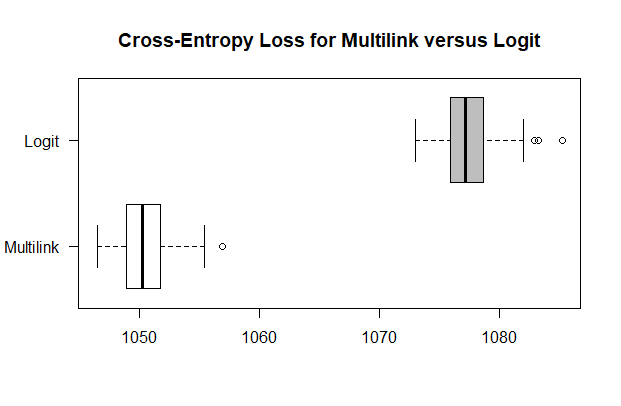}
\caption{Boxplots of Cross-Entropy Loss by 5-Fold Cross-Validations over $200$ Random Partitions for Modeling Extended Trauma Data}
\label{fig:box_multilink}
\end{figure}

To further show that the improvement is significant, we use five-fold cross-validations with cross-entropy loss \citep{hastie2009elements,dousti2023variable} over $200$ randomly generated partitions. As showed in Figure~\ref{fig:box_multilink}, our mixed-link model (or multi-link model) is significantly better than the original logit model in terms of prediction accuracy.

\subsection{Metabolic syndrome dataset with NA responses}\label{sec:metabolic_syndrome}

In this section, we use a metabolic syndrome dataset discussed by \cite{musa2023data} to illustrate that the proposed two-group models (see Example~\ref{ex:two_group_s}) can be used for analyzing data with missing categorical responses.

For this metabolic syndrome dataset, the goal is to explore the association between {\tt FBS} (fasting blood sugar) and three covariates, namely {\tt hpt} (hypertension status, yes or no), {\tt cholesterol} (total cholesterol, floored to 0,1,\ldots,23 in mmol/L), and {\tt weight} (body weight, floored to 30, 40,\ldots,190 in kilogram). In \cite{musa2023data}, the response {\tt FBS} was treated as a categorical variable with categories Normal (less than 6.1 mmol/L), IFG (Impaired Fasting Glucose, between 6.1 mmol/L and 6.9 mmol/L), DM (Diabetis Mellitus, 7.0 mmol/L or higher), as well as 251 NA's among the 4,282 observations. 

Having removed the observations with NA responses, a main-effects baseline-category logit model \eqref{eq:model_baseline} with npo was used in \cite{musa2023data} as an illustration. According to $\AIC$ (see Section~\ref{sec:model_selection}), the best main-effects multinomial logistic model without the NA category is actually a continuation-ratio logit model \eqref{eq:continuation_model} with npo and its natural order $\{$Normal, IFG, DM$\}$. We call it the {\it Model without NA} for this dataset.   

To check whether the conclusions are consistent if the 251 observations with NA responses are included, we look for the best model for all 4,282 observations. We first follow \cite{wang2023identifying} and use $\AIC$ to choose the most appropriate order for the four categories including NA, called a working order. The best main-effects model chosen by $\AIC$ is a continuation-ratio npo model with the working order $\{$Normal, IFG, DM, NA$\}$, whose $\AIC$ value is $930.40$ with the cross-entropy loss $3539.40$ based on a five-fold cross-validation \citep{hastie2009elements,dousti2023variable}.  For illustration purposes, we then find the best two-group model (Example~\ref{ex:two_group_s}) with npo  and logit link (that is, $g_1 = g_2 = g_3 = {\rm logit}$), which assumes a baseline-category sub-model \eqref{eq:model_baseline} on one group $\{$DM, IFG$\}$ and a continuation-ratio sub-model \eqref{eq:continuation_model} on the other group $\{$Normal, IFG, NA$\}$. It has $\AIC$ value $927.56$ and cross-entropy loss $3537.97$. According to \cite{burnham2004aic}, the chosen two-group model is significantly better than \cite{wang2023identifying}'s model with the working order $\{$Normal, IFG, DM, NA$\}$. We call the selected two-group model the {\it Model with NA} for this dataset.

\begin{figure}[ht]
\centering
\includegraphics[scale=0.4]{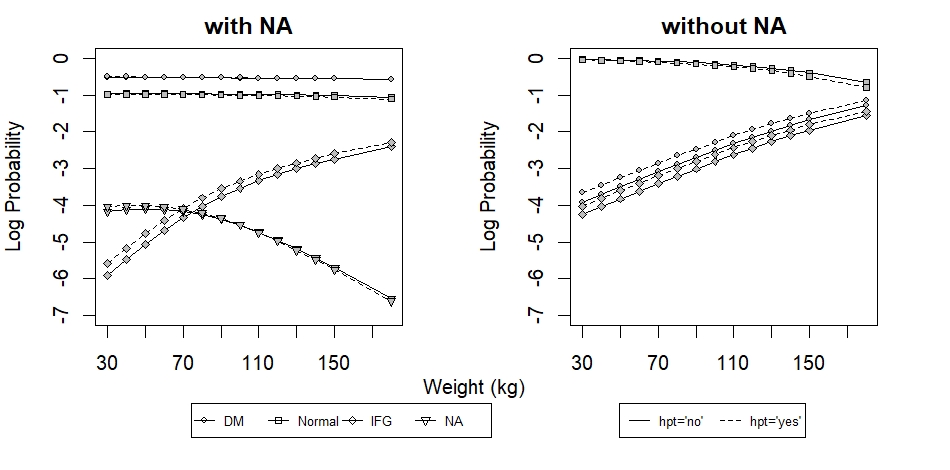}
\caption{Log-scale Categorical Probability against Weight Based on Models with or without NA Category for the Metabolic Syndrome Dataset}
\label{fig:logprob_vs_weight}
\end{figure}

\begin{figure}[ht]
\centering
\includegraphics[scale=0.4]{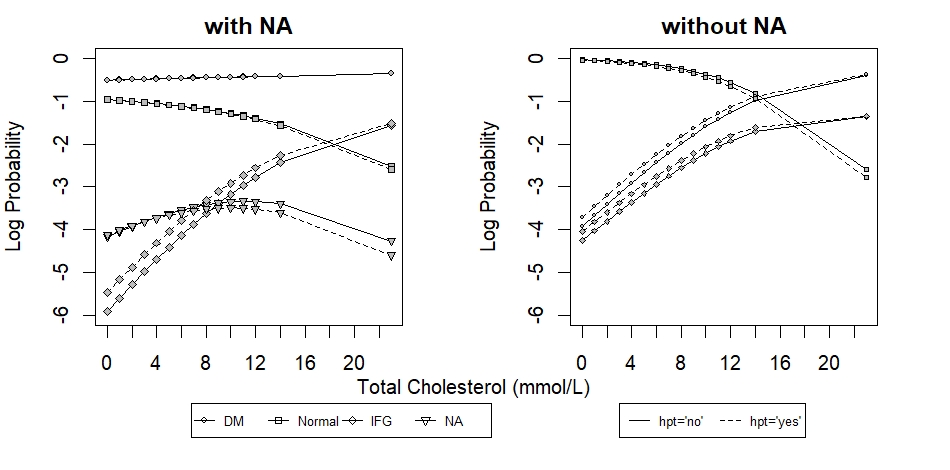}
\caption{Log-scale Categorical Probability against Cholesterol  Based on Models with or without NA Category for the Metabolic Syndrome Dataset}
\label{fig:logprob_vs_chol}
\end{figure}

Figure~\ref{fig:logprob_vs_weight} shows how $\log\hat\pi_{ij}$ changes against {\tt weight} based on the fitted Model with NA and Model without NA, respectively. When {\tt weight} increases, the probability of Normal or IFG category changes with a similarly pattern with or without NA. However, the patterns of DM category are quite different. Based on the fitted Model without NA, the conclusion is that the risk of DM increases significantly along with {\tt weight}; while according to the fitted Model with NA, the risk of DM is fairly flat and seems not so relevant to {\tt weight}. Similar inconsistency occurs as well for the risk of DM against {\tt cholesterol} with or without NA (see Figure~\ref{fig:logprob_vs_chol}). 

In other words, if we remove all observations with NA responses, we may conclude that {\tt cholesterol} and {\tt weight} heavily affect the risk of both IFG and DM; while with the complete data, their effects are still important on IFG, but not that important on DM. One possible explanation is that according to the Log Probability of NA against {\tt weight} (see Figure~\ref{fig:logprob_vs_weight}, left panel), the chance of NA clearly decreases as {\tt weight} increases. That is, the responses were not missing at random. In Appendix~\ref{sec:more_metabolic_syndrome_two_group_model}, we further apply Algorithm~\ref{algo:mixture_model} in Appendix~\ref{sec:po_npo_mixture_algorithm} to Models with or without NA, and obtain the most appropriate po-npo mixture models (see Section~\ref{sec:po_npo_mixture_model}), respectively. The corresponding conclusions are essentially the same.

\subsection{Six cities data with dichotomous conditional logit model}\label{sec:six_cities_cond_link_model}

In this section, we use a six cities data to show how the dichotomous conditional link model (Example~\ref{ex:binary_cond_link_model}) described in Section~\ref{sec:multinomial_conditional_link} works for longitudinal binary responses.

The six cities data, provided by \cite{fitzmaurice1993likelihood}, consist of observations of 537 children from Steubenville, Ohio. The only explanatory variable $x\in \{1,2\}$ indicates the mother's smoking habits ($1$ for non-smoking, $2$ for smoking) during the first year of the study. Four binary responses, denoted by $Z_1, Z_2, Z_3, Z_4 \in \{0,1\}$, indicate the presence of wheeze at ages 7, 8, 9, and 10 years, respectively, for the children under study. The goal is to study whether the maternal smoking habit increases the risk of the child's respiratory illness.

A mixed logistic model was used by \cite{zeger1988models} to model each $Z_i$'s with subject-specific random effects; a likelihood-based method was proposed by \cite{fitzmaurice1993likelihood} for modeling $Z_i$'s individually, along with assumptions on the associations between responses; and a series of multivariate logistic models were proposed by \cite{pmcc1995} and a final model was selected based on incremental deviances for the six cities data. All the three methods discovered some effect of smoking on the children's wheeze status, but none of them was statistically significant.

Following the description in Section~\ref{sec:multinomial_conditional_link}, we use a multinomial response ${\mathbf Y}_i = (Y_{i1}, \ldots,$ $ Y_{i,16})^T$ with the number of categories $J=16$ to represent the outcomes of the four binary variables. The categories $j=1, \ldots, 16$ correspond to  $(Z_{i1}, Z_{i2}, Z_{i3}, Z_{i4}) =(1,0,0,0)$, $(0,1,0,0)$, $(1,1,0,0)$, $\ldots, (1,1,1,1)$, $(0,0,0,0)$, 
respectively, along with the categorical probabilities $\pi_{i1}, \ldots, \pi_{i,16} \in (0,1)$ at $x_i \in \{1,2\}$.
Inspired by \cite{evans2013marginal}, we propose a dichotomous conditional link model with logit link for the six cities data (the corresponding $\mathbf{L}, \mathbf{R}, \mathbf{b}$ as in Model~\eqref{eq:mlm_in_matrix} are provided in Appendix~\ref{sec:more_six_cities_cond_link_model}). To find the best po-npo mixture model (Example~\ref{ex:po_ppo_mixture}) as described in Section~\ref{sec:po_npo_mixture_model}, we apply Algorithm~\ref{algo:mixture_model} in Appendix~\ref{sec:po_npo_mixture_algorithm} with intercepts included. The fitted parameters of the dichotomous conditional logit model with selected po-npo mixture are listed in Table~\ref{tab:six_cities_param:mixture}. Following R outputs, we use, for example, `***' to indicate that the $p$-value for the corresponding significance test (i.e., whether the estimated parameter is significantly different from zero) is between $0$ and $0.001$. The corresponding $\AIC$ value of our final model is $119.98$, which is significantly better than $127.83$ of the corresponding multivariate logistic model in \cite{pmcc1995}. 

More importantly, by allowing the regression coefficients of the smoking habit $x_i$ to be different across scenarios, our fitted model implies that the effect of maternal smoking habit varies across the age and medical history of the children. Furthermore, our fitted model (see Table~\ref{tab:six_cities_param:mixture}) indicates that such an effect is statistically significant at the age of 8, but not at 7; and its significance and magnitude also depend on the medical history of the children.

\begin{table}[h]    \centering
    \caption{Estimated Parameters for Six Cities Data under Dichotomous Conditional Logit Model}
    \label{tab:six_cities_param:mixture}
    \footnotesize
    \begin{tabular}{c|l|ll}
    \hline
     $j$ & Probability under Logit Transformation & Intercept & Smoking Habit\\
     & & $\hat\beta_{j1}$ & $\hat\beta_{j2}$\\
     \hline
     1 & $\text{logit}(P(Z_{i1}=1))$ & -1.611*** & 0\\
     2 & $\text{logit}(P(Z_{i2}=1|Z_{i1}=0))$ & -2.383*** & 0.555**\\
     3 & $\text{logit}(P(Z_{i2}=1|Z_{i1}=1))$ & -0.506** & 0.555**\\
     4 & $\text{logit}(P(Z_{i3}=1|Z_{i1}=0,Z_{i2}=0))$ & -2.383*** & 0\\
     5 & $\text{logit}(P(Z_{i3}=1|Z_{i1}=1,Z_{i2}=0))$ & -1.611*** & 0.555**\\
     6 & $\text{logit}(P(Z_{i3}=1|Z_{i1}=0,Z_{i2}=1))$ & -0.506** & 0.555**\\
     7 & $\text{logit}(P(Z_{i3}=1|Z_{i1}=1,Z_{i2}=1))$ & $\hspace{0.1cm}$0.671** & 0\\
     8 & $\text{logit}(P(Z_{i4}=1|Z_{i1}=0,Z_{i2}=0,Z_{i3}=0))$ & -3.100*** & 0\\
     9 & $\text{logit}(P(Z_{i4}=1|Z_{i1}=1,Z_{i2}=0,Z_{i3}=0))$ & -2.383*** & 1.539*\\
     10 & $\text{logit}(P(Z_{i4}=1|Z_{i1}=0,Z_{i2}=1,Z_{i3}=0))$ & -2.383*** & 0\\
     11 & $\text{logit}(P(Z_{i4}=1|Z_{i1}=1,Z_{i2}=1,Z_{i3}=0))$ & -1.611*** & 0.555**\\
     12 & $\text{logit}(P(Z_{i4}=1|Z_{i1}=0,Z_{i2}=0,Z_{i3}=1))$ & -1.611*** & 0\\
     13 & $\text{logit}(P(Z_{i4}=1|Z_{i1}=1,Z_{i2}=0,Z_{i3}=1))$ & -0.506** & 0\\
     14 & $\text{logit}(P(Z_{i4}=1|Z_{i1}=0,Z_{i2}=1,Z_{i3}=1))$ & -0.506** & 0.555**\\
     15 & $\text{logit}(P(Z_{i4}=1|Z_{i1}=1,Z_{i2}=1,Z_{i3}=1))$ & $\hspace{0.1cm}$0.671** & 0\\
    \hline
    \end{tabular}\\
    \normalsize
    \footnotesize{Notes: Signif. codes for $p$-value: 0 ‘***’ 0.001 ‘**’ 0.01 ‘*’ 0.05 ‘.’ 0.1 ‘ ’ 1}\\
\end{table}

\subsection{Police data with po-npo mixture model}\label{sec:police_po-npo_mixture_model}

In this section, we use a police data discussed by \cite{wang2023identifying} to show that the po-npo mixture model (see Example~\ref{ex:po_ppo_mixture}) can be significantly better than the traditional ppo models. 

The police data \citep{wang2023identifying} consist of summarized information of $n=12,483$ suspects' Armed status (gun, other, or unarmed), Gender (0 or 1), Flee (0 or 1), Mental illness (0 or 1), as well as the responses of police with four categories, Tasered, Shot, Shot \& Tasered, and Other, which do not have a natural ordering. According to \cite{wang2023identifying}, a continuation-ratio logit model~\eqref{eq:continuation_model} with npo and the working order $\{$Tasered, Shot, Other, Shot \& Tasered$\}$ chosen by $\AIC$ is significantly better than the baseline-category model~\eqref{eq:model_baseline} or multiclass logistic model. The best model reported in \cite{wang2023identifying} has the $\AIC$ value $192.01$.

To find the most appropriate po-npo mixture model, we run Algorithm~\ref{algo:mixture_model} including intercepts with iterations $t=1, \ldots, 6$. The corresponding $\AIC$ values after the six iterations are $190.04, 188.16, 186.49, $ $186.37, 186.27, 189.69$, respectively. Since the $6$th iteration leads to an increased $\AIC$ value, then we report the fitted po-npo mixture model right after the $5$th iteration. The fitted parameters are tabularized in Table~\ref{tab:param:mixture}.

\begin{table}[ht]
    \centering
    \caption{Estimated Parameters for Police Data under po-npo Mixture Model}
    \label{tab:param:mixture}
    \begin{tabular}{c|llllll}
    \hline
     $j$ & Intercept & Armed Status & Armed Status & Gender & Flee & Mental Illness \\
     & $\hat\beta_{j1}$ & $\hat\beta_{j2}$ (Other) & $\hat\beta_{j3}$ (Unarmed) & $\hat\beta_{j4}$ & $\hat\beta_{j5}$ & $\hat\beta_{j6}$\\
     \hline
     1 & -5.974*** & -0.624** & $\,\, $2.013*** & $\,\,$1.172*** & -20.892 & $\,\,$1.326***\\
     2 & $\,\,$5.908*** & -2.424*** & -1.086*** & $\,\,$0 & -1.555*** & -0.573***\\
     3 & $\,\,$0 & -0.624** & $\,\,$2.013*** & -2.581*** & -9.093 & $\,\,$1.326***\\
    \hline
    \end{tabular}\\
    \normalsize
    \footnotesize{Notes: Signif. codes for $p$-value: 0 ‘***’ 0.001 ‘**’ 0.01 ‘*’ 0.05 ‘.’ 0.1 ‘ ’ 1}\\
\end{table}

Compared with the $\AIC$ value $192.01$ of \cite{wang2023identifying}'s npo model, the $\AIC$ value $186.27$ of the reported po-npo mixture model is significantly better according to \cite{burnham2004aic}.

\subsection{House flies experiment with predictor selection}\label{sec:MLM_example_fly}

In this section, we use a house flies data discussed by \cite{atkinson1999} to show how the confidence intervals and hypothesis tests described in Section~\ref{sec:confidence_test} can be used for predictor or variable selection.

Reported by \cite{itepan1995}, the emergence of house flies data consist of summarized responses from $n=3,500$ pupae under a radiation experiment with the only covariate $x_i$~, Dose of radiation.  There are $J=3$ possible response categories, Unopened, Opened but died (before completing emergence), and Completed emergence, which have a nested or hierarchical structure.  

A continuation-ratio logit model was proposed by \cite{atkinson1999} for the emergence of house flies data as follows (see also \cite{bu2020}):
\begin{equation}\label{eq:dropx_atkinson}
\log\left(\frac{\pi_{i1}}{\pi_{i2} + \pi_{i3}}\right) = \beta_{11} + \beta_{12} x_i + \beta_{13} x_i^2, \>\>\>
\log\left(\frac{\pi_{i2}}{\pi_{i3}}\right) = \beta_{21} + \beta_{22} x_i\ ,
\end{equation}
where $x_i \in [80, 200]$ is the radiation level in Gy unit. By utilizing Algorithm~\ref{algo:fisher_scoring}, we obtain the fitted parameters $\hat{\boldsymbol\theta} = (\hat\beta_{11}, \hat\beta_{12}, \hat\beta_{13}, \hat\beta_{21}, \hat\beta_{22})^T = (-1.935, -0.02642, 0.0003174,$ $-9.159, 0.06386)^T$, which is consistent with the values reported in \cite{atkinson1999}.

As described in Section~\ref{sec:confidence_test}, we further compute the Fisher information matrix ${\mathbf F}(\hat{\boldsymbol{\theta}})$ and the confidence intervals of $\boldsymbol{\theta}$. We find out that only the $95\%$ confidence interval $(-0.0541, 0.0013)$ for $\beta_{12}$ contains $0$, which implies a reduced continuation-ratio model as follows:
\begin{equation}\label{eq:dropx_MLM}
\log\left(\frac{\pi_{i1}}{\pi_{i2} + \pi_{i3}}\right) = \beta_{11} + \beta_{13} x_i^2, \>\>\>
\log\left(\frac{\pi_{i2}}{\pi_{i3}}\right) = \beta_{21} + \beta_{22} x_i\ .
\end{equation}
It is significantly better than model~\eqref{eq:dropx_atkinson}. In terms of $\BIC$ values (see Section~\ref{sec:model_selection}), Model~\eqref{eq:dropx_MLM}'s $108.17$ is also better than Model~\eqref{eq:dropx_atkinson}'s $112.91$.

\section{Conclusion}
\label{sec:conc}

The proposed multinomial link models are not only more flexible than existing multinomial regression models and their extensions in the literature by allowing separate link functions (Example~\ref{ex:mixed_link_ppo}) and more flexible model structures (Example~\ref{ex:po_ppo_mixture}), but also include two brand-new classes of models. The two-group models (Example~\ref{ex:two_group_s}) allow users to incorporate NA or Unknown as a regular category into data analysis and can correct misleading conclusions due to missing-not-at-random. The  dichotomous conditional link models (Example~\ref{ex:binary_cond_link_model}) provide users with more powerful tools for analyzing longitudinal binary responses. Different from the multivariate logistic models \citep{pmcc1995}, a dichotomous conditional link model can follow the timeline more naturally when modeling longitudinal binary responses. It is also easier to interpret than mixed generalized linear models \citep{zeger1988models} for longitudinal responses.

The algorithms for finding the MLE and the most appropriate model are developed for the unified form of model~\eqref{eq:mlm_in_matrix} or \eqref{eq:mlm_j}, and are therefore applicable for all multinomial link models. It supports model selection among a fairly general family of multinomial regression models, and facilitates users to analyze categorical data with a handy toolbox.

The proposed algorithms with theoretical justifications solve the infeasibility issue of cumulative link models in existing statistical software. We also provide easy-to-use conditions~\eqref{eq:Theta_in_general} for general multinomial link models and simplified conditions (see Theorem~\ref{thm:cumulative_models_assumption_1234}) for cumulative-related models, which cover the classical cumulative link models as a special case.

%%%%%%%%%%%%%%%%%%%%%%%%%%%%%%%%%%%%%%%%%%%%%%
%% Example with multiple Appendixes:        %%
%%%%%%%%%%%%%%%%%%%%%%%%%%%%%%%%%%%%%%%%%%%%%%
\appendix
\renewcommand{\thesection}{\Alph{section}.\arabic{section}}
\setcounter{section}{0}

\begin{appendices}
\section{List of notations in the main text}\label{subsec:list}

\begin{list}{}{
		\setlength{\labelwidth}{0.2in}
		\setlength{\leftmargin}{0.6in}
		\setlength{\labelsep}{.3in}
		\setlength{\rightmargin}{0in}
	}
\item[$\circ$\hfill] Element-wise product, also known as Hadamard product, e.g., if ${\mathbf A = (a_{ij})_{ij}}$ and ${\mathbf B} = (b_{ij})_{ij}$, then $A\circ B = (a_{ij}b_{ij})_{ij}$
\item[$\oslash$\hfill] Element-wise division, also known as Hadamard division, e.g., if ${\mathbf A = (a_{ij})_{ij}}$ and ${\mathbf B} = (b_{ij})_{ij}$, then ${\mathbf A} \oslash {\mathbf B} = (a_{ij}/b_{ij})_{ij}$
\item[$\preceq$\hfill] Element-wise $\leq$, e.g., $0 \preceq (a_1, \ldots, a_n)^T$ if and only if $0\leq a_i$ for each $i$; and $(c_1, \ldots, c_n)^T \preceq (a_1, \ldots, a_n)^T$ if and only if $c_i\leq a_i$ for each $i$
\item[${\mathbf 0}_J$\hfill] Vector of $0$'s, ${\mathbf 0}_J = (0, \ldots, 0)^T \in \mathbb{R}^J$
\item[${\mathbf 0}_{k\times J}$\hfill] Matrix of $0$'s in $\mathbb{R}^{k\times J}$
\item[${\mathbf 1}_J$\hfill] Vector of $1$'s, ${\mathbf 1}_J = (1, \ldots, 1)^T \in \mathbb{R}^J$
\item[${\mathbf b}$\hfill] Constant vector, ${\mathbf b} = (b_1, \ldots, b_{J-1})^T \in \mathbb{R}^{J-1}$, see \eqref{eq:mlm_in_matrix} 
\item[$b_j$\hfill] The $j$th coordinate of ${\mathbf b}$, $j=1, \ldots, J-1$
\vskip 1mm
\item[$\bar{\mathbf C}$\hfill] $(2J-1)\times J$ constant matrix for multinomial logistic models, see \eqref{eq:logitunifiedmodel}
\item[${\mathbf C}_i$\hfill] $J\times (J-1)$ matrix at ${\mathbf x}_i$~, ${\mathbf C}_i = ({\mathbf c}_{i1}, \ldots, {\mathbf c}_{i,J-1}) = 	{\mathbf E}_i {\mathbf D}_i^{-1} \cdot {\rm diag}({\mathbf L}\boldsymbol\pi_i) \cdot {\rm diag} (\boldsymbol\rho_i^{-2}) \cdot {\rm diag}\left(({\mathbf g}^{-1})'(\boldsymbol\eta_i)\right)$, see \eqref{eq:C_i}, $i=1, \ldots, m$
\item[${\mathbf c}_{ij}$\hfill] The $j$th column vector of ${\mathbf C}_i$~, in $\mathbb{R}^J$, see \eqref{eq:C_i}, $i=1, \ldots, m$; $j=1, \ldots, J-1$
\vskip 1mm
\item[$d$\hfill] Total number of design factors, $d\geq 1$
\item[${\mathbf D}_i$\hfill] ${\mathbf D}_i = {\rm diag}(\boldsymbol\rho_i^{-1}) {\mathbf L} - {\mathbf R} \in \mathbb{R}^{(J-1)\times (J-1)}$, where ${\rm diag}(\boldsymbol\rho_i^{-1}) = {\rm diag}\{\rho_{i1}^{-1}, \ldots, \rho_{i,J-1}^{-1}\}$ in $\mathbb{R}^{(J-1)\times (J-1)}$, $i=1, \ldots, m$
\item[${\mathbf E}_i$\hfill] ${\mathbf E}_i = 
\left[\begin{array}{c}
	{\mathbf I}_{J-1}\\ {\mathbf 0}_{J-1}^T
\end{array}\right] -
\bar{\boldsymbol{\pi}}_i {\mathbf 1_{J-1}^T}\quad \in \mathbb{R}^{J\times (J-1)}$, see \eqref{eq:E_i}
\item[${\mathbf F}$\hfill] Fisher information matrix of the regression model with parameter(s) $\boldsymbol{\theta}$, ${\mathbf F} = {\mathbf F}(\boldsymbol{\theta}) = \sum_{i=1}^m n_i {\mathbf F}_i \in \mathbb{R}^{p\times p}$, see \eqref{eq:Fisher_F_i}
\item[${\mathbf F}_i$\hfill] Fisher information matrix at the $i$th covariate setting ${\mathbf x}_i$~, see \eqref{eq:F_i}
\item[${\mathbf f}_j$\hfill] Vector of $p$ predictor functions for the $j$th response category, e.g., ${\mathbf f}_j({\mathbf x}_i) = (f_{j1}({\mathbf x}_i),$ $\ldots, f_{jp}({\mathbf x}_i))^T$
\item[$f_{jl}$\hfill] The $l$th predictor function in ${\mathbf f}_j$~, $j = 1, \ldots, J-1$; $l=1, \ldots, p$
\item[${\mathbf F}_{jl}$\hfill] ${\mathbf F}_{jl} = (f_{jl}({\mathbf x}_1), \ldots, f_{jl}({\mathbf x}_m))^T \in \mathbb{R}^m$, $j=1, \ldots, J-1$ and $l=1, \ldots, p$, see \eqref{eq:H_general}
\item[$F_\nu$\hfill] Cumulative distribution function of $t$-distribution with the number $\nu$ of degrees of freedom, see Table~\ref{tab:link_functions}
\item[$f_\nu$\hfill] Probability density function of $t$-distribution with the number $\nu$ of degrees of freedom, see Table~\ref{tab:link_functions}
\item[${\mathbf g}$\hfill] Vector of $J-1$ link functions, ${\mathbf g} = (g_1, \ldots, g_{J-1})^T$, see \eqref{eq:mlm_in_matrix} and \eqref{eq:mlm_j}
\item[$g$\hfill] Link function, see \eqref{eq:clm}
\vskip 1mm
\item[$({\mathbf g}^{-1})'(\boldsymbol\eta_i)$\hfill] Element-wise operations, $({\mathbf g}^{-1})'\left(\boldsymbol\eta_i\right) = ((g_1^{-1})'(\eta_{i1}), \ldots, (g_{J-1}^{-1})'(\eta_{i,J-1}))^T$ in $\mathbb{R}^{J-1}$, $i=1, \ldots, m$, see \eqref{eq:p_pi_p_theta}
\item[$g_j$\hfill] Link function for the $j$th response category, see \eqref{eq:mlm_j}, assuming $g_j^{-1}$ exists and maps $\eta \in (-\infty, \infty)$ to $\rho \in (0,1)$
\item[${\mathbf H}$\hfill] $p\times m(J-1)$ matrix, 
see \eqref{eq:H_general}
\item[${\mathbf H}_c$\hfill] $p_c\times m$ matrix for the common component of $J-1$ categories in a general ppo model (see Example~\ref{ex:ppo_model_H}), ${\mathbf H}_c = ({\mathbf h}_c({\mathbf x}_1), \ldots, {\mathbf h}_c({\mathbf x}_m))$ 
\item[${\mathbf h}_c({\mathbf x}_i)$\hfill] Vector of $p_c$ predictors associated with the $p_c$ parameters ${\boldsymbol\zeta} = (\zeta_1, \ldots, \zeta_{p_c})^T$ that are common for response categories $1, \ldots, J-1$ at ${\mathbf x}_i$~, ${\mathbf h}_c({\mathbf x}_i) = (h_1({\mathbf x}_i), \ldots,$ $h_{p_c}({\mathbf x}_i))^T$, $h_l$'s are known predictor functions, $i=1, \ldots, m$; $l=1, \ldots, p_c$
\item[${\mathbf H}_{cj}$\hfill] $p_c\times m$ matrix in a po-ppo mixture model (see Example~\ref{ex:po_ppo_mixture}), ${\mathbf H}_{cj} = ({\mathbf h}_{cj}({\mathbf x}_1), \ldots,$ ${\mathbf h}_{cj}({\mathbf x}_m))$ 
\item[${\mathbf h}_{cj}({\mathbf x}_i)$\hfill] Vector of $p_c$ predictors associated with the $p_c$ parameters ${\boldsymbol\zeta} = (\zeta_1, \ldots, \zeta_{p_c})^T$ for the $j$th category at ${\mathbf x}_i$ in a po-npo mixture model (see Example~\ref{ex:po_ppo_mixture}), ${\mathbf h}_{cj}({\mathbf x}_i) = (h_{cj1}({\mathbf x}_i), \ldots, h_{cjp_c}({\mathbf x}_i))^T$, $h_{cjl}$'s are known predictor functions, $i=1, \ldots, m$; $j=1, \ldots, J-1$; $l=1, \ldots, p_c$
\item[$h_{cjl}({\mathbf x}_i)$\hfill] The $l$th predictor function in ${\mathbf h}_{cj}$ at ${\mathbf x}_i$~, $l=1, \ldots, p_c$
\item[${\mathbf H}_j$\hfill] $p_j\times m$ matrix only for the $j$th category in a general ppo model (see Example~\ref{ex:ppo_model_H}), ${\mathbf H}_j = ({\mathbf h}_j({\mathbf x}_1), \ldots, {\mathbf h}_j({\mathbf x}_m))$, $j=1, \ldots, J-1$
\item[${\mathbf h}_j({\mathbf x}_i)$\hfill] Vector of $p_j$ predictors associated with the $p_j$ parameters $\boldsymbol\beta_j = (\beta_{j1}, \ldots, \beta_{jp_j})^T$ for the $j$th response category at ${\mathbf x}_i$~,  ${\mathbf h}_j({\mathbf x}_i) = (h_{j1}({\mathbf x}_i),$ $\ldots,$ $h_{jp_j}({\mathbf x}_i))^T$,  $h_{jl}$'s are known predictor functions, $j=1, \ldots, J-1$; $i=1, \ldots, m$; $l=1, \ldots, p_j$
\item[$h_{jl}({\mathbf x}_i)$\hfill] The $l$th predictor function in ${\mathbf h}_{j}$ at ${\mathbf x}_i$~, $l=1, \ldots, p_j$
\item[$h_{l}({\mathbf x}_i)$\hfill] The $l$th predictor function in ${\mathbf h}_{c}$ at ${\mathbf x}_i$~, $l=1, \ldots, p_c$
\item[${\mathbf I}_J$\hfill] The identity matrix of order $J$, ${\mathbf I}_{J} = {\rm diag}({\mathbf 1}_{J})$
\item[${\cal I}_t$\hfill] Collection of response categories at the $t$th period, ${\cal I}_t = \{0, 1\}$ for Example~\ref{ex:binary_cond_link_model}
\item[$J$\hfill] Total number of response categories, $J\geq 2$
\item[${\cal J}$\hfill] Collection of all response categories, ${\cal J} = \{1, \ldots, J\}$
\item[$k$\hfill] $k+1$ is the number of categories in the first group of a two-group model, see \eqref{eq:rho_ij_two_group_s}
\item[${\mathbf L}$\hfill] Constant $(J-1)\times (J-1)$ matrix, see \eqref{eq:mlm_in_matrix}, ${\mathbf L} = ({\mathbf L}_1, \ldots, {\mathbf L}_{J-1})^T$
\item[$\bar{\mathbf L}$\hfill] $(2J-1)\times J$ constant matrix for multinomial logistic models, see \eqref{eq:logitunifiedmodel}
\item[${\mathbf L}_j$\hfill] The $j$th row vector of ${\mathbf L}$, ${\mathbf L}_j = (L_{j1}, \ldots, L_{j,J-1})^T$, $j=1, \ldots, J-1$
\item[$l_j$\hfill] Index of intercept for the $j$th response category, i.e., $f_{jl_j}({\mathbf x}_i)\equiv 1$, see Algorithm~\ref{algo:initial_theta_supplement} 
\item[$L_{jl}$\hfill] The $l$th coordinate in ${\mathbf L}_j$, $j=1, \ldots, J-1$; $l=1, \ldots, J-1$, see Example~\ref{ex:binary_cond_link_model}
\item[$L(\boldsymbol\theta)$\hfill] Likelihood function of the multinomial link models~\eqref{eq:mlm_in_matrix}, $\boldsymbol{\theta} \in \boldsymbol{\Theta}$
\item[$l(\boldsymbol\theta)$\hfill] Log-likelihood of the multinomial link models~\eqref{eq:mlm_in_matrix}, \\ $l(\boldsymbol\theta) = \log L(\boldsymbol{\theta}) = \sum_{i=1}^m {\mathbf Y}_i^T\log \bar{\boldsymbol\pi}_i + \sum_{i=1}^m \log (n_i!) - \sum_{i=1}^m\sum_{j=1}^J \log (Y_{ij}!)$
\item[$m$\hfill] Total number of distinct experimental settings or covariate settings, $m\geq 2$
\item[$n$\hfill] Total number of observations, $n=\sum_{i=1}^m n_i > 0$
\item[$n_i$\hfill] Number of observations at the $i$th experimental setting ${\mathbf x}_i$~, $n_i > 0$, $i=1, \ldots, m$
\item[$p$\hfill] Total number of parameters for a multinomial link model, $p \geq 1$
\item[${\mathbf P}_0$\hfill] Sub-collection of $\boldsymbol{\rho}_i$'s, ${\mathbf P}_0 = \{(\rho_1, \ldots, \rho_{J-1}) \in \mathbb{R}^{J-1} \mid \rho_j = \frac{{\mathbf L}_j^T \boldsymbol{\pi}}{{\mathbf R}_j^T \boldsymbol{\pi} + (1-{\mathbf 1}_{J-1}^T \boldsymbol{\pi}) b_j}, j=1, \ldots, J-1; \boldsymbol{\pi}\in \boldsymbol{\Pi}_0\}$
for given ${\mathbf L}, {\mathbf R}$ and ${\mathbf b}$, see Assumption~\ref{as:P_0_to_Pi_0_assumption}
\item[$p_c$\hfill] Number of common parameters for response categories $1, \ldots, J-1$ in a ppo model (see Examples~\ref{ex:po_ppo_mixture} or \ref{ex:ppo_model_H}), $p_c \geq 0$
\item[$p_j$\hfill] Number of parameters for the $j$th response category in a ppo model (see Examples~\ref{ex:po_ppo_mixture} or \ref{ex:ppo_model_H}) , $p_j \geq 1$
\item[${\mathbf R}$\hfill] Constant $(J-1)\times (J-1)$ matrix, see \eqref{eq:mlm_in_matrix}, ${\mathbf R} = ({\mathbf R}_1, \ldots, {\mathbf R}_{J-1})^T$
\item[${\mathbf R}_j$\hfill] The $j$th row vector of ${\mathbf R}$, ${\mathbf R}_j = (R_{j1}, \ldots, R_{j,J-1})^T$, $j=1, \ldots, J-1$
\item[$R_{jl}$\hfill] The $l$th coordinate in ${\mathbf R}_j$, $j=1, \ldots, J-1$; $l=1, \ldots, J-1$, see Example~\ref{ex:binary_cond_link_model}
\item[$s$\hfill] Baseline category of the first group in a two-group model, see \eqref{eq:rho_ij_two_group_s}
\item[$T$\hfill] Total number of periods for longitudinal models, see Example~\ref{ex:binary_cond_link_model}, $T\geq 2$
\item[${\cal T}$\hfill] Collection of all possible outcomes in a longitudinal categorical model, ${\cal T} = {\cal I}_1 \times \cdots \times {\cal I}_T$
\item[${\mathbf U}$\hfill] $m(J-1)\times m(J-1)$ block matrix, ${\mathbf U} = ({\mathbf U}_{st})_{s,t=1,\ldots, J-1}$
\item[${\mathbf U}_i$\hfill] $(J-1)\times (J-1)$ matrix at ${\mathbf x}_i$~, ${\mathbf U}_i = \left(u_{st}(\boldsymbol\pi_i)\right)_{s,t=1,\ldots, J-1} = {\mathbf C}_i^T {\rm diag} (\bar{\boldsymbol\pi}_i)^{-1} {\mathbf C}_i$~, $i=1, \ldots, m$
\item[${\mathbf U}_{st}$\hfill] $m\times m$ diagonal matrix, ${\mathbf U}_{st} = {\rm diag}\{n_1 u_{st}({\boldsymbol\pi}_1), \ldots, n_m u_{st}({\boldsymbol\pi}_m)\}$, $s,t = 1, \ldots, J-1$
\item[$u_{st}(\boldsymbol\pi_i)$\hfill] The $(s,t)$th entry of ${\mathbf U}_i$~,  $u_{st}(\boldsymbol\pi_i) = {\mathbf c}_{is}^T {\rm diag} (\bar{\boldsymbol\pi}_i)^{-1} {\mathbf c}_{it}$, $s,t=1, \ldots, J-1$, $i=1, \ldots, m$
\item[$W$\hfill] Wald statistic for testing $H_0: \boldsymbol\theta = \boldsymbol\theta_0$~, $W = (\hat{\boldsymbol\theta} - \boldsymbol\theta_0)^T {\mathbf F}(\hat{\boldsymbol\theta}) (\hat{\boldsymbol\theta} - \boldsymbol\theta_0)$
\item[${\mathbb X}$\hfill] $m(J-1)\times p$ matrix, ${\mathbb X} = ({\mathbf X}_1^T, \ldots, {\mathbf X}_m^T)^T$, see Algorithm~\ref{algo:initial_theta}
\item[${\mathbf x}_i$\hfill] The $i$th distinct experimental setting or covariate setting, ${\mathbf x}_i = (x_{i1}, \ldots, x_{id})^T$, $i=1, \ldots, m$
\item[${\mathbf X}_i$\hfill] $(J-1)\times p$ model matrix at ${\mathbf x}_i$~, ${\mathbf X}_i= ({\mathbf f}_1({\mathbf x}_i), \ldots, {\mathbf f}_{J-1} ({\mathbf x}_i))^T$ in general, see \eqref{eq:mlm_in_matrix} and \eqref{eq:general_X_i}
\item[$\bar{\mathbf X}_i$\hfill] $J\times p$ model matrix for multinomial logistic models, see \eqref{eq:logitunifiedmodel}
\item[$x_{il}$\hfill] Value of the $l$th covariate at the $i$th setting ${\mathbf x}_i$~, $i=1, \ldots, m$; $l=1, \ldots, d$
\item[${\mathbb Y}$\hfill] Vector in $\mathbb{R}^{m(J-1)}$ for finding an initial estimate of $\boldsymbol{\theta}$, see Algorithm~\ref{algo:initial_theta}
\item[${\mathbf Y}_i$\hfill] Multinomial response (random vector) observed at ${\mathbf x}_i$~, ${\mathbf Y}_i = (Y_{i1}, \ldots, Y_{iJ})^T \sim {\rm Multinomial}(n_i; \pi_{i1}, \ldots, \pi_{iJ})$, $i=1, \ldots, m$
\item[${\mathbf y}_i$\hfill] Observed multinomial response (counts) at ${\mathbf x}_i$~, ${\mathbf y}_i = (y_{i1}, \ldots, y_{yJ})^T$ with $n_i = \sum_{j=1}^J y_{ij}$~, $i=1, \ldots, m$
\item[$Y_{ij}$\hfill] Number of observations with ${\mathbf x}_i$ and response category $j$, $i=1, \ldots, m$; $j=1, \ldots, J$
\item[$y_{ij}$\hfill] Observed number of observations with ${\mathbf x}_i$ and response category $j$, $i=1, \ldots, m$; $j=1, \ldots, J$
\item[$Z_{it}$\hfill] Categorical response at the $t$th period with ${\mathbf x}_i$~, see Example~\ref{ex:binary_cond_link_model}
\item[$\boldsymbol\beta_{j}$\hfill] Vector of parameters for the $j$th response category in a ppo model (see Examples~\ref{ex:po_ppo_mixture} or \ref{ex:ppo_model_H}), $\boldsymbol\beta_{j} = (\beta_{j1}, \ldots, \beta_{jp_j})^T$, $j=1, \ldots, J-1$
\item[$\beta_{jl}$\hfill] The $l$th parameter in $\boldsymbol{\beta}_j$~, $j=1, \ldots, J-1$; $l=1, \ldots, p_j$
\vskip 1mm
\item[$\delta$\hfill] Step length in $(0, 1]$ in Algorithm~\ref{algo:fisher_scoring}, e.g., $\delta=0.5$
\item[$\epsilon$\hfill] Tolerance level of relative error in Algorithm~\ref{algo:fisher_scoring}, e.g., $\epsilon = 10^{-6}$
\item[$\boldsymbol\zeta$\hfill] Vector of common parameters for response categories $1, \ldots, J-1$ in a ppo model (see Examples~\ref{ex:po_ppo_mixture} or \ref{ex:ppo_model_H}), $\boldsymbol\zeta = (\zeta_1, \ldots, \zeta_{p_c})^T$  
\item[$\zeta_l$\hfill] The $l$th parameter in $\boldsymbol{\zeta}$, $l=1, \ldots, p_c$
\item[$\eta$\hfill] Linear predictor, $\eta = g(\rho) \in (-\infty, \infty)$, see Table~\ref{tab:link_functions}
\item[$\boldsymbol\eta_{i}$\hfill] Vector of linear predictors at ${\mathbf x}_i$~, ${\boldsymbol\eta}_i = (\eta_{i1}, \ldots, \eta_{i,J-1})^T = {\mathbf X}_i \boldsymbol\theta \in \mathbb{R}^{J-1}$ in general, see \eqref{eq:mlm_in_matrix}, $i=1, \ldots, m$
\item[$\eta_{ij}$\hfill] $\eta_{ij} = {\mathbf f}_j^T({\mathbf x}_i) \boldsymbol{\theta}$ in general, the $j$th coordinate of $\boldsymbol{\eta}_i$~, $j=1, \ldots, J-1$, see \eqref{eq:mlm_j}
\item[$\boldsymbol\Theta$\hfill] Feasible parameter space, the collection of all feasible parameter vectors, see \eqref{eq:Theta_in_general}
\item[$\boldsymbol\theta$\hfill] Vector of all parameters, $\boldsymbol\theta = (\theta_1, \ldots, \theta_p)^T \in \mathbb{R}^p$ in general
\item[$\hat{\boldsymbol\theta}$\hfill] Maximum likelihood estimate (MLE) of $\boldsymbol{\theta}$, $\hat{\boldsymbol{\theta}} = (\hat{\theta}_1, \ldots, \hat{\theta}_p)^T \in \mathbb{R}^p$ 
\item[$\theta_l$\hfill] The $l$th parameter in $\boldsymbol{\theta}$, $l=1, \ldots, p$
\vskip 1mm
\item[$\hat{\theta}_l$\hfill] Maximum likelihood estimate (MLE) of $\theta_l$~, $l=1, \ldots, p$
\vskip 1mm
\item[$\boldsymbol{\theta}^{(t)}$\hfill] Iterated parameter vector at the $t$th iteration, see Algorithm~\ref{algo:fisher_scoring}
\item[$\Lambda$\hfill] Likelihood ratio test statistic, $\Lambda = -2 \log (\max_{\boldsymbol\theta_2} L(\boldsymbol\theta_1 = {\mathbf 0}_r, \boldsymbol\theta_2)/ \max_{\boldsymbol\theta} L(\boldsymbol\theta))$
\item[$\lambda_0$\hfill] Predetermined threshold for the minimum eigenvalue of ${\mathbf F}(\boldsymbol{\theta}^{(t)})$, e.g., $\lambda_0 = 10^{-6}$, see the context of Algorithm~\ref{algo:fisher_scoring}
\item[$\nu$\hfill] Number of degrees of freedom for $t$-distribution, see Table~\ref{tab:link_functions}
\item[$\boldsymbol{\Pi}_0$\hfill] Collection of all $\boldsymbol{\pi}_i$ under our consideration, $\boldsymbol{\Pi}_0 = \{(\pi_1, \ldots, \pi_{J-1})^T \in \mathbb{R}^{J-1} \mid \pi_j > 0, j=1, \ldots, J-1; \sum_{j=1}^{J-1} \pi_j < 1\}$
\item[$\boldsymbol\pi_i$\hfill] Vector of response categorical probabilities at ${\mathbf x}_i$~, $\boldsymbol\pi_i = (\pi_{i1}, \ldots, \pi_{i,J-1})^T \in \mathbb{R}^{J-1}$, $\pi_{iJ} = 1 - \sum_{j=1}^{J-1} \pi_{ij}$~, $i=1, \ldots, m$
\item[$\bar{\boldsymbol\pi}_i$\hfill] Vector of all response categorical probabilities at ${\mathbf x}_i$~, $\bar{\boldsymbol\pi}_i = (\pi_{i1}, \ldots, \pi_{i,J-1}, \pi_{iJ})^T \in \mathbb{R}^{J}$,  $i=1, \ldots, m$
\item[$\pi_{ij}$\hfill] Probability that the response falls into the $j$th category at the $i$th experimental setting, $\sum_{j=1}^J \pi_{ij} = 1$, assuming $0<\pi_{ij}<1$, $i=1, \ldots, m$; $j=1, \ldots, J$
\vskip 1mm
\item[$\hat{\pi}_{ij}$\hfill] Estimated $\pi_{ij}$~, $\hat{\pi}_{ij} = \pi_{ij}(\hat{\boldsymbol{\theta}})$, $i=1, \ldots, m$; $j=1, \ldots, J$ 
\item[$\rho$\hfill] Transformed linear predictor, $\rho = g^{-1}(\eta) \in (0,1)$, see Table~\ref{tab:link_functions}
\item[$\boldsymbol\rho_i$\hfill] Vector of transformed response category probabilities or linear predictors at ${\mathbf x}_i$~, $\boldsymbol\rho_i = (\rho_{i1}, \ldots, \rho_{i,J-1})^T \in \mathbb{R}^{J-1}$, see \eqref{eq:rho_ij_general}, $i=1, \ldots, m$ 
\item[$\frac{\boldsymbol\rho_i}{1-\boldsymbol\rho_i}$\hfill] Element-wise operations, $\boldsymbol\rho_i/(1-\boldsymbol\rho_i) = \left(\frac{\rho_{i1}}{1-\rho_{i1}}, \ldots, \frac{\rho_{i,J-1}}{1-\rho_{i,J-1}}\right)^T \in \mathbb{R}^{J-1}$, $i=1, \ldots, m$
\item[$\boldsymbol\rho_i^{-1}$\hfill] Element-wise operation, $\boldsymbol\rho_i^{-1} = (\rho_{i1}^{-1}, \ldots, \rho_{i,J-1}^{-1})^T \in \mathbb{R}^{J-1}$, $i=1, \ldots, m$
\item[$\rho_{ij}$\hfill] The $j$th coordinate of $\boldsymbol\rho_i$~, see \eqref{eq:rho_ij_general}, $\rho_{ij} = g_j^{-1}(\eta_{ij}) \in (0,1)$, $i=1, \ldots, m$; $j=1, \ldots, J-1$
\item[$\sigma$\hfill] Predetermined one-to-one correspondence from ${\cal T}$ to ${\cal J}$, $\sigma^{-1}(j)$ is denoted by $(s_{j1}, \ldots, s_{jT})^T \in {\cal T}$ for each $j\in {\cal T}$, see Example~\ref{ex:binary_cond_link_model}
\item[$\hat\sigma_{ij}$\hfill] The $(i,j)$th entry of  ${\mathbf F}(\hat{\boldsymbol\theta})^{-1}$, i.e., ${\mathbf F}(\hat{\boldsymbol\theta})^{-1} = \left(\hat\sigma_{ij}\right)_{i,j=1,\ldots, p}$
\vskip 1mm
\item[$\Phi$\hfill] Cumulative distribution function of standard normal distribution, see Table~\ref{tab:link_functions}
\item[$\phi$\hfill] Probability density function of standard normal distribution, see Table~\ref{tab:link_functions}
\end{list}

\section{More on mixed-link models}
\label{subsec:m_link_m}
\normalsize

The mixed-link models~\eqref{eq:mixed_link_ppo}+\eqref{eq:rho_ij} introduced in Example~\ref{ex:mixed_link_ppo} include four classes of models, namely baseline-category mixed-link models, cumulative mixed-link models, adjacent-categories mixed-link models, and continuation-ratio mixed-link models.

In this section, we provide the technical details that make the mixed-link models~\eqref{eq:mixed_link_ppo}+\eqref{eq:rho_ij} a special class of the multinomial link models~\eqref{eq:mlm_in_matrix} or \eqref{eq:mlm_j}.

By letting the model matrix ${\mathbf X}_i$ in \eqref{eq:mlm_in_matrix} or \eqref{eq:mlm_j}
take the following specific form
\begin{equation}\label{eqn:Xi_ppo_J-1}
{\mathbf X}_i= \begin{pmatrix}
{\mathbf h}_1^T({\mathbf x}_i) &   & & {\mathbf h}_c^T({\mathbf x}_i)\\
&   \ddots &  & \vdots \\
&   & {\mathbf h}_{J-1}^T({\mathbf x}_i) & {\mathbf h}_c^T({\mathbf x}_i)
\end{pmatrix}\> \in \mathbb{R}^{(J-1) \times p}
\end{equation}
with the regression parameter vector $\boldsymbol\theta=(\boldsymbol\beta_{1}^T,\cdots,\boldsymbol\beta_{J-1}^T,\boldsymbol\zeta^T)^T$ consists of $p=p_1+\cdots+p_{J-1}+p_c$ unknown parameters in total,  model~\eqref{eq:mlm_in_matrix} with ppo can be written as
\begin{equation}\label{eq:mlm_ppo_j}
g_j\left(\frac{{\mathbf L}^T_j \boldsymbol\pi_i}{{\mathbf R}^T_j \boldsymbol\pi_i + \pi_{iJ} b_j}\right) = \eta_{ij} = {\mathbf h}_j^T({\mathbf x}_i)\boldsymbol\beta_j+{\mathbf h}_c^T({\mathbf x}_i)\boldsymbol\zeta, \quad j=1, \ldots, J-1.
\end{equation}

In the rest of this section, we specify the $(J-1)\times (J-1)$ matrices ${\mathbf L}$, ${\mathbf R}$ and the vector ${\mathbf b}$ in model~\eqref{eq:mlm_in_matrix} (or equivalently the vectors ${\mathbf L}_j$~, ${\mathbf R}_j$~, and the numbers $b_j$ in model~\eqref{eq:mlm_j}) for each of the four classes of mixed-link models. 

To facilitate the readers (see Step~$3^\circ$ of Algorithm~\ref{algo:fisher_matrix} in Appendix~\ref{subsec:calculate_gradient_fisher}), we also provide the explicit formulae for ${\mathbf D}_i^{-1}$ and ${\mathbf D}_i^{-1} {\mathbf b}$, which are critical for computing the Fisher information matrix and the fitted categorical probabilities, where ${\mathbf D}_i = {\rm diag}(\boldsymbol{\rho}_i^{-1}) {\mathbf L} - {\mathbf R}$.

\subsection{Baseline-category mixed-link models}\label{sec:baseline-category}  
In this case, ${\mathbf L} = {\mathbf R} = {\mathbf I}_{J-1}$~, the identity matrix of order $J-1$, and ${\mathbf b} = {\mathbf 1}_{J-1}$~, the vector of all ones with length $J-1$. A special case is $g_1 = \cdots = g_{J-1} = g$. Then 
\begin{eqnarray*}
{\mathbf D}_i^{-1} &=& {\rm diag}\left(\boldsymbol\rho_i/(1-\boldsymbol\rho_i)\right) = {\rm diag}\left\{\frac{\rho_{i1}}{1-\rho_{i1}}, \ldots, \frac{\rho_{i,J-1}}{1-\rho_{i,J-1}}\right\}\ , \\
{\mathbf D}_i^{-1} {\mathbf b} &=& \boldsymbol\rho_i/(1-\boldsymbol\rho_i) = \left(\frac{\rho_{i1}}{1-\rho_{i1}}, \ldots, \frac{\rho_{i,J-1}}{1-\rho_{i,J-1}}\right)^T\ . %\\
\end{eqnarray*}

A special case is when $J=2$, $\boldsymbol\rho_i = \boldsymbol\pi_i = \pi_{i1} \in \mathbb{R}$.

\subsection{Cumulative mixed-link models}\label{sec:cumulative}

In this case, 
\[
{\mathbf L} = \left[\begin{array}{cccc}
1 &   &   & \\
1 & 1 &   & \\
\vdots & \vdots & \ddots & \\
1 & 1 & \cdots & 1
\end{array}\right] \quad \in \mathbb{R}^{(J-1)\times (J-1)}\ ,
\]
${\mathbf R} = {\mathbf 1}_{J-1} {\mathbf 1}_{J-1}^T$~, ${\mathbf b} = {\mathbf 1}_{J-1}$~.  A special case is $g_1 = g_2 = \cdots = g_{J-1} = g$. Then 
\[
{\mathbf D}_i^{-1} = \left[\begin{array}{cccccc}
\rho_{i1} & 0  & 0  & \cdots & 0 &  \frac{\rho_{i,J-1}\rho_{i1}}{1-\rho_{i,J-1}} \\
-\rho_{i1} & \rho_{i2} & 0  & \cdots & 0 & \frac{\rho_{i,J-1}(\rho_{i2} - \rho_{i1})}{1-\rho_{i,J-1}} \\
0 & -\rho_{i2} & \rho_{i3} & \cdots & 0 & \frac{\rho_{i,J-1}(\rho_{i3} - \rho_{i2})}{1-\rho_{i,J-1}}\\
\vdots & \vdots & \vdots & \ddots & \vdots & \vdots \\
0 & 0 & 0 & \cdots & \rho_{i,J-2} & \frac{\rho_{i,J-1}(\rho_{i,J-2} - \rho_{i,J-3})}{1-\rho_{i,J-1}}\\
0 & 0 & 0 & \cdots & -\rho_{i,J-2} & \frac{\rho_{i,J-1}(1 - \rho_{i,J-2})}{1-\rho_{i,J-1}}
\end{array}\right] \quad \in \mathbb{R}^{(J-1)\times (J-1)}
\]
exists, ${\mathbf D}_i^{-1} {\mathbf b} = (1-\rho_{i,J-1})^{-1} (\rho_{i1}, \rho_{i2} - \rho_{i1}, \ldots,$ $\rho_{i,J-2} - \rho_{i,J-3},$ $\rho_{i,J-1} - \rho_{i,J-2})^T$, and ${\mathbf 1}_{J-1}^T {\mathbf D}_i^{-1} {\mathbf b} = \rho_{i,J-1}/(1-\rho_{i,J-1})$.

One special case is when $J=2$, $\boldsymbol\rho_i = \boldsymbol\pi_i = \pi_{i1} \in \mathbb{R}$.

Another special case is when $J=3$, 
\[
{\mathbf D}_i^{-1} = \left[\begin{array}{cc}
\rho_{i1} &  \frac{\rho_{i2}\rho_{i1}}{1-\rho_{i2}} \\
-\rho_{i1} & \frac{\rho_{i2}(1 - \rho_{i1})}{1-\rho_{i2}}
\end{array}\right] \quad \in \mathbb{R}^{2\times 2}
\]
exists, and ${\mathbf D}_i^{-1} {\mathbf b} = (1-\rho_{i2})^{-1} (\rho_{i1}, \rho_{i2} - \rho_{i1})^T$.

\subsection{Adjacent-categories mixed-link models}\label{sec:adjacent-categories}

In this case, ${\mathbf L} = {\mathbf I}_{J-1}$~, 
\[
{\mathbf R} = \left[\begin{array}{ccccc}
1 & 1 &    &   &  \\
  & 1 & 1  &   &  \\
  &   & \ddots  & \ddots & \\
  &   & & 1 & 1\\
  &   &       & & 1
\end{array}\right] \in \mathbb{R}^{(J-1)\times (J-1)}, \quad {\mathbf b} = \left[\begin{array}{c}
0 \\ 0\\ \vdots \\ 0 \\ 1
\end{array}\right] \in \mathbb{R}^{J-1}\ .
\]
A special case is $g_1 = \cdots = g_{J-1} = g$.  Then ${\mathbf D}_i^{-1} = (a_{st})_{s,t=1,\ldots, J-1}$ exists with 
\[
a_{st} = \left\{\begin{array}{cl}
\prod_{l=s}^t \frac{\rho_{il}}{1-\rho_{il}} & \mbox{ if } s \leq t\ ;\\
0 & \mbox{ if } s > t\ .
\end{array}\right.
\]
All elements of 
\[
{\mathbf D}_i^{-1} {\mathbf b} = \left(\prod_{l=1}^{J-1} \frac{\rho_{il}}{1-\rho_{il}}, \ \prod_{l=2}^{J-1} \frac{\rho_{il}}{1-\rho_{il}}, \ \ldots, \ \prod_{l=J-1}^{J-1} \frac{\rho_{il}}{1-\rho_{il}}\right)^T \in \mathbb{R}^{J-1}
\]
are positive.

One special case is when $J=2$, ${\mathbf L} = {\mathbf R} = {\mathbf b} = 1$ and then $\boldsymbol\rho_i = \boldsymbol\pi_i = \pi_{i1} \in \mathbb{R}$.

\begin{remark}\label{remark:adjacent_vglm} \quad {\rm
For adjacent-categories logit models, the {\tt vglm} function in the R package {\tt VGAM} calculates $\log \left(\frac{\pi_{i,j+1}}{\pi_{ij}}\right)$ instead of $\log \left(\frac{\pi_{ij}}{\pi_{i,j+1}}\right)$. As a result, the $\boldsymbol{\theta}$ discussed in this paper is different from the $\boldsymbol{\theta}_{\rm vglm}$ calculated from the {\tt vglm} function. Nevertheless, $\boldsymbol{\hat{\pi}}_i$ and thus the maximum log-likelihood based on  $\hat{\boldsymbol{\theta}}$ or $\hat{\boldsymbol{\theta}}_{\rm vglm}$ still match, respectively (see also Appendix~\ref{sec:existing_issue_subsection}).
\hfill{$\Box$}
}\end{remark}

\subsection{Continuation-ratio mixed-link models}\label{sec:continuation-ratio}

In this case, ${\mathbf L} = {\mathbf I}_{J-1}$~, 
\[
{\mathbf R} = \left[\begin{array}{cccc}
1 & 1 & \cdots  & 1 \\
  & 1 & \cdots  & 1 \\
  &   & \ddots  & \vdots \\
  &   &        & 1
\end{array}\right] \in \mathbb{R}^{(J-1)\times (J-1)}\ ,
\]
${\mathbf b} = {\mathbf 1}_{J-1}$~. A special case is $g_1 = \cdots = g_{J-1} = g$. 
Then ${\mathbf D}_i^{-1} = (a_{st})_{s,t=1,\ldots, J-1}$ exists with 
\[
a_{st} = \left\{\begin{array}{cl}
\rho_{is} \rho_{it} \prod_{l=s}^t (1-\rho_{il})^{-1} & \mbox{ if }s < t\ ;\\
\rho_{is} (1-\rho_{is})^{-1} & \mbox{ if } s = t\ ; \\
0 & \mbox{ if } s > t\ .
\end{array}\right.
\] 
All elements of 
\[
{\mathbf D}_i^{-1} {\mathbf b} = \left(\frac{\rho_{i1}}{\prod_{l=1}^{J-1} (1-\rho_{il})}, \ \frac{\rho_{i2}}{\prod_{l=2}^{J-1} (1-\rho_{il})}, \ \ldots, \ \frac{\rho_{i,J-1}}{\prod_{l=J-1}^{J-1} (1-\rho_{il})}\right)^T \in \mathbb{R}^{J-1}
\]
are positive. It can be verified that ${\mathbf 1}_{J-1}^T {\mathbf D}_i^{-1} {\mathbf b} = \prod_{l=1}^{J-1} (1-\rho_{il})^{-1} - 1$.

One special case is when $J=2$, $\boldsymbol\rho_i = \boldsymbol\pi_i = \pi_{i1} \in \mathbb{R}$.

\section{More on two-group models}
\label{subsec:more_two_group}

In this section, we show the technical details that make the two-group models~\eqref{eq:mixed_link_ppo}+\eqref{eq:rho_ij_two_group_s} introduced in Example~\ref{ex:two_group_s} a special class of the multinomial link models~\eqref{eq:mlm_in_matrix} or \eqref{eq:mlm_j}.

Similarly to Appendix~\ref{subsec:m_link_m} for Example~\ref{ex:mixed_link_ppo}, the model matrix ${\mathbf X}_i$ in \eqref{eq:mlm_in_matrix} or \eqref{eq:mlm_j} 
takes the form of \eqref{eqn:Xi_ppo_J-1}; and the regression parameter vector $\boldsymbol\theta=(\boldsymbol\beta_{1}^T,\cdots,\boldsymbol\beta_{J-1}^T,\boldsymbol\zeta^T)^T$ consists of $p=p_1+\cdots+p_{J-1}+p_c$ unknown parameters in total. Then  model~\eqref{eq:mlm_in_matrix} can be written as \eqref{eq:mlm_ppo_j}. 

In the rest of this section, we specify the $(J-1)\times (J-1)$ matrices ${\mathbf L}$, ${\mathbf R}$ and the vector ${\mathbf b}$ in model~\eqref{eq:mlm_in_matrix} (or equivalently the vectors ${\mathbf L}_j$~, ${\mathbf R}_j$~, and the numbers $b_j$ in model~\eqref{eq:mlm_j}) for each of the three classes of two-group models. Similarly to Appendix~\ref{subsec:m_link_m} for Example~\ref{ex:mixed_link_ppo}, we also provide the explicit formulae for ${\mathbf D}_i^{-1}$ and ${\mathbf D}_i^{-1} {\mathbf b}$.

First, we focus on a special class of two-group models whose two groups share the same baseline category $J$. That is, $s=J$ in this case, which leads to simplified notation.

\begin{example}\label{ex:two_group_J} {\bf Two-group models with shared baseline category}\quad {\rm
Under \eqref{eq:mixed_link_ppo}, the same form as in the mixed-link models (Example~\ref{ex:mixed_link_ppo}), we further assume that there exists an integer $k$, such that, $1\leq k \leq J-3$ and
\begin{equation}\label{eq:rho_ij_two_group_J}
\rho_{ij} = \left\{
\begin{array}{cl}
\frac{\pi_{ij}}{\pi_{ij} + \pi_{iJ}} & \mbox{ for }j=1, \ldots, k\ ;\\
\frac{\pi_{i,k+1} + \cdots + \pi_{ij}}{\pi_{i,k+1} + \cdots + \pi_{iJ}} & \mbox{ for baseline-cumulative and }j=k+1, \ldots, J-1\ ;\\
\frac{\pi_{ij}}{\pi_{ij} + \pi_{i,j+1}} & \mbox{ for baseline-adjacent and }j=k+1, \ldots, J-1\ ;\\
\frac{\pi_{ij}}{\pi_{ij} +\cdots + \pi_{iJ}} & \mbox{ for baseline-continuation and }j=k+1, \ldots, J-1\ .
\end{array}
\right.
\end{equation}
It indicates that the response categories form two groups, $\{1, \ldots, k, J\}$ and $\{k+1, \ldots, J-1, J\}$, which share the same baseline category $J$.

The two-group models~\eqref{eq:mixed_link_ppo}+\eqref{eq:rho_ij_two_group_J} also consist of three classes, namely baseline-cumulative, baseline-adjacent, and baseline-continuation mixed-link models with shared baseline category, which are all special cases of the multinomial link model \eqref{eq:mlm_in_matrix} or \eqref{eq:mlm_j}
(see Appendices~\ref{sec:b-c_link_model}, \ref{sec:b-a-c_link_model}, and \ref{sec:b-c-r_link_model}).
\hfill{$\Box$}
}\end{example}

\subsection{Baseline-cumulative mixed-link models with shared baseline category}\label{sec:b-c_link_model}

There are two groups of response categories in this model. One group of $k + 1 \geq 2$ categories are controlled by a baseline-category mixed-link model and the other group of $J-k \geq 3$ categories are controlled by a cumulative mixed-link model. The two groups share the same baseline category $J$. More specifically, let $1\leq k \leq J-3$ and
\[
\rho_{ij} = \left\{\begin{array}{cl}
\frac{\pi_{ij}}{\pi_{ij} + \pi_{iJ}} & \mbox{ for }j=1, \ldots, k\ ;\\
\frac{\pi_{i,k+1} + \cdots + \pi_{ij}}{\pi_{i,k+1} + \cdots + \pi_{iJ}} & \mbox{ for }j=k+1, \ldots, J-1\ .
\end{array}\right.
\]
As for link functions, a special case is $g_1 = \cdots = g_k = g_a$ and $g_{k+1} = \cdots = g_{J-1}=g_b$~. Then
\[
{\mathbf L} = \left[\begin{array}{ccccc}
{\mathbf I}_k & & & & \\
& 1 &   &   & \\
& 1 & 1 &   & \\
& \vdots & \vdots & \ddots & \\
& 1 & 1 & \cdots & 1
\end{array}\right], \>\>\> 
{\mathbf R} = \left[\begin{array}{ccccc}
{\mathbf I}_k & & & & \\
& 1 & 1 & \cdots  & 1 \\
& 1 & 1 & \cdots  & 1 \\
& \vdots & \vdots & \ddots & \vdots \\
& 1 & 1 & \cdots & 1
\end{array}\right]\ ,
\]
${\mathbf b} = {\mathbf 1}_{J-1}$~. Then 
\footnotesize
\[
{\mathbf D}_i^{-1} = \left[\begin{array}{ccccccccc}
\frac{\rho_{i1}}{1-\rho_{i1}} & & & & & & & & \\
& \ddots & & & & & & & \\
& & \frac{\rho_{ik}}{1-\rho_{ik}} & & & & & & \\
& & & \rho_{i,k+1} & 0  & 0  & \cdots & 0 & \frac{\rho_{i,J-1}\rho_{i,k+1}}{1-\rho_{i,J-1}} \\
& & & -\rho_{i,k+1} & \rho_{i,k+2} & 0  & \cdots & 0 & \frac{\rho_{i,J-1}(\rho_{i,k+2} - \rho_{i,k+1})}{1-\rho_{i,J-1}} \\
& & & 0 & -\rho_{i,k+2} & \rho_{i,k+3} & \cdots & 0 & \frac{\rho_{i,J-1}(\rho_{i,k+3} - \rho_{i,k+2})}{1-\rho_{i,J-1}}\\
& & & \vdots & \vdots & \vdots & \ddots & \vdots & \vdots \\
& & & 0 & 0 & 0 & \cdots & \rho_{i,J-2} & \frac{\rho_{i,J-1}(\rho_{i,J-2} - \rho_{i,J-3})}{1-\rho_{i,J-1}}\\
& & & 0 & 0 & 0 & \cdots & -\rho_{i,J-2} & \frac{\rho_{i,J-1}(1 - \rho_{i,J-2})}{1-\rho_{i,J-1}}
\end{array}\right]\ ,
\]
\normalsize
\begin{eqnarray*}
{\mathbf D}_i^{-1} {\mathbf b} &=& \left(\frac{\rho_{i1}}{1-\rho_{i1}}, \ldots, 
\frac{\rho_{ik}}{1-\rho_{ik}}, \frac{\rho_{i,k+1}}{1-\rho_{i,J-1}}, \frac{\rho_{i,k+2}-\rho_{i,k+1}}{1-\rho_{i,J-1}}, \ldots, \frac{\rho_{i,J-1} - \rho_{i,J-2}}{1-\rho_{i,J-1}}\right)^T\ ,\\
{\mathbf 1}_{J-1}^T {\mathbf D}_i^{-1} {\mathbf b} &=& \sum_{l=1}^k \frac{\rho_{il}}{1-\rho_{il}} + \frac{\rho_{i,J-1}}{1-\rho_{i,J-1}}\ .
\end{eqnarray*}

One special case is when $J=4$ and $k=1$, 
\[
{\mathbf D}_i^{-1} = \left[\begin{array}{ccc}
\frac{\rho_{i1}}{1-\rho_{i1}} & & \\
& \rho_{i2} &  \frac{\rho_{i3}\rho_{i2}}{1-\rho_{i3}} \\
 & -\rho_{i2} & \frac{\rho_{i3}(1 - \rho_{i2})}{1-\rho_{i3}}
\end{array}\right]\ .
\]

\subsection{Baseline-adjacent mixed-link models with shared baseline category}\label{sec:b-a-c_link_model}

There are two groups of response categories in this model. One group of $k + 1 \geq 2$ categories are controlled by a baseline-category mixed-link model and the other group of $J-k \geq 3$ categories are controlled by an adjacent-categories mixed-link model. The two groups share the same baseline category $J$. More specifically, let $1\leq k \leq J-3$ and
\[
\rho_{ij} = \left\{\begin{array}{cl}
\frac{\pi_{ij}}{\pi_{ij} + \pi_{iJ}} & \mbox{ for }j=1, \ldots, k\ ;\\
\frac{\pi_{ij}}{\pi_{ij} + \pi_{i,j+1}} & \mbox{ for }j=k+1, \ldots, J-1\ .
\end{array}\right.
\]
As for link functions, a special case is $g_1 = \cdots = g_k = g_a$ and $g_{k+1} = \cdots = g_{J-1}=g_b$~. Then
${\mathbf L} = {\mathbf I}_{J-1}$, 
\[
{\mathbf R} = \left[\begin{array}{cccccc}
{\mathbf I}_k & & & & & \\
& 1 & 1 &    &   &  \\
&  & 1 & 1  &   &  \\
&  &   & \ddots  & \ddots & \\
&  &   & & 1 & 1\\
&  &   &       & & 1
\end{array}\right] \in \mathbb{R}^{(J-1)\times (J-1)}, \quad {\mathbf b} = \left[\begin{array}{c}
{\mathbf 1}_k \\ 0\\ \vdots \\ 0 \\ 1
\end{array}\right] \in \mathbb{R}^{J-1}\ .
\]
Then 
\footnotesize
\[
{\mathbf D}_i^{-1} = 
\left[\begin{array}{cccccccc}
\frac{\rho_{i1}}{1-\rho_{i1}} & & & & & & & \\
& \ddots & & & & & & \\
& & \frac{\rho_{ik}}{1-\rho_{ik}} & & & & & \\
& & & \prod_{l=k+1}^{k+1} \frac{\rho_{il}}{1-\rho_{il}} & \prod_{l=k+1}^{k+2} \frac{\rho_{il}}{1-\rho_{il}} & \cdots & \prod_{l=k+1}^{J-2} \frac{\rho_{il}}{1-\rho_{il}} & \prod_{l=k+1}^{J-1} \frac{\rho_{il}}{1-\rho_{il}} \\
& & & & \prod_{l=k+2}^{k+2} \frac{\rho_{il}}{1-\rho_{il}} & \cdots & \prod_{l=k+2}^{J-2} \frac{\rho_{il}}{1-\rho_{il}} & \prod_{l=k+2}^{J-1} \frac{\rho_{il}}{1-\rho_{il}} \\
& & & & & \ddots & \vdots & \vdots \\
& & & & & & \prod_{l=J-2}^{J-2} \frac{\rho_{il}}{1-\rho_{il}} & \prod_{l=J-2}^{J-1} \frac{\rho_{il}}{1-\rho_{il}} \\
& & & & & & & \prod_{l=J-1}^{J-1} \frac{\rho_{il}}{1-\rho_{il}}
\end{array}\right]\ .
\]
\normalsize
All elements of 
\[
{\mathbf D}_i^{-1} {\mathbf b} = \left( \frac{\rho_{i1}}{1-\rho_{i1}}, \ \cdots,\ \frac{\rho_{ik}}{1-\rho_{ik}},\ \prod_{l=k+1}^{J-1} \frac{\rho_{il}}{1-\rho_{il}}, \ \ldots, \ \prod_{l=J-1}^{J-1} \frac{\rho_{il}}{1-\rho_{il}}\right)^T \in \mathbb{R}^{J-1}
\]
are positive.

\subsection{Baseline-continuation mixed-link models with shared baseline category}\label{sec:b-c-r_link_model}

There are two groups of response categories in this model. One group of $k + 1 \geq 2$ categories are controlled by a baseline-category mixed-link model and the other group of $J-k \geq 3$ categories are controlled by a continuation-ratio mixed-link model. The two groups share the same baseline category $J$. More specifically, let $1\leq k \leq J-3$ and
\[
\rho_{ij} = \left\{\begin{array}{cl}
\frac{\pi_{ij}}{\pi_{ij} + \pi_{iJ}} & \mbox{ for }j=1, \ldots, k\ ;\\
\frac{\pi_{ij}}{\pi_{ij} +\cdots + \pi_{iJ}} & \mbox{ for }j=k+1, \ldots, J-1\ .
\end{array}\right.
\]
As for link functions, a special case is $g_1 = \cdots = g_k = g_a$ and $g_{k+1} = \cdots = g_{J-1}=g_b$~. Then
${\mathbf L} = {\mathbf I}_{J-1}$~, 
\[
{\mathbf R} = \left[\begin{array}{ccccc}
{\mathbf I}_k & & & & \\
& 1 & 1 & \cdots  & 1 \\
&  & 1 & \cdots  & 1 \\
&  &   & \ddots  & \vdots \\
&  &   &        & 1
\end{array}\right] \in \mathbb{R}^{(J-1)\times (J-1)}\ ,\> {\mathbf b} = {\mathbf 1}_{J-1}\ ,\> \mbox{ and }
\]
\footnotesize
\[
{\mathbf D}_i^{-1} = 
\left[\begin{array}{cccccccc}
\frac{\rho_{i1}}{1-\rho_{i1}} & & & & & & & \\
& \ddots & & & & & & \\
& & \frac{\rho_{ik}}{1-\rho_{ik}} & & & & & \\
& & & \frac{\rho_{i,k+1}}{\prod_{l=k+1}^{k+1} (1-\rho_{il})} & \frac{\rho_{i,k+1} \rho_{i,k+2}}{\prod_{l=k+1}^{k+2} (1-\rho_{il})} & \cdots & \frac{\rho_{i,k+1} \rho_{i,J-2}}{\prod_{l=k+1}^{J-2} (1-\rho_{il})} & \frac{\rho_{i,k+1} \rho_{i,J-1}}{\prod_{l=k+1}^{J-1} (1-\rho_{il})} \\
& & & & \frac{\rho_{i,k+2}}{\prod_{l=k+2}^{k+2} (1-\rho_{il})} & \cdots & \frac{\rho_{i,k+2} \rho_{i,J-2}}{\prod_{l=k+2}^{J-2} (1-\rho_{il})} & \frac{\rho_{i,k+2} \rho_{i,J-1}}{\prod_{l=k+2}^{J-1} (1-\rho_{il})} \\
& & & & & \ddots & \vdots & \vdots \\
& & & & & & \frac{\rho_{i,J-2}}{\prod_{l=J-2}^{J-2} (1-\rho_{il})} & \frac{\rho_{i,J-2} \rho_{i,J-1}}{\prod_{l=J-2}^{J-1} (1-\rho_{il})} \\
& & & & & & & \frac{\rho_{i,J-1}}{\prod_{l=J-1}^{J-1} (1-\rho_{il})}
\end{array}\right]\ .
\]
\normalsize
All elements of 
\[
{\mathbf D}_i^{-1} {\mathbf b} = \left(\frac{\rho_{i1}}{1-\rho_{i1}}, \ \ldots,\ \frac{\rho_{ik}}{1-\rho_{ik}},\ \frac{\rho_{i,k+1}}{\prod_{l=k+1}^{J-1} (1-\rho_{il})}, \ \ldots, \ \frac{\rho_{i,J-1}}{\prod_{l=J-1}^{J-1} (1-\rho_{il})}\right)^T \in \mathbb{R}^{J-1}
\]
are positive. It can be verified that 
\[
{\mathbf 1}_{J-1}^T {\mathbf D}_i^{-1} {\mathbf b} = \sum_{l=1}^k \frac{\rho_{il}}{1-\rho_{il}} + \prod_{l=k+1}^{J-1} (1-\rho_{il})^{-1} - 1\ .
\]

\subsection{More on two-group models in Example~\ref{ex:two_group_s}}\label{sec:more_two_group_s}

To simplify the notations, we first rewrite ${\mathbf L}$, ${\mathbf R}$ and ${\mathbf b}$ for the two-group models with shared baseline category $J$ in Appendices~\ref{sec:b-c_link_model}, \ref{sec:b-a-c_link_model}, and \ref{sec:b-c-r_link_model} as follows:
\[
{\mathbf L} = \left[\begin{array}{cc}
{\mathbf I}_k & \\
 & {\mathbf L}_{(2)}
 \end{array}\right],\>\>\>
{\mathbf R} = \left[\begin{array}{cc}
{\mathbf I}_k & \\
 & {\mathbf R}_{(2)}
 \end{array}\right],\>\>\>
{\mathbf b} = \left[\begin{array}{c}
{\mathbf 1}_k \\
 {\mathbf b}_{(2)}
 \end{array}\right], 
\]
where ${\mathbf L}_{(2)}, {\mathbf R}_{(2)}$ stand for the corresponding $(J-1-k)\times (J-1-k)$ matrices, and ${\mathbf b}_{(2)}$ stands for the corresponding vector of length $J-1-k$ for the second group of response categories other than $J$, as described in Appendices~\ref{sec:b-c_link_model}, \ref{sec:b-a-c_link_model} and \ref{sec:b-c-r_link_model}, respectively. Note that ${\mathbf L}_{(2)}, {\mathbf R}_{(2)}, {\mathbf b}_{(2)}$ take different forms for baseline-cumulative (see \eqref{eq:L_R_b_(2)_baseline_cumulative_two_group}), baseline-adjacent (see \eqref{eq:L_R_b_(2)_baseline_adjacent_two_group}), and baseline-continuation (see \eqref{eq:L_R_b_(2)_baseline_continuation_two_group}) mixed-link models. We denote $\boldsymbol{\rho}_{i(1)} = (\rho_{i1}, \ldots, \rho_{ik})^T$, $\boldsymbol{\rho}_{i(2)} = (\rho_{i,k+1}, \ldots, \rho_{i,J-1})^T$, and ${\mathbf D}_{i(2)} = {\rm diag}(\boldsymbol{\rho}_{i(2)}^{-1}) {\mathbf L}_{(2)} - {\mathbf R}_{(2)} \in \mathbb{R}^{(J-1-k)\times (J-1-k)}$. Then  $\boldsymbol{\rho}_i = (\boldsymbol{\rho}_{i(1)}^T, \boldsymbol{\rho}_{i(2)}^T)^T$, and
\[
{\mathbf D}_i = {\rm diag}(\boldsymbol{\rho}_i^{-1}) {\mathbf L} - {\mathbf R} = 
\left[\begin{array}{ccc}
{\rm diag}(\boldsymbol{\rho}_{i(1)}^{-1}) - {\mathbf I}_k & \hspace{0.3cm} & \\
 & & {\mathbf D}_{i(2)}
\end{array}
\right]\ .
\]
According to Appendices~\ref{sec:b-c_link_model}, \ref{sec:b-a-c_link_model} and \ref{sec:b-c-r_link_model}, ${\mathbf D}_{i(2)}$ is invertible for baseline-cumulative (see Appendix~\ref{sec:b-c_link_model}), baseline-adjacent (see Appendix~\ref{sec:b-a-c_link_model}), and baseline-continuation (see Appendix~\ref{sec:b-c-r_link_model}) mixed-link models. Therefore, we can rewrite
\[
{\mathbf D}_i^{-1} =  
\left[\begin{array}{ccc}
{\rm diag}\left(\frac{\boldsymbol{\rho}_{i(1)}}{1-\boldsymbol{\rho}_{i(1)}}\right)  & \hspace{0.3cm} & \\
 & & {\mathbf D}_{i(2)}^{-1}
\end{array}
\right], \>\>\>
{\mathbf D}_i^{-1} {\mathbf b} =  
\left[\begin{array}{c}
\frac{\boldsymbol{\rho}_{i(1)}}{1-\boldsymbol{\rho}_{i(1)}}\\{\mathbf D}_{i(2)}^{-1} {\mathbf b}_{(2)}
\end{array}
\right],
\]
whose more detailed forms can be found in Appendices~\ref{sec:b-c_link_model}, \ref{sec:b-a-c_link_model} and \ref{sec:b-c-r_link_model} accordingly.

\medskip
Now we are ready to cover other two-group models introduced in Example~\ref{ex:two_group_s}. That is, the first group is $\{1, \ldots, k, s\}$ with baseline-category $s\neq J$, and the second group is $\{k+1, \ldots, J\}$ with baseline-category $J$. In this case, the ${\mathbf L}$ matrix is exactly the same as in Appendices~\ref{sec:b-c_link_model}, \ref{sec:b-a-c_link_model} and \ref{sec:b-c-r_link_model} for the corresponding mixed-link models.
The ${\mathbf R}$ matrix can be obtained by adding more $1$'s to the corresponding ${\mathbf R}$ matrix in Appendices~\ref{sec:b-c_link_model}, \ref{sec:b-a-c_link_model} and \ref{sec:b-c-r_link_model}. More specifically, we only need to change the $0$'s at the $(1,s), (2, s), \ldots, (k,s)$th entries of ${\mathbf R}$ to $1$'s. The vector ${\mathbf b} = (b_1, \ldots, b_{J-1})^T$ is quite different though. Actually, in this case,
\[
{\mathbf b} = \left\{\begin{array}{cl}
({\mathbf 0}_k^T, {\mathbf 1}_{J-k-1}^T)^T & \mbox{ for  baseline-cumulative;}
\\
({\mathbf 0}_{J-2}^T, 1)^T & \mbox{ for   baseline-adjacent;}
\\
({\mathbf 0}_k^T, {\mathbf 1}_{J-k-1}^T)^T & \mbox{ for  baseline-continuation,}
\end{array}\right.
\]
where ${\mathbf 0}_k = (0, \ldots, 0)^T \in \mathbb{R}^k$ and ${\mathbf 1}_k = (1, \ldots, 1)^T \in \mathbb{R}^k$.

In other words, when $s\neq J$, the two-group models in Example~\ref{ex:two_group_s} are built by
\begin{equation}\label{eq:L_R_b_two_group_s}
{\mathbf L} = \left[\begin{array}{cc}
{\mathbf I}_k & \\
 & {\mathbf L}_{(2)}
 \end{array}\right],\>\>\>
{\mathbf R} = \left[\begin{array}{cc}
{\mathbf I}_k & {\mathbf E}_{k(s)}\\
 & {\mathbf R}_{(2)}
 \end{array}\right],\>\>\>
{\mathbf b} = \left[\begin{array}{c}
{\mathbf 0}_k \\
 {\mathbf b}_{(2)}
 \end{array}\right], 
\end{equation}
where ${\mathbf E}_{k(s)} = [{\mathbf 0}_{k\times (s-k-1)}, {\mathbf 1}_k, {\mathbf 0}_{k\times (J-1-s)}] \in \mathbb{R}^{k\times (J-1-k)}$, $s=k+1, \ldots, J-1$.

As an illustrative example, a baseline-cumulative mixed-link model with $J=5$, $k=1$, and $s=3$ has two groups of categories $\{1,3\}$ (with baseline $3$) and $\{2, 3, 4, 5\}$ (with baseline $5$), as well as 
\[
{\mathbf L} = \left[\begin{array}{cccc}
1 & & & \\
& 1   &   & \\
& 1 & 1   & \\
& 1 & 1 &  1
\end{array}\right], \>\>\> 
{\mathbf R} = \left[\begin{array}{cccc}
1 & & 1 & \\
& 1 & 1 &  1 \\
& 1 & 1 &  1 \\
& 1 & 1 & 1
\end{array}\right], \>\>\>
{\mathbf b} = \left(\begin{array}{c}
0 \\ 1\\ 1\\ 1
\end{array}\right)\ .
\]

In general with $s\neq J$, it can be verified that
\begin{eqnarray}
{\mathbf D}_i &=& {\rm diag}(\boldsymbol{\rho}_i^{-1}) {\mathbf L} - {\mathbf R} = 
\left[\begin{array}{ccc}
{\rm diag}(\boldsymbol{\rho}_{i(1)}^{-1}) - {\mathbf I}_k & \hspace{0.3cm} & -{\mathbf E}_{k(s)}\\
{\mathbf 0}_{(J-1-k)\times k} & & {\mathbf D}_{i(2)}
\end{array}
\right]\ ,\label{eq:D_i_two_group_s}\\
{\mathbf D}_i^{-1} &=&  
\left[\begin{array}{ccc}
{\rm diag} \left(\frac{\boldsymbol{\rho}_{i(1)}}{1-\boldsymbol{\rho}_{i(1)}}\right)  & \hspace{0.3cm} & \frac{\boldsymbol{\rho}_{i(1)}}{1-\boldsymbol{\rho}_{i(1)}} \cdot ({\mathbf D}_{i(2)}^{-1})_{[s-k]}\\
{\mathbf 0}_{(J-1-k)\times k} & & {\mathbf D}_{i(2)}^{-1}
\end{array}
\right]\ ,\label{eq:D_i_inverse_two_group_s}\\
{\mathbf D}_i^{-1} {\mathbf b} &=&  
\left[\begin{array}{c}
\frac{\boldsymbol{\rho}_{i(1)}}{1-\boldsymbol{\rho}_{i(1)}} \cdot ({\mathbf D}_{i(2)}^{-1})_{[s-k]}\cdot {\mathbf b}_{(2)}\\
{\mathbf D}_{i(2)}^{-1} {\mathbf b}_{(2)}
\end{array}
\right]\ ,\label{eq:D_i_b_two_group_s}
\end{eqnarray}
where $({\mathbf D}_{i(2)}^{-1})_{[s-k]} \in \mathbb{R}^{1\times (J-1-k)}$ denotes the $(s-k)$th row of ${\mathbf D}_{i(2)}^{-1}$~, and
\[
({\mathbf D}_{i(2)}^{-1})_{[s-k]}\cdot {\mathbf b}_{(2)} = \left\{\begin{array}{cl}
\frac{\rho_{i,k+1}}{1-\rho_{i,J-1}}\ , & \mbox{ for baseline-cumulative and }s=k+1;\\
& \\
\frac{\rho_{is}-\rho_{i,s-1}}{1-\rho_{i,J-1}}\ , & \mbox{ for baseline-cumulative and }s=k+2, \ldots, J-1;\\
& \\
\prod_{l=s}^{J-1} \frac{\rho_{il}}{1-\rho_{il}}\ , & \mbox{ for baseline-adjacent and }s=k+1, \ldots, J-1;\\
& \\
\frac{\rho_{is}}{\prod_{l=s}^{J-1} (1-\rho_{il})}\ , & \mbox{ for baseline-continuation and }s=k+1, \ldots, J-1.
\end{array}\right.
\]
The explicit form of ${\mathbf D}_{i(2)}^{-1} {\mathbf b}_{(2)}$ can be found in Appendices~\ref{sec:b-c_link_model}, \ref{sec:b-a-c_link_model} and \ref{sec:b-c-r_link_model} accordingly.

\section{More on dichotomous conditional link models}
\label{sec:more_on_dichotomous_condi_link_model}

In this section, we provide more technical details about Example~\ref{ex:binary_cond_link_model}.
Recall that in this case, ${\cal I}_1 = \cdots = {\cal I}_T = \{0,1\}$, $J=2^T$, and the one-to-one correspondence $\sigma : {\cal T} = {\cal I}_1 \times \cdots \times {\cal I}_T \rightarrow {\cal J} = \{1, \ldots, J\}$ is defined as follows:
\[
\sigma(z_1, \ldots, z_T) = \left\{\begin{array}{cl}
\sum_{t=1}^T z_t 2^{t-1}\ , & \mbox{ if } \sum_{t=1}^T z_t\geq 1;\\
2^T\ , & \mbox{ if }\sum_{t=1}^T z_t = 0.
\end{array}\right.
\]

For a general dichotomous conditional link model, we denote $\sigma^{-1}(j) = (s_{j1}, \ldots, s_{jT})^T \in {\cal T}$ for each $j\in {\cal J}$. For the model described by Example~\ref{ex:binary_cond_link_model}, it can be verified that
\begin{equation}\label{eq:s_jt_binary_clm}
s_{jt} \equiv \left\lfloor\frac{j}{2^{t-1}}\right\rfloor \mod 2
\end{equation}
for all $t=1, \ldots, T$ and $j=1, \ldots, J$, where ``$\lfloor x\rfloor$'' stands for the largest integer that is no more than $x$ (see, e.g., \cite{APL1962}). Based on \eqref{eq:s_jt_binary_clm}, we have the following results (see also Example~\ref{ex:binary_cond_link_model}):

\begin{lemma}\label{lem:L_R_b_binary_clm}
For the dichotomous conditional link model described by Example~\ref{ex:binary_cond_link_model}, ${\mathbf L} = (L_{jl})_{(J-1)\times (J-1)}$, ${\mathbf R} = (R_{jl})_{(J-1)\times (J-1)}$, ${\mathbf b} = (b_1, \ldots, b_{J-1})^T$ can be determined as follows:
\begin{itemize}
\item[(i)] $L_{1l} = 1$ for odd $l$, and $0$ for even $l$; $R_{1l}=1$ for each $l=1, \ldots, J-1$; $b_1=1$;
\item[(ii)] Given $t=2, \ldots, T$ and $z_1, \ldots, z_{t-1}\in \{0,1\}$, we let $j=\sigma(z_1, \ldots, z_{t-1}, 1, 0, \ldots, 0) = \sum_{r=1}^{t-1} z_r 2^{r-1} + 2^{t-1}$. Then
\begin{eqnarray*}
L_{jl} &=& \left\{\begin{array}{cl}
1\ , & \mbox{ if }l \equiv \sum_{r=1}^{t-1} z_r 2^{r-1} + 2^{t-1} \mod 2^t;\\
0\ , & \mbox{ otherwise},
\end{array}\right.\\
R_{jl} &=& \left\{\begin{array}{cl}
1\ , & \mbox{ if }l \equiv \sum_{r=1}^{t-1} z_r 2^{r-1}  \mod 2^{t-1};\\
0\ , & \mbox{ otherwise},
\end{array}\right.\\
b_{j} &=& \left\{\begin{array}{cl}
1\ , & \mbox{ if } z_1=\cdots = z_{t-1} = 0;\\
0\ , & \mbox{ otherwise}.
\end{array}\right.
\end{eqnarray*}
\end{itemize}
\end{lemma}

To explore ${\mathbf D}_i$~, ${\mathbf D}_i^{-1}$, and ${\mathbf D}_i^{-1} {\mathbf b}$ for dichotomous conditional link models, we derive the following lemma based on Lemma~\ref{lem:L_R_b_binary_clm} to facilitate programming and deriving further formulae:

\begin{lemma}\label{lem:L_R_b_iterative_binary_clm}
For the dichotomous conditional link model described by Example~\ref{ex:binary_cond_link_model}, ${\mathbf L}, {\mathbf R}, {\mathbf b}$ can be built up iteratively as follows:
\begin{itemize}
\item[(i)] Denote ${\mathbf L}^{(1)} = {\mathbf R}^{(1)} = \{1\}$ as $1\times 1$ matrices, and ${\mathbf b}^{(1)} = \{1\}$ as a length-$1$ vector.
\item[(ii)] Given $t=1, \ldots, T-1$, ${\mathbf L}^{(t)}, {\mathbf R}^{(t)} \in \mathbb{R}^{(2^t-1)\times (2^t-1)}$ and ${\mathbf b}^{(t)} \in \mathbb{R}^{2^t-1}$, we construct
\begin{eqnarray*}
{\mathbf L}^{(t+1)} &=& \left[\begin{array}{c|c}
{\mathbf L}^{(t)} & \> {\mathbf 0}_{2^t-1}\ ,\> {\mathbf L}^{(t)}\\ \hline
{\mathbf 0}_{2^t\times (2^t-1)}  & {\mathbf I}_{2^t}
\end{array}\right] \in \mathbb{R}^{(2^{t+1}-1)\times (2^{t+1}-1)}\ ,\\
{\mathbf R}^{(t+1)} &=& \left[\begin{array}{c|c}
{\mathbf R}^{(t)} & \> {\mathbf b}^{(t)}\ ,\> {\mathbf R}^{(t)}\\ \hline
\begin{array}{c}
{\mathbf 0}_{2^t-1}^T\\
{\mathbf I}_{2^t-1}
\end{array} & {\mathbf I}_{2^t}
\end{array}\right] \in \mathbb{R}^{(2^{t+1}-1)\times (2^{t+1}-1)}\ ,\\
{\mathbf b}^{(t+1)} &=& (({\mathbf b}^{(t)})^T, 1, 0, \ldots, 0)^T \in \mathbb{R}^{2^{t+1}-1}\ .
\end{eqnarray*}
\end{itemize}
Then ${\mathbf L} = {\mathbf L}^{(T)}, {\mathbf R} = {\mathbf R}^{(T)}$ and ${\mathbf b} = {\mathbf b}^{(T)}$.
\end{lemma}

Based on Lemma~\ref{lem:L_R_b_iterative_binary_clm} and ${\mathbf D}_i = {\rm diag}(\boldsymbol{\rho}_i^{-1}) {\mathbf L} - {\mathbf R}$, we obtain the following theorem:

\begin{theorem}\label{thm:Di_inverse_binary_clm}
For the dichotomous conditional link model described by Example~\ref{ex:binary_cond_link_model}, ${\mathbf D}_i$ is invertible. Furthermore, ${\mathbf D}_i$~,  ${\mathbf D}_i^{-1}$, and ${\mathbf D}_i^{-1} {\mathbf b}$ can be constructed iteratively as follows:
\begin{itemize}
\item[(i)] Denote ${\mathbf D}_i^{(1)} = \{\frac{1-\rho_{i1}}{\rho_{i1}}\}$ and $({\mathbf D}_i^{(1)})^{-1} = \{\frac{\rho_{i1}}{1-\rho_{i1}}\}$ as $1\times 1$ matrices, and $({\mathbf D}_i^{(1)})^{-1} {\mathbf b}^{(1)} = \{\frac{\rho_{i1}}{1-\rho_{i1}}\}$ as a length-$1$ vector.
\item[(ii)] Given $t=1, \ldots, T-1$, ${\mathbf D}_i^{(t)}, ({\mathbf D}_i^{(t)})^{-1} \in \mathbb{R}^{(2^t-1)\times (2^t-1)}$ and $({\mathbf D}_i^{(t)})^{-1} {\mathbf b}^{(t)} \in \mathbb{R}^{2^t-1}$, we construct
\[
{\mathbf D}_i^{(t+1)} = \left[\begin{array}{c|c|c}
{\mathbf D}_i^{(t)} &  -{\mathbf b}^{(t)} & {\mathbf D}_i^{(t)}\\ \hline
{\mathbf 0}_{2^t-1}^T & \frac{1-\rho_{i,2^t}}{\rho_{i,2^t}} & {\mathbf 0}_{2^t-1}^T\\ \hline
-{\mathbf I}_{2^t-1} & {\mathbf 0}_{2^t-1} & {\rm diag}\left(\frac{1-\boldsymbol{\rho}_{i(2)}^{(t+1)}}{\boldsymbol{\rho}_{i(2)}^{(t+1)}}\right)
\end{array}\right] \in \mathbb{R}^{(2^{t+1}-1)\times (2^{t+1}-1)}\ ,
\]
$({\mathbf D}_i^{(t+1)})^{-1} =$
\footnotesize
\[
 \left[\begin{array}{c|c|c}
{\rm diag}(1- \boldsymbol{\rho}_{i(2)}^{(t+1)})({\mathbf D}_i^{(t)})^{-1} & \frac{\rho_{i,2^t}}{1-\rho_{i,2^t}} \cdot {\rm  diag}(1-\boldsymbol{\rho}_{i(2)}^{(t+1)}) \cdot ({\mathbf D}_i^{(t)})^{-1} {\mathbf b}^{(t)} & -{\rm diag}(\boldsymbol{\rho}_{i(2)}^{(t+1)})\\ \hline
{\mathbf 0}_{2^t-1}^T & \frac{\rho_{i,2^t}}{1-\rho_{i,2^t}} & {\mathbf 0}_{2^t-1}^T\\ \hline
{\rm diag}(\boldsymbol{\rho}_{i(2)}^{(t+1)})({\mathbf D}_i^{(t)})^{-1} & \frac{\rho_{i,2^t}}{1-\rho_{i,2^t}} \cdot {\rm  diag}(\boldsymbol{\rho}_{i(2)}^{(t+1)}) \cdot ({\mathbf D}_i^{(t)})^{-1} {\mathbf b}^{(t)} & {\rm diag}(\boldsymbol{\rho}_{i(2)}^{(t+1)})
\end{array}\right]\ ,
\]
\normalsize
\[
({\mathbf D}_i^{(t+1)})^{-1}{\mathbf b}^{(t+1)} = \left[\begin{array}{c}
\frac{1}{1-\rho_{i,2^t}} \cdot {\rm  diag}(1-\boldsymbol{\rho}_{i(2)}^{(t+1)}) \cdot ({\mathbf D}_i^{(t)})^{-1} {\mathbf b}^{(t)}\\ \hline
\frac{\rho_{i,2^t}}{1-\rho_{i,2^t}}\\ \hline
\frac{1}{1-\rho_{i,2^t}} \cdot {\rm  diag}(\boldsymbol{\rho}_{i(2)}^{(t+1)}) \cdot ({\mathbf D}_i^{(t)})^{-1} {\mathbf b}^{(t)}
\end{array}\right] \in \mathbb{R}^{2^{t+1}-1}\ ,
\]
where $\boldsymbol{\rho}_{i(2)}^{(t+1)} = (\rho_{i,2^t+1}, \ldots, \rho_{i,2^{t+1}-1})^T \in \mathbb{R}^{2^t-1}$.
\end{itemize}  
Then ${\mathbf D}_i = {\mathbf D}_i^{(T)}, {\mathbf D}_i^{-1} = ({\mathbf D}_i^{(T)})^{-1}$ and ${\mathbf D}_i^{-1} {\mathbf b} = ({\mathbf D}_i^{(T)})^{-1} {\mathbf b}^{(T)}$.
\end{theorem}

As a direct corollary of Theorem~\ref{thm:Di_inverse_binary_clm}, we obtain the following interesting results:

\begin{corollary}\label{cor:|Di|_binary_clm}
For the dichotomous conditional link model described by Example~\ref{ex:binary_cond_link_model}, we have
\begin{itemize}
\item[(i)] $|D_i| = \prod_{l=1}^{J-1} \rho_{il}^{-1} \cdot \prod_{t=1}^T (1-\rho_{i,2^t-1}) > 0$~;
\item[(ii)] $1+{\mathbf 1}_{J-1}^T {\mathbf D}_i^{-1} {\mathbf b} = \prod_{t=1}^T (1-\rho_{i,2^{t-1}})^{-1} > 0$~;
\item[(iii)] All the $J-1$ coordinates of ${\mathbf D}_i^{-1} {\mathbf b}$ are strictly positive.
\end{itemize}
\end{corollary}

Corollary~\ref{cor:|Di|_binary_clm} clears the case of dichotomous conditional link models in Theorem~\ref{thm:models_assumption_1234}. That is, $\boldsymbol{\Theta} = \mathbb{R}^p$ for dichotomous conditional link models.
Furthermore,  Lemma~\ref{lem:L_R_b_iterative_binary_clm}, Theorem~\ref{thm:Di_inverse_binary_clm}, and Corollary~\ref{cor:|Di|_binary_clm} provide detailed technical supports on computations related to dichotomous conditional link models (see Step~$3^\circ$ of Algorithm~\ref{algo:fisher_matrix} in Appendix~\ref{subsec:calculate_gradient_fisher}).   

\section{Dichotomous conditional link models and others}
\label{sec:more_on_multi_condi_link_model}

In this section, we use a toy example to explain the connections and differences between the log-linear models, the multivariate logistic models in the literature \citep{pmcc1995}, and the dichotomous conditional link model proposed in Section~\ref{sec:multinomial_conditional_link}.

Suppose there are two binary responses $Z_{i1}, Z_{i2} \in \{0,1\}$ at each of the $i$th covariate vector ${\mathbf x}_i$~, $i=1, \ldots, m$. In our notations, we denote the categorical probabilities $\pi_{i1} = P(Z_{i1}=1, Z_{i2}=0)$, $\pi_{i2} = P(Z_{i1}=0, Z_{i2} = 1)$, $\pi_{i3} = P(Z_{i1}=1, Z_{i2}=1)$, and $\pi_{i4} = P(Z_{i1}=0, Z_{i2}=0)$.

According to \cite{pmcc1995}, a multivariate logistic model or a bivariate logistic model (see also \cite{pmcc1989}) takes the form (with their ``2'' replaced by our ``0'') of
\begin{equation}\label{eq:bivariate_logistic}
\left\{\begin{array}{rcl}
\log\left(\frac{\pi_{i1}+\pi_{i3}}{\pi_{i2}+\pi_{i4}}\right) & = & \eta_{i1}\ =\ {\rm logit }(P(Z_{i1}=1))\ ;\\
\log\left(\frac{\pi_{i2}+\pi_{i3}}{\pi_{i1}+\pi_{i4}}\right) & = & \eta_{i2}\ =\ {\rm logit }(P(Z_{i2}=1))\ ;\\
\log\left(\frac{\pi_{i3} \cdot \pi_{i4}}{\pi_{i1}\cdot \pi_{i2}}\right) & = & \eta_{i3}\ .
\end{array}\right.
\end{equation}
It can be verified that we must have $\eta_{i3}\equiv 0$ in \eqref{eq:bivariate_logistic} if $Z_{i1}$ and $Z_{i2}$ are independent. In other words, the multivariate logistic model~\eqref{eq:bivariate_logistic} is equivalent to two marginal univariate logistic regression models under an independence assumption of $Z_{i1}$ and $Z_{i2}$~.

As for log-linear models, a typical one \citep{pmcc1995} takes the form of 
\begin{equation}\label{eq:bivariate_log_linear}
\left\{\begin{array}{rcl}
\log\left(\frac{\pi_{i3}}{\pi_{i2}}\right) & = & \eta_{i1}\ ;\\
\log\left(\frac{\pi_{i3}}{\pi_{i1}}\right) & = & \eta_{i2}\ ;\\
\log\left(\frac{\pi_{i3} \cdot \pi_{i4}}{\pi_{i1}\cdot \pi_{i2}}\right) & = & \eta_{i3}\ .
\end{array}\right.
\end{equation}
If $Z_{i1}$ and $Z_{i2}$ are independent, we must have $\eta_{i3}\equiv 0$ in \eqref{eq:bivariate_log_linear}. 
It can be verified that in this case, $\log(\pi_{i1}/\pi_{i4}) = {\rm logit }(P(Z_{i1}=1))$, and $\log(\pi_{i2}/\pi_{i4}) = {\rm logit }(P(Z_{i2}=1))$. That is, the log-linear model~\eqref{eq:bivariate_log_linear} is also equivalent to two marginal univariate logistic regression models under an independence assumption of $Z_{i1}$ and $Z_{i2}$~.

Following Example~\ref{ex:binary_cond_link_model}, a dichotomous conditional link model with logit link takes the form of
\begin{equation}\label{eq:bivariate_multinomial_cond_link_model}
\left\{\begin{array}{rcl}
\log\left(\frac{\pi_{i1}+\pi_{i3}}{\pi_{i2}+\pi_{i4}}\right) & = & \eta_{i1}\ =\ {\rm logit }(P(Z_{i1}=1))\ ;\\
\log\left(\frac{\pi_{i2}}{\pi_{i4}}\right) & = & \eta_{i2}\ =\ {\rm logit }(P(Z_{i2}=1\mid Z_{i1}=0))\ ;\\
\log\left(\frac{\pi_{i3}}{\pi_{i1}}\right) & = & \eta_{i3}\ =\ {\rm logit }(P(Z_{i2}=1\mid Z_{i1}=1))\ .
\end{array}\right.
\end{equation}
If $Z_{i1}$ and $Z_{i2}$ are independent, then $P(Z_{i2}=1\mid Z_{i1}=0) =  P(Z_{i2}=1\mid Z_{i1}=1) = P(Z_{i2}=1)$, which implies ${\rm logit }(P(Z_{i2}=1)) = \eta_{i2}\equiv \eta_{i3}$. In other words, in this case, the dichotomous conditional logit model is also  equivalent to two marginal univariate logistic regression models.

As a conclusion, when $Z_{i1}$ and $Z_{i2}$ are independent, these three models are all equivalent to two marginal logistic models.

\medskip
However, in general when $Z_{i1}$ and $Z_{i2}$ are not independent, these three models are different in nature. According to the first two equations in the multivariate logistic model~\eqref{eq:bivariate_logistic}, the marginal models for $Z_{i1}$ and $Z_{i2}$ are still univariate logistic models.

For the log-linear model~\eqref{eq:bivariate_log_linear}, it can be verified that
\begin{eqnarray*}
    {\rm logit }(P(Z_{i1}=1)) &=& \eta_{i1} + \log\left(\frac{1+\exp\{\eta_{i2}\}}{\exp\{\eta_{i3}\} + \exp\{\eta_{i2}\}}\right)\ ;\\
    {\rm logit }(P(Z_{i2}=1)) &=& \eta_{i2} + \log\left(\frac{1+\exp\{\eta_{i1}\}}{\exp\{\eta_{i3}\} + \exp\{\eta_{i1}\}}\right)\ .    
\end{eqnarray*}
Neither of these two equations belong to a univariate logistic model in general.

As for the dichotomous conditional logit model~\eqref{eq:bivariate_multinomial_cond_link_model}, its first equation indicates that the marginal model for $Z_{i1}$ is still a univariate logistic one. However, it can be verified that
\[
{\rm logit }(P(Z_{i2}=1)) = \eta_{i3} + \log\left(\frac{\exp\{\eta_{i1} + \eta_{i2}\} + \exp\{\eta_{i1}\} + \exp\{\eta_{i2}\} + \exp\{\eta_{i2} - \eta_{i3}\}}{\exp\{\eta_{i1}+\eta_{i2}\} + \exp\{\eta_{i1}\} + \exp\{\eta_{i3}\} + 1}\right)
\]
is not a univariate logistic model in general.

\section{Summary of notations for specifying a multinomial link model} \label{sec:summary_notation}

In this section, we summarize the notations for specifying a multinomial link model proposed in Section~\ref{sec:mlm}.

A general multinomial link model takes its matrix form as in \eqref{eq:mlm_in_matrix} or its equation form as in \eqref{eq:mlm_j}. It consists of two components. The left hand side $g_j\left(\frac{{\mathbf L}^T_j \boldsymbol\pi_i}{{\mathbf R}^T_j \boldsymbol\pi_i + \pi_{iJ} b_j}\right)$ of \eqref{eq:mlm_j} indicates that the model is a baseline-category mixed-link model (see Appendix~\ref{sec:baseline-category}), a cumulative mixed-link model (see Appendix~\ref{sec:cumulative}), an adjacent-categories mixed-link model (see Appendix~\ref{sec:adjacent-categories}), a continuation-ratio mixed-link model (see Appendix~\ref{sec:continuation-ratio}), a baseline-cumulative (two-group) mixed-link model (see Appendix~\ref{sec:b-c_link_model}), a baseline-adjacent (two-group) mixed-link model (see Appendix~\ref{sec:b-a-c_link_model}), a baseline-continuation (two-group) mixed-link model (see Appendix~\ref{sec:b-c-r_link_model}), a dichotomous conditional link model (see Example~\ref{ex:binary_cond_link_model}), or others.
The right hand side 
${\mathbf f}_j^T({\mathbf x}_i) \boldsymbol{\theta}$ of \eqref{eq:mlm_j} indicates that the model is with proportional odds (po, with $\beta_j+{\mathbf h}_c^T({\mathbf x}_i)\boldsymbol\zeta$), nonproportional odds (npo, with ${\mathbf h}_j^T({\mathbf x}_i)\boldsymbol\beta_j$), partial proportional odds (ppo, with ${\mathbf h}_j^T({\mathbf x}_i)\boldsymbol\beta_j+{\mathbf h}_c^T({\mathbf x}_i)\boldsymbol\zeta$, see Example~\ref{ex:ppo_model_H}), po-npo mixture (with ${\mathbf h}_j^T({\mathbf x}_i)\boldsymbol\beta_j+{\mathbf h}_{cj}^T({\mathbf x}_i)\boldsymbol\zeta$, see Example~\ref{ex:po_ppo_mixture}), or other structures. Overall, the model is called, for example, a cumulative mixed-link model with proportional odds, a baseline-cumulative (two-group) mixed-link model with po-npo mixture, etc. To specify such a model, we need to know 
\begin{itemize}
    \item[(i)] Constant matrices ${\mathbf L}, {\mathbf R} \in \mathbb{R}^{(J-1)\times (J-1)}$ and constant vector ${\mathbf b} \in \mathbb{R}^{J-1}$;
    \item[(ii)] Link functions $g_1, \ldots, g_{J-1}$~, as well as their inverses $g_1^{-1}, \ldots, g_{J-1}^{-1}$ and the corresponding first-order derivatives $(g_1^{-1})', \ldots, (g_{J-1}^{-1})'$ (see Table~\ref{tab:link_functions} for relevant formulae);
    \item[(iii)] Predictor functions ${\mathbf f}_j({\mathbf x}_i) = (f_{j1}({\mathbf x}_i), \ldots, f_{jp}({\mathbf x}_i))^T$ with $\boldsymbol{\theta} = (\theta_1, \ldots, \theta_p)^T$ in general;  ${\mathbf h}_j({\mathbf x}_i) = (h_{j1}({\mathbf x}_i), \ldots, h_{jp_j}({\mathbf x}_i))^T$, $j=1, \ldots, J-1$ and ${\mathbf h}_c({\mathbf x}_i) = (h_1({\mathbf x}_i), \ldots, h_{p_c}({\mathbf x}_i))^T$ or ${\mathbf h}_{cj}({\mathbf x}_i) = (h_{cj1}({\mathbf x}_i), \ldots, h_{cjp_c}({\mathbf x}_i))^T$ with parameters $\boldsymbol\theta = (\boldsymbol\beta_1^T,$ $ \ldots, \boldsymbol\beta_{J-1}^T, \boldsymbol\zeta^T)^T \in \mathbb{R}^{p_1 + \cdots + p_{J-1} + p_c}$ for ppo model or po-npo mixture model; ${\mathbf h}_j({\mathbf x}_i) \equiv 1$, $j=1, \ldots, J-1$, $p_1=\cdots = p_{J-1}=1$ with $\boldsymbol\theta = (\beta_1, \ldots, \beta_{J-1},$ $\boldsymbol\zeta^T)^T \in \mathbb{R}^{J-1+p_c}$  for po models; ${\mathbf h}_c({\mathbf x}_i) \equiv 0$, $p_c=0$ and $\boldsymbol\theta = (\boldsymbol\beta_1^T, \ldots, \boldsymbol\beta_{J-1}^T)^T \in \mathbb{R}^{p_1 + \cdots + p_{J-1}}$ for npo models.
\end{itemize}
Once (i), (ii) and (iii) are given, the model is specified. We can further calculate 
\begin{itemize}
\item[(iv)] The model matrix ${\mathbf X}_i \in \mathbb{R}^{(J-1)\times p}$ according to \eqref{eq:general_X_i} for general models. Special cases include \eqref{eq:po_npo_Xi}
for po-npo mixture model (see Example~\ref{ex:po_ppo_mixture}), \eqref{eqn:Xi_ppo_J-1} for ppo model,
\[
	{\mathbf X}_i= \begin{pmatrix}
		1 &   & & {\mathbf x}_i^T\\
		&   \ddots &  & \vdots \\
		&   & 1 & {\mathbf x}_i^T
	\end{pmatrix}\> \in \mathbb{R}^{(J-1) \times (d+J-1)}
\]
for main-effects po models (in this case, $p_1 = \cdots = p_{J-1}=1$, $p_c = d$, $p=d+J-1$), 
\begin{equation}\label{eq:main_effects_npo_Xi}
	{\mathbf X}_i= \begin{pmatrix}
		1\ {\mathbf x}_i^T &   &  \\
		&   \ddots &   \\
		&   & 1\ {\mathbf x}_i^T
	\end{pmatrix}\> \in \mathbb{R}^{(J-1) \times (d+1)(J-1)}
\end{equation}
for main-effects npo models (in this case, $p_1 = \cdots = p_{J-1} = d+1$, $p_c=0$, $p=(d+1)(J-1)$). 
\end{itemize}

\section{More on Fisher information matrix}\label{subsec:supp_fisher_mat}

In this section, we provide more technical details about Section~\ref{sec:fisher_information} on Fisher information matrix.

Recall that given distinct ${\mathbf x}_i, i=1,\cdots,m$, we have independent multinomial responses
\[
{\mathbf Y}_i=(Y_{i1},\cdots,Y_{iJ})^T \sim {\rm Multinomial}(n_i; \pi_{i1},\cdots,\pi_{iJ})\ ,
\]
where $n_i=\sum_{j=1}^J Y_{ij} > 0$ and $0 < \pi_{ij} < 1$. The log-likelihood for the multinomial model is
\[
	l(\boldsymbol\theta) =\log \left(\prod_{i=1}^m \frac{{n_i}!}{{Y_{i1}}!\cdots{Y_{iJ}}!} \pi_{i1}^{Y_{i1}}\cdots\pi_{iJ}^{Y_{iJ}}\right) =\sum_{i=1}^m {\mathbf Y}_i^T\log \bar{\boldsymbol\pi}_i + \sum_{i=1}^m \log (n_i!) - \sum_{i=1}^m\sum_{j=1}^J \log (Y_{ij}!)\ ,
\]
where $\bar{\boldsymbol\pi}_i = (\pi_{i1}, \ldots, \pi_{iJ})^T = (\boldsymbol\pi_i^T, \pi_{iJ})^T$, and $\log \bar{\boldsymbol\pi}_i=(\log \pi_{i1}, \cdots,\log \pi_{iJ})^T$.  

Recall that $g_1^{-1}, \ldots, g_{J-1}^{-1}$ are all differentiable. Then the score vector
\[
	\frac{\partial l}{\partial \boldsymbol\theta^T} = \sum_{i=1}^{m}{\mathbf Y}_i^T{\rm diag}(\bar{\boldsymbol\pi}_i)^{-1}\frac{\partial \bar{\boldsymbol\pi}_i}{\partial \boldsymbol\theta^T}
\]
with
\[
	\frac{\partial \bar{\boldsymbol\pi}_i}{\partial \boldsymbol\theta^T}
	= \frac{\partial \bar{\boldsymbol\pi}_i}{\partial \boldsymbol\rho_i^T} \cdot \frac{\partial \boldsymbol\rho_i}{\partial \boldsymbol\eta_i^T}  \cdot \frac{\partial \boldsymbol\eta_i}{\partial \boldsymbol\theta^T}\\
	= \frac{\partial \bar{\boldsymbol\pi}_i}{\partial \boldsymbol\rho_i^T} \cdot {\rm diag}\left(\left({\mathbf g}^{-1}\right)'\left(\boldsymbol\eta_i\right)\right) \cdot {\mathbf X}_i\ ,
\]
where $\boldsymbol\rho_i = (\rho_{i1}, \ldots, \rho_{i,J-1})^T \in \mathbb{R}^{J-1}$, $\boldsymbol\eta_i = (\eta_{i1},$ $\ldots,$ $\eta_{i,J-1})^T$ $\in $ $\mathbb{R}^{J-1}$, ${\rm diag}\left(\left({\mathbf g}^{-1}\right)'\left(\boldsymbol\eta_i\right)\right)$ $ = {\rm diag}\left\{\left(g_1^{-1}\right)'\left(\eta_{i1}\right), \ldots, \left(g_{J-1}^{-1}\right)'\left(\eta_{i, J-1}\right)\right\} \in \mathbb{R}^{(J-1)\times (J-1)}$. As for $\partial \bar{\boldsymbol\pi}_i/\partial \boldsymbol\rho_i^T$, Lemma~\ref{lem:part_pi_part_rho} provides a formula for $\partial \boldsymbol\pi_i/\partial \boldsymbol\rho_i^T$~. 

\begin{lemma}\label{lem:part_pi_part_rho}
Suppose $\boldsymbol{\theta}\in \boldsymbol{\Theta}$. Then 
\[
\frac{\partial\boldsymbol\pi_i}{\partial \boldsymbol\rho_i^T} = \left({\mathbf I}_{J-1} - \boldsymbol\pi_i {\mathbf 1}_{J-1}^T\right) {\mathbf D}_i^{-1} \cdot {\rm diag}\left({\mathbf L}\boldsymbol\pi_i\right) \cdot {\rm diag} \left(\boldsymbol\rho_i^{-2} \right)\ .
\]
\hfill{$\Box$}
\end{lemma}

\noindent
{\bf Proof of Lemma~\ref{lem:part_pi_part_rho}:} 
Applying the chain rule of vector differentiation and matrix differentials (see, for example, Chapter~17 in \cite{seber2008}) to \eqref{eq:pi_from_rho}, we obtain
\begin{eqnarray*}
\frac{\partial\boldsymbol\pi_i}{\partial \boldsymbol\rho_i^T} &=& \frac{\partial\boldsymbol\pi_i}{\partial ({\mathbf D}_i^{-1} {\mathbf b})^T} \cdot \frac{\partial ({\mathbf D}_i^{-1} {\mathbf b})}{\partial \boldsymbol\rho_i^T}\\
&=& \frac{1}{1 + {\mathbf 1}^T_{J-1} {\mathbf D}_i^{-1} {\mathbf b}} \left({\mathbf I}_{J-1} - \boldsymbol\pi_i {\mathbf 1}_{J-1}^T\right) \cdot {\mathbf D}_i^{-1} {\rm diag} \left({\mathbf L} {\mathbf D}_i^{-1}{\mathbf b}\right) {\rm diag} \left(\boldsymbol{\rho}_i^{-2}\right) \\
&=& \left({\mathbf I}_{J-1} - \boldsymbol\pi_i {\mathbf 1}_{J-1}^T\right) \cdot {\mathbf D}_i^{-1} {\rm diag} \left({\mathbf L} \boldsymbol{\pi}_i\right) {\rm diag} \left(\boldsymbol{\rho}_i^{-2}\right)\ .
\end{eqnarray*}
\hfill{$\Box$}

Since $\pi_{iJ} = 1 - {\mathbf 1}_{J-1}^T \boldsymbol{\pi}_i$~, 
\[
\frac{\partial \pi_{iJ}}{\partial \boldsymbol\rho_i^T} = -{\mathbf 1}_{J-1}^T \frac{\partial\boldsymbol\pi_i}{\partial \boldsymbol\rho_i^T} = \left({\mathbf 0}_{J-1}^T - \pi_{iJ} {\mathbf 1}_{J-1}^T\right) {\mathbf D}_i^{-1} \cdot {\rm diag}\left({\mathbf L}\boldsymbol\pi_i\right) \cdot {\rm diag} \left(\boldsymbol\rho_i^{-2} \right) \ .
\]
By combining $\partial \boldsymbol\pi_i/\partial \boldsymbol\rho_i^T$ and $\partial \pi_{iJ}/\partial \boldsymbol\rho_i^T$, we obtain
\begin{equation}\label{eq:part_bar_pi_part_rho}
\frac{\partial \bar{\boldsymbol{\pi}}_i}{\partial \boldsymbol\rho_i^T} = {\mathbf E}_i {\mathbf D}_i^{-1} \cdot {\rm diag}\left({\mathbf L}\boldsymbol\pi_i\right) \cdot {\rm diag} \left(\boldsymbol\rho_i^{-2} \right) \ ,
\end{equation}
where ${\mathbf E}_i$ is defined in \eqref{eq:E_i}.

\begin{lemma}\label{lemma:for_partial_l}
$ 
\bar{\boldsymbol\pi}_i^T {\rm diag}(\bar{\boldsymbol\pi}_i)^{-1}{\mathbf E}_i= {\mathbf 1}_J^T {\mathbf E}_i = {\mathbf 0}  
$.
\hfill{$\Box$}
\end{lemma}

\noindent
{\bf Proof of Lemma~\ref{lemma:for_partial_l}:}
Since $\pi_{i1} + \cdots + \pi_{iJ} =1$, then $\bar{\boldsymbol\pi}_i^T {\rm diag}(\bar{\boldsymbol\pi}_i)^{-1}{\mathbf E}_i = {\mathbf 1}_J^T {\mathbf E}_i = {\mathbf 1}_{J-1}^T - {\mathbf 1}_{J-1}^T = {\mathbf 0}_{J-1}^T$~.
\hfill{$\Box$}

Since the product of two diagonal matrices is exchangeable, then
\[
{\rm diag} \left({\mathbf L} \boldsymbol{\pi}_i\right) \cdot {\rm diag} \left(\boldsymbol{\rho}_i^{-2}\right) = {\rm diag}\left((1-\boldsymbol\rho_i)/\boldsymbol\rho_i\right) \cdot {\rm diag} \left({\mathbf L} \boldsymbol{\pi}_i\right) \cdot {\rm diag} \left(\boldsymbol\rho_i (1-\boldsymbol\rho_i)\right)^{-1}\ ,
\]
where ${\rm diag}\left((1-\boldsymbol\rho_i)/\boldsymbol\rho_i\right) = {\rm diag}\left\{(1-\rho_{i1})/\rho_{i1}, \ldots, (1-\rho_{i,J-1})/\rho_{i,J-1}\right\}$ and\\ ${\rm diag} \left(\boldsymbol\rho_i (1-\boldsymbol\rho_i)\right)^{-1} = {\rm diag}\left\{\rho_{i1}^{-1} (1-\rho_{i1})^{-1}, \ldots, \rho_{i,J-1}^{-1} (1-\rho_{i,J-1})^{-1}\right\}$.
Thus an equivalent formula of \eqref{eq:part_bar_pi_part_rho} is
\[
\frac{\partial\bar{\boldsymbol\pi}_i}{\partial \boldsymbol\rho_i^T} = {\mathbf E}_i {\mathbf D}_i^{-1} \cdot {\rm diag}\left(\frac{1-\boldsymbol\rho_i}{\boldsymbol\rho_i}\right) \cdot {\rm diag}\left({\mathbf L}\boldsymbol\pi_i\right) \cdot {\rm diag} \left(\boldsymbol\rho_i (1-\boldsymbol\rho_i)\right)^{-1}\ .
\]
It can be verified that ${\mathbf E}_i {\mathbf D}_i^{-1} \cdot {\rm diag}\left((1-\boldsymbol\rho_i)/\boldsymbol\rho_i\right) \cdot {\rm diag}\left({\mathbf L}\boldsymbol\pi_i\right)$ is consistent with the first $(J-1)$ columns of $({\mathbf C}^T {\mathbf D}_i^{-1} {\mathbf L})^{-1}$ in \cite{bu2020} for multinomial logitistic models, although their ${\mathbf D}_i$ and ${\mathbf L}$ are different from here. Therefore, Lemma~S.5 in the Supplementary Material of \cite{bu2020} is a direct conclusion of Lemma~\ref{lemma:for_partial_l} here.

As another direct conclusion of Lemma~\ref{lemma:for_partial_l},
\[
	E\left(\frac{\partial l}{\partial \boldsymbol\theta^T}\right) = \sum_{i=1}^{m} n_i \bar{\boldsymbol{\pi}}_i^T{\rm diag}(\bar{\boldsymbol\pi}_i)^{-1}\frac{\partial \bar{\boldsymbol\pi}_i}{\partial \boldsymbol\theta^T} = {\mathbf 0}\ .
\]

\section{Algorithm for calculating gradient and Fisher information matrix}\label{subsec:calculate_gradient_fisher}

In this section, we provide detailed formulae for Step~$2$ of Algorithm~\ref{algo:fisher_scoring} in Section~\ref{sec:fisher_scoring}, which are summarized in Algorithm~\ref{algo:fisher_matrix}.

\begin{algorithm}
Calculating gradient and Fisher information matrix at $\boldsymbol\theta$
\label{algo:fisher_matrix}
\begin{itemize}
        \item[$0^\circ$] Input: ${\mathbf X}_i \in \mathbb{R}^{(J-1)\times p}$ according to \eqref{eq:general_X_i}, ${\mathbf y}_i \in \mathbb{R}^J$, $i=1, \ldots, m$; an feasible $\boldsymbol\theta \in \boldsymbol\Theta$.
	\item[$1^\circ$] Obtain $\boldsymbol\eta_i = {\mathbf X}_i \boldsymbol\theta$, $i=1, \ldots, m$. Here $\boldsymbol\eta_i = (\eta_{i1}, \ldots, \eta_{i,J-1})^T \in \mathbb{R}^{J-1}$.
	\item[$2^\circ$] Obtain $\rho_{ij} = g_j^{-1}(\eta_{ij})$, $j=1, \ldots, J-1$ and $\boldsymbol\rho_i = (\rho_{i1}, \ldots, \rho_{i,J-1})^T \in \mathbb{R}^{J-1}$, $i=1, \ldots, m$.
	\item[$3^\circ$] Calculate ${\mathbf D}_i^{-1} = \left[{\rm diag}\left(\boldsymbol\rho_i^{-1}\right) {\mathbf L} - {\mathbf R} \right]^{-1} \in \mathbb{R}^{(J-1)\times (J-1)}$ (avoid calculating the inverse matrix directly whenever an explicit formula is available) and ${\mathbf D}_i^{-1} {\mathbf b}$, $i=1, \ldots, m$.
	\item[$4^\circ$] Calculate $\boldsymbol\pi_i = ({\mathbf D}_i^{-1} {\mathbf b})/(1 + {\mathbf 1}^T_{J-1} {\mathbf D}_i^{-1} {\mathbf b}) \in \mathbb{R}^{J-1}$, 
	$\pi_{iJ} = \left(1 + {\mathbf 1}^T_{J-1} {\mathbf D}_i^{-1} {\mathbf b}\right)^{-1} \in \mathbb{R}$ and thus $\bar{\boldsymbol\pi}_i = (\boldsymbol\pi_i^T, \pi_{iJ})^T = (\pi_{i1}, \ldots, \pi_{i,J-1}, \pi_{iJ})^T \in \mathbb{R}^J$, $i=1, \ldots, m$.
 
    \item[$5^\circ$] Calculate ${\mathbf C}_i = {\mathbf E}_i {\mathbf D}_i^{-1} \cdot {\rm diag}\left( ({\mathbf L}\boldsymbol\pi_i) \circ (\boldsymbol\rho_i^{-2})  \circ \left({\mathbf g}^{-1}\right)'(\boldsymbol\eta_i)\right) \in \mathbb{R}^{J\times (J-1)}$, where 
	\begin{eqnarray*}
		{\mathbf E}_i &=& 
		\left[\begin{array}{c}
			{\mathbf I}_{J-1}\\ {\mathbf 0}_{J-1}^T
		\end{array}\right] -
		\bar{\boldsymbol{\pi}}_i {\mathbf 1_{J-1}^T} \> \in \> \mathbb{R}^{J\times (J-1)}\ ,\\
		\boldsymbol\rho_i^{-2} &=& \left(\rho_{i1}^{-2}, \ldots, \rho_{i,J-1}^{-2}\right)^T \>\in\> \mathbb{R}^{J-1} \ ,\\
		\left({\mathbf g}^{-1}\right)'\left(\boldsymbol\eta_i\right) &=& \left( (g_1^{-1})'(\eta_{i1}), \ldots, (g_{J-1}^{-1})'(\eta_{i,J-1})\right)^T \>\in \> \mathbb{R}^{J-1}\ , 
	\end{eqnarray*}
${\mathbf L}\boldsymbol\pi_i \in \mathbb{R}^{J-1}$, ``$\circ$'' denotes the element-wise product (also known as Hadamard product), ${\mathbf I}_{J-1}$ is the identity matrix of order $J-1$, ${\mathbf 0}_{J-1}$ is the vector of $J-1$ zeros, and ${\mathbf 1}_{J-1}$ is the vector of $J-1$ ones, $i=1, \ldots, m$. 
\item[$6^\circ$] Calculate the gradient $\partial l/\partial \boldsymbol\theta^T = \sum_{i=1}^{m} {\mathbf y}_i^T {\rm diag}(\bar{\boldsymbol\pi}_i)^{-1} {\mathbf C}_i {\mathbf X}_i$
at $\boldsymbol{\theta}$.
\item[$7^\circ$] Calculate the Fisher information ${\mathbf F}_i = {\mathbf X}_i^T {\mathbf C}_i^T {\rm diag}(\bar{\boldsymbol\pi}_i)^{-1} {\mathbf C}_i {\mathbf X}_i$ at ${\mathbf x}_i$~, $i=1, \ldots, m$ and then the Fisher information matrix ${\mathbf F}(\boldsymbol{\theta})=\sum_{i=1}^{m}n_i{\mathbf F}_i$ at $\boldsymbol{\theta}$.
\item[$8^\circ$]  Report $\partial l/\partial \boldsymbol\theta^T$ and ${\mathbf F}(\boldsymbol{\theta})$.
\end{itemize}
\end{algorithm}

For Step~3 of Algorithm~\ref{algo:fisher_matrix}, instead of calculating the numerical inverse of matrix ${\mathbf D}_i$~, we recommend using the explicit formulae for calculating ${\mathbf D}_i^{-1}$ directly (see Appendices~\ref{subsec:m_link_m}, \ref{subsec:more_two_group} and \ref{sec:more_on_dichotomous_condi_link_model}).

\section{Algorithm for finding the most appropriate po-npo mixture model}\label{sec:po_npo_mixture_algorithm}

Inspired by the backward selection strategy for selecting a subset of covariates \citep{hastie2009elements, dousti2023variable}, in this section we provide a backward selection algorithm for finding the most appropriate po-npo mixture model (see Example~\ref{ex:po_ppo_mixture}) for a given dataset. It aims to identify a good (if not the best) po-npo mixture model by iteratively merging the closest pair of parameters or dropping off the parameter that has the smallest absolute value.

\begin{algorithm}\label{algo:mixture_model}
Backward selection for the most appropriate po-npo mixture model
\begin{itemize}
    \item[$1^\circ$] First fit the corresponding main-effects npo model (see \eqref{eq:main_effects_npo_Xi}) and tabularize the fitted parameters as $\hat{\boldsymbol{\beta}}^{(0)} = (\hat{\beta}_{jl}^{(0)}) \in \mathbb{R}^{(J-1)\times (d+1)}$ with the corresponding $\AIC$ value $\textsc{AIC}^{(0)}$. The rows of $\hat{\boldsymbol{\beta}}^{(0)}$ represent the model equations labelled by $j=1, \ldots, J-1$, and the columns represent the intercepts ($l=1$) and the $d$ covariates ($l=2, \ldots, d+1$). Denote the initial set of constraints on $\hat{\boldsymbol{\beta}}^{(0)}$ as ${\mathcal C}^{(0)} = \emptyset$.

    \item[$2^\circ$] For $t\geq 1$, given the previous estimated parameters $\hat{\boldsymbol{\beta}}^{(t-1)} = (\hat{\beta}_{jl}^{(t-1)})$ and its set of constraints ${\mathcal C}^{(t-1)}$, consider a predetermined set ${\cal L} \subseteq \{1, 2, \ldots, d+1\}$ of column indices (e.g., ${\cal L} = \{2, \ldots, d+1\}$ for excluding operations on the intercepts), do
    \begin{itemize}
        \item[1)] Find the pair $1\leq a<b\leq J-1$ and $l\in {\cal L}$ such that $|\hat\beta_{al}^{(t-1)} - \hat\beta_{bl}^{(t-1)}|$ attains the minimum among all nonzero differences, denote ${\mathcal C}_{\text{merge}}^{(t)} = \{\hat\beta_{al}^{(t)} = \hat\beta_{bl}^{(t)}\}$, record the fitted parameters $\hat{\boldsymbol{\beta}}_{\text{merge}}^{(t)}$ under the constraints ${\mathcal C}^{(t-1)} \cup {\mathcal C}_{\text{merge}}^{(t)}$, and the corresponding $\AIC$ value  $\text{AIC}_{\text{merge}}^{(t)}$;

        \item[2)] Find $1\leq a\leq J-1$ and $l\in {\cal L}$ such that by adding constraint ${\mathcal C}_{\text{drop}}^{(t)} = \{\hat\beta_{al}^{(t)} = 0\}$, the fitted parameters  $\hat{\boldsymbol{\beta}}_{\text{drop}}^{(t)}$ achieves the smallest $\AIC$ value $\text{AIC}_{\text{drop}}^{(t)}$;
    \end{itemize}
    
    \item[$3^\circ$] Compare the $\AIC$ values  $\text{AIC}_{\text{merge}}^{(t)}$ and $\text{AIC}_{\text{drop}}^{(t)}$ obtained  in Step~$2^\circ$. If $\text{AIC}_{\text{merge}}^{(t)} < \text{AIC}_{\text{drop}}^{(t)}$, record $\hat{\boldsymbol{\beta}}^{(t)}=\hat{\boldsymbol{\beta}}_{\text{merge}}^{(t)}$ and ${\mathcal C}^{(t)} = {\mathcal C}^{(t-1)} \cup {\mathcal C}_{\text{merge}}^{(t)}$; otherwise, record $\hat{\boldsymbol{\beta}}^{(t)}=\hat{\boldsymbol{\beta}}_{\text{drop}}^{(t)}$ and ${\mathcal C}^{(t)} = {\mathcal C}^{(t-1)} \cup {\mathcal C}_{\text{drop}}^{(t)}$~. In other words, $\textsc{AIC}^{(t)} = \min\{\text{AIC}_{\text{merge}}^{(t)},\  \text{AIC}_{\text{drop}}^{(t)}\}$.

    \item[$4^\circ$]  
    If $\textsc{AIC}^{(t)} < \textsc{AIC}^{(t-1)}$, let $t \leftarrow t+1$ and go to Step~$2^\circ$; otherwise, go to Step~$5^\circ$.
    
    \item[$5^\circ$] Report the po-npo mixture model corresponding to constraints ${\mathcal C}^{(t-1)}$ as the most appropriate model with fitted parameters  $\hat{\boldsymbol{\beta}}^{(t-1)}$ and $\AIC$ value $\textsc{AIC}^{(t-1)}$.
\end{itemize}
\end{algorithm}

In Step~$2^\circ$ of Algorithm~\ref{algo:mixture_model}, the users have the option of allowing or not the intercepts to be merged or dropped, that is, ${\cal L} = \{1, 2, \ldots, d+1\}$ or $\{2, \ldots, d+1\}$. In many applications, it is common to assume an unconstrained intercept for each model equation (see Algorithm~\ref{algo:initial_theta_supplement}). However, one may gain more degrees of freedom for significance tests on model parameters by simplifying the structure of intercepts (see Section~\ref{sec:six_cities_cond_link_model} for an example) when parameter feasibility is not a concern (see Section~\ref{sec:feasible_parameter}).

\section{More on existing infeasibility issue}\label{sec:existing_issue_subsection}

In this section, we provide more technical details about the comparison study performed in Section~\ref{sec:MLM_existing_issue}, including the error/warning messages when using SAS \texttt{proc} {\tt logistic} and R function \texttt{vglm}. Recall that we fit a main-effects cumulative logit model with po on 1,000 bootstrapped datasets generated from the 802 observations summarized in Table V of \cite{chuang1997}, using SAS, R package {\tt VGAM}, and our algorithms for multinomial link models (MLM), respectively.

\subsection{SAS error information}
\label{sec:SAS_error_information}

When using SAS \texttt{proc} {\tt logistic} (SAS studio version 3.81, Enterprise Edition), negative fitted probabilities are detected for 44 out of the 1,000 bootstrapped datasets. The detailed warning message is, ``Negative individual predicted probabilities were identified in the final model fit.  You may want to modify your UNEQUALSLOPES specification. The LOGISTIC procedure continues in spite of the above warning. Results shown are based on the last maximum likelihood iteration. Validity of the model fit is questionable''. Among the 44 bootstrapped datasets, 40 of them involve a single negative predictive probability, and 4 have two negative predictive probabilities, which lead to 48 negative predictive probabilities in total. 

As a summary of the 48 negative predictive probabilities, the minimum is $-0.1186$, the maximum is $-7.6\times 10^{-5}$, the mean is $-0.0585$, and the median is $-0.0701$. In other words, the negative predictive probabilities cannot be simply treated as numerical errors. 

\subsection{R error information}
\label{sec:R_error_information}

When using the \texttt{vglm} function in R package \texttt{VGAM} (version 1.1-11, published on 2024-05-15), the function stops running on 4 datasets with error message ``NA/NaN/Inf in foreign function call (arg 1)'', indicating missing fitted probabilities during fitting the model. Besides those 4 cases, 38 datasets involve negative predictive probabilities with 44 warning messages. Among them, 23 datasets involve a single negative probability, and 15 cases have two negative probabilities.

As a summary of the 53 negative predictive probabilities, the minimum is $-0.0247$, the maximum is $-5.5\times 10^{-5}$, the mean is $-0.0055$, and the median is $-0.0033$. Those negative prbabilities cannot be explained as numerical errors either.

\subsection{Comparison among fitted models}
\label{sec:comparison_among_fitted_models}

In this section, instead of comparing the fitted parameter values based on different software, we  compare the fitted predictive probabilities $\hat{\pi}_{ij}$'s and the maximum log-likelihood $l(\hat{\boldsymbol{\theta}})$ (see Remark~\ref{remark:adjacent_vglm} in Appendix~\ref{sec:adjacent-categories}). 

\begin{figure}[ht]
\centering
\includegraphics[scale=0.5]{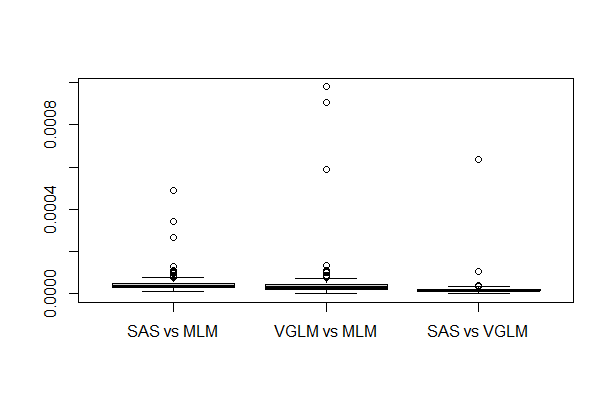}
\caption{Boxplots of Root Mean Squared Differences of Fitted Predictive Probabilities}
\label{fig:box_MSE_fitprob}
\end{figure}

We first calculate the root mean squared differences of the fitted probabilities $(m^{-1}J^{-1} $ $\sum_{i=1}^m \sum_{j=1}^J (\hat\pi_{ij}^{(a)}-\hat\pi_{ij}^{(b)})^2)^{1/2}$, where $a,b=1,2,3$ stand for SAS \texttt{proc} {\tt logistic}, R \texttt{vglm}, and our programs for MLM, respectively. In Figure~\ref{fig:box_MSE_fitprob}, we display the pairwise root mean squared differences only when both methods under comparison obtain feasible parameter estimates. Roughly speaking, the root mean squared differences are tiny, and the predictive probabilities are consistent across SAS, R and our algorithms when the fitted model parameters are feasible. 

\begin{figure}[ht]
\centering
\includegraphics[scale=0.5]{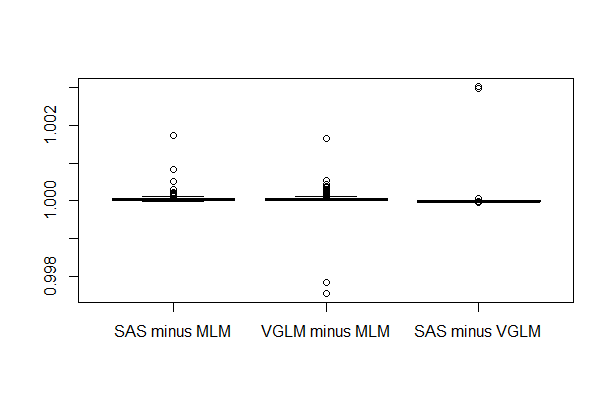}
\caption{Boxplots of Ratios of Maximum Likelihoods}
\label{fig:box_diff_lik}
\end{figure}

Next we compare the ratio of maximum likelihoods $\exp\{l(\hat{\boldsymbol{\theta}}^{(a)})-l(\hat{\boldsymbol{\theta}}^{(b)})\}$ for each pair of the three methods. If the ratio is about $1$, then the two maximum likelihoods are about the same, which indicates comparable performance in term of maximizing the likelihood. The boxplots of pairwise ratios, when both estimates are feasible, are displayed in Figure~\ref{fig:box_diff_lik}. The ratios are fairly close to $1$ using the three different methods. In other words, when the estimated parameters are feasible, the three methods roughly obtain consistent maximum likelihood. 

Both SAS \texttt{proc} {\tt logistic} and R \texttt{vglm} have infeasibility issues for about $4\%$ bootstrapped datasets in this example, while our algorithms work well for all the 1,000 cases. Overall, our algorithms outperform the current SAS and R package {\tt VGAM}.

\section{More on metabolic syndrome data analysis}\label{sec:more_metabolic_syndrome_two_group_model}

In this section, we continue our analysis in Section~\ref{sec:metabolic_syndrome} on the metabolic syndrome data described by \cite{musa2023data}. 

In Section~\ref{sec:metabolic_syndrome}, we obtain two fitted models selected by \AIC among the main-effects multinomial logistic models with npo. One is a continuation-ratio logit model \eqref{eq:continuation_model} with npo fitted on the data after removing the observations with NA responses, called the Model without NA (see Model 1 in Table~\ref{tab:param:metabolic_withoutna_nodrop}). The other is a baseline-continuation (two-group, namely $\{$DM, IFG$\}$ and $\{$Normal, IFG, NA$\}$) logistic model with npo (see also Example~\ref{ex:two_group_s}), called the Model with NA (see Model 2 in Table~\ref{tab:param:metabolic_withna_nodrop}). From the significance codes for $p$-values, we can see that not all estimated parameters are significantly different from zero, which suggests that further variable selections are necessary.

\begin{table}[ht]
    \centering
    \caption{Estimated Parameters of the Model without NA for Metabolic Syndrome Data}
    \label{tab:param:metabolic_withoutna_nodrop}
    \begin{tabular}{c|llll}
    \hline
    \multicolumn{5}{c}{Model 1: Continuation-ratio Logit Model with npo}\\ \hline
     & Intercept & Cholesterol & Weight & Hypertension \\
     $\log(\cdot)$ & $\hat\beta_{j1}$ & $\hat\beta_{j2}$ & $\hat\beta_{j3}$ & $\hat\beta_{j4}$\\
     \hline
     Normal/(IFG+DM) & $\;$3.995*** & -0.254*** & -0.022*** & -0.888***\\
     IFG/DM & -0.359*** & $\;$0.027*** & $\;$0.001 & $\;$0.038\\
    \hline
    \multicolumn{5}{c}{Model 1A: Continuation-ratio Logit Model with Selected po-npo Structure}\\ \hline
      & Intercept & Cholesterol & Weight & Hypertension \\
     $\log(\cdot)$ & $\hat\beta_{j1}$ & $\hat\beta_{j2}$ & $\hat\beta_{j3}$ & $\hat\beta_{j4}$\\
     \hline
     Normal/(IFG+DM) & $\;$3.988*** & -0.252*** & -0.022*** & -0.891***\\
     IFG/DM & -0.490*** & $\;$0 & $\;$0 & $\;$0\\
    \hline
    \end{tabular}\\
    \normalsize
    \footnotesize{Notes: Signif. codes for $p$-value: 0 ‘***’ 0.001 ‘**’ 0.01 ‘*’ 0.05 ‘.’ 0.1 ‘ ’ 1}\\
\end{table}

By applying Algorithm~\ref{algo:mixture_model} in Appendix~\ref{sec:po_npo_mixture_algorithm} to Model 1 in Table~\ref{tab:param:metabolic_withoutna_nodrop}, we find a significantly better model with a po-npo mixture (see Section~\ref{sec:po_npo_mixture_model}) in terms of {\AIC}  values (dropped from 727.97 to 722.56), denoted as Model~1A in Table~\ref{tab:param:metabolic_withoutna_nodrop}. The row of ``IFG/DM'' with zero coefficients on all covariates indicates $\log(\pi_{i,\mbox{\tiny IFG}}/\pi_{i,\mbox{\tiny DM}}) \equiv -0.490$. In other words, if we remove all observations with NA responses, we would conclude that the levels of total cholesterol, body weight, and hypertension status do not help with distinguishing DM patients from  IFG people.

\begin{table}[ht]
    \centering
    \caption{Estimated Parameters of the Model with NA for Metabolic Syndrome Data}
    \label{tab:param:metabolic_withna_nodrop}
    \begin{tabular}{c|llll}
    \hline
    \multicolumn{5}{c}{Model 2: Baseline-continuation Two-group Logit Model with npo}\\ \hline
      & Intercept & Cholesterol & Weight & Hypertension \\
     $\log(\cdot)$ & $\hat\beta_{j1}$ & $\hat\beta_{j2}$ & $\hat\beta_{j3}$ & $\hat\beta_{j4}$\\
     \hline
     DM/IFG & $\;$0.441*** & $\;$0.019*** & -0.001* & $\;$0.050.\\
     Normal/(IFG+NA) & $\;$3.379*** & -0.176*** & -0.011*** & -0.539***\\
     IFG/NA & -2.915*** & $\;$0.192*** & $\;$0.039*** &  $\;$1.235***\\
    \hline
    \multicolumn{5}{c}{Model 2A: Baseline-continuation Two-group Logit Model with Selected po-npo Structure}\\ \hline
      & Intercept & Cholesterol & Weight & Hypertension \\
     $\log(\cdot)$ & $\hat\beta_{j1}$ & $\hat\beta_{j2}$ & $\hat\beta_{j3}$ & $\hat\beta_{j4}$\\
     \hline
     DM/IFG & $\;$0.447*** & $\;$0.019*** & -0.001* & $\;$0\\
     Normal/(IFG+NA) & $\;$3.379*** & -0.176*** & -0.011*** & -0.564***\\
     IFG/NA & -2.916*** & $\;$0.192*** & $\;$0.039*** &  $\;$1.266***\\
    \hline
    \end{tabular}\\
    \normalsize
    \footnotesize{Notes: Signif. codes for $p$-value: 0 ‘***’ 0.001 ‘**’ 0.01 ‘*’ 0.05 ‘.’ 0.1 ‘ ’ 1}\\
\end{table}

Similarly, we apply Algorithm~\ref{algo:mixture_model} to Model 2 in Table~\ref{tab:param:metabolic_withna_nodrop}. The ended po-npo mixture model is denoted by Model~2A in Table~\ref{tab:param:metabolic_withna_nodrop}. Only the coefficient of \texttt{hpt} (hypertension status) is dropped for comparing DM against IFG. Although the difference between {\AIC} values (dropped from 927.56 to 925.65) is not significant according to \cite{burnham2004aic}, we still prefer Model~2A over Model~2 since it is also supported by the significance test on $\hat{\beta}_{14}$~. Note that the {\AIC} values of Model~1A and Model~2A are not comparable, since they are modeling different datasets (with or without the observations with NA).

According to the fitted Model~2A in Table~\ref{tab:param:metabolic_withna_nodrop}, we conclude that the hypertension status does not have a significant effect on distinguishing DM from IFG, which is the same as the one based on Model~1A. However, Model~2A suggests that the other covariates, namely total cholesterol and body weight, have significant effects on the classification between DM and IFG, which are different from Model~1A. It seems that Model~2A is more powerful in detecting even tiny effects by including the observations with NA responses.

\begin{figure}[ht]
\centering
\includegraphics[scale=0.35]{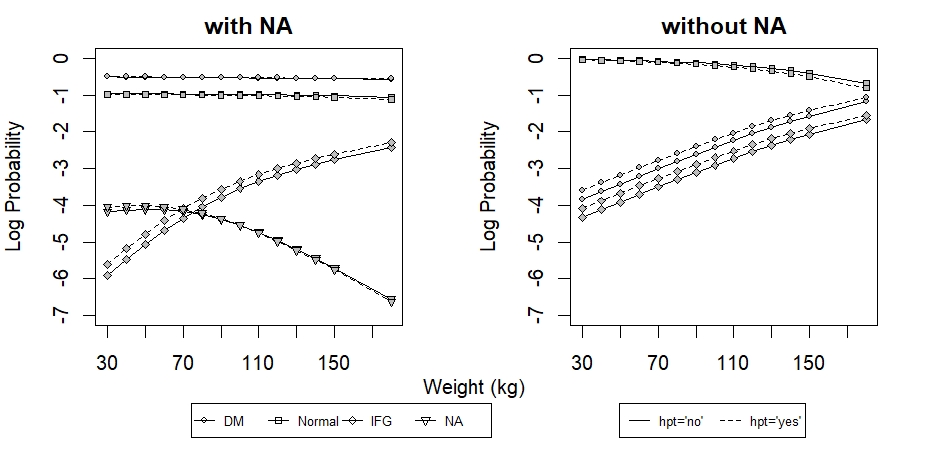}
\caption{Log-scale Categorical Probability against Weight Based on Models with or without NA Category for the Metabolic Syndrome Dataset, after po-npo model selection}
\label{fig:logprob_vs_weight_algo5}
\end{figure}

\begin{figure}[ht]
\centering
\includegraphics[scale=0.35]{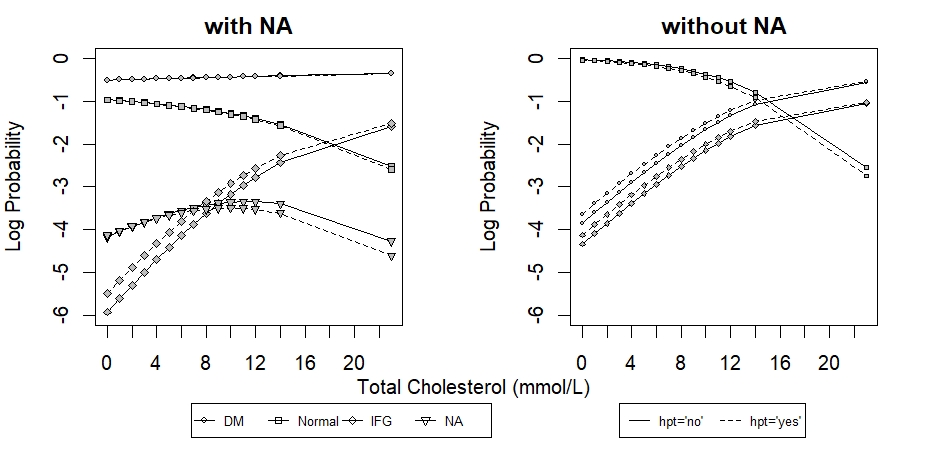}
\caption{Log-scale Categorical Probability against Cholesterol Based on Models with or without NA Category for the Metabolic Syndrome Dataset, after po-npo model selection}
\label{fig:logprob_vs_chol_algo5}
\end{figure}

Similarly to Figures~\ref{fig:logprob_vs_weight} and \ref{fig:logprob_vs_chol} for Model~1 and Model~2, we plot Model~1A (right panels) and Model~2A (left panels) in Figures~\ref{fig:logprob_vs_weight_algo5} and \ref{fig:logprob_vs_chol_algo5}. In terms of the relationships between the categorical probabilities and total cholesterol or body weight, the patterns are fairly similar to those displayed in Figures~\ref{fig:logprob_vs_weight} and \ref{fig:logprob_vs_chol}. If we remove the observations with NA responses, we would conclude that the log categorical probabilities of DM and IFG increase apparently and parallelly (i.e., up to a constant) along with body weight (right panel in Figure~\ref{fig:logprob_vs_weight_algo5}) and total cholesterol (right panel in Figure~\ref{fig:logprob_vs_chol}). However, if we keep those observations with NA responses in our analysis, it seems that the risk of DM slightly decreases along with body weight (left panel in Figure~\ref{fig:logprob_vs_weight_algo5}), and slightly increases along with total cholesterol (left panel in Figure~\ref{fig:logprob_vs_chol}). We also conclude that the chance of NA decreases along with body weight (left panel in Figure~\ref{fig:logprob_vs_weight_algo5}). In other words, our conclusions are essentially the same as the ones we conclude in Section~\ref{sec:metabolic_syndrome}.

\section{More on six cities data with dichotomous conditional link model}\label{sec:more_six_cities_cond_link_model}

In this section, we provide more technical details for applying the dichotomous conditional link model described in Section~\ref{sec:multinomial_conditional_link} to the six cities data (see also Section~\ref{sec:six_cities_cond_link_model}).

Different from \cite{pmcc1995}, we use binary variables $Z_{i1}, Z_{i2}, Z_{i3}, Z_{i4} \in \{0,1\}$ to denote the presence of wheeze of the $i$th child at ages 7, 8, 9, and 10 years, respectively. 
A multinomial response ${\mathbf Y}_i = (Y_{i1}, \ldots, Y_{i,16})^T$ with the number of categories $J=16$ is used to represent the outcomes of the four binary variables. Here the categories $j=1, \ldots, 16$ correspond to  $(Z_{i1}, Z_{i2}, Z_{i3}, Z_{i4}) = (1, 0, 0, 0),(0, 1, 0, 0), (1, 1, 0, 0), (0, 0, 1,$ $0),$ $ 
(1, 0, 1, 0),$ $(0, 1, 1, 0),$ $(1, 1, 1, 0),$ $ (0, 0, 0, 1), $ $(1, 0, 0, 1),$ $(0, 1, 0, 1),$ $(1, 1, 0, 1), $ $(0, 0, 1, $ $1), $ $
(1, 0, 1, 1), (0, 1, 1, 1), (1, 1, 1, 1), (0, 0, 0, 0),$ 
respectively, along with the categorical probabilities $\pi_{i1}, \ldots, \pi_{i,16} \in (0,1)$ at $x_i \in \{1,2\}$, which indicates the smoking status of the mothers.

For this study, we use the following $2^4-1 = 15$ equations to build up a multinomial conditional link model with $\eta_{ij} = \beta_{j1} + \beta_{j2} x_i$~, $j=1, \ldots, 15$. Note that we use logit link function for all equations for illustration purpose (see Example~\ref{ex:binary_cond_link_model} for more general expressions).

\begin{eqnarray*}
\text{logit}(P(Z_{i1}=1))=\text{logit}(\pi_{i1}+\pi_{i3}+ \cdots +\pi_{i,15}) &=& \eta_{i1}\ ;\\
\text{logit}(P(Z_{i2}=1\mid Z_{i1}=0))= \text{logit}\left(\frac{\pi_{i2}+\pi_{i6}+\pi_{i,10}+\pi_{i,14}}{\pi_{i2}+\pi_{i4}+ \pi_{i6} + \cdots +\pi_{i,16}}\right) &=& \eta_{i2}\ ;\\
\text{logit}(P(Z_{i2}=1\mid Z_{i1}=1))=\text{logit}\left(\frac{\pi_{i3}+\pi_{i7}+\pi_{i,11}+\pi_{i,15}}{\pi_{i1}+\pi_{i3}+\pi_{i5}+ \cdots +\pi_{i,15}}\right) &=& \eta_{i3}\ ;\\
\text{logit}(P(Z_{i3}=1\mid Z_{i1}=0, Z_{i2}=0))=\text{logit}\left(\frac{\pi_{i4}+\pi_{i,12}}{\pi_{i4}+\pi_{i8}+\pi_{i,12}+\pi_{i,16}}\right) &=& \eta_{i4}\ ;\\
\text{logit}(P(Z_{i3}=1\mid Z_{i1}=1, Z_{i2}=0))=\text{logit}\left(\frac{\pi_{i5}+\pi_{i,13}}{\pi_{i1}+\pi_{i5}+\pi_{i9}+\pi_{i,13}}\right) &=& \eta_{i5}\ ;\\
\text{logit}(P(Z_{i3}=1\mid Z_{i1}=0, Z_{i2}=1))=\text{logit}\left(\frac{\pi_{i6}+\pi_{i,14}}{\pi_{i2}+\pi_{i6}+\pi_{i,10}+\pi_{i,14}}\right) &=& \eta_{i6}\ ;\\
\text{logit}(P(Z_{i3}=1\mid Z_{i1}=1, Z_{i2}=1))=\text{logit}\left(\frac{\pi_{i7}+\pi_{i,15}}{\pi_{i3}+\pi_{i7}+\pi_{i,11}+\pi_{i,15}}\right) &=& \eta_{i7}\ ;\\
\text{logit}(P(Z_{i4}=1\mid Z_{i1}=0, Z_{i2}=0, Z_{i3}=0))=\text{logit}\left(\frac{\pi_{i8}}{\pi_{i8}+\pi_{i,16}}\right) &=& \eta_{i8}\ ;\\
\text{logit}(P(Z_{i4}=1\mid Z_{i1}=1, Z_{i2}=0, Z_{i3}=0))=\text{logit}\left(\frac{\pi_{i9}}{\pi_{i1}+\pi_{i9}}\right) &=& \eta_{i9}\ ;\\
\text{logit}(P(Z_{i4}=1\mid Z_{i1}=0, Z_{i2}=1, Z_{i3}=0))=\text{logit}\left(\frac{\pi_{i,10}}{\pi_{i2}+\pi_{i,10}}\right) &=& \eta_{i,10}\ ;\\
\text{logit}(P(Z_{i4}=1\mid Z_{i1}=1, Z_{i2}=1, Z_{i3}=0))=\text{logit}\left(\frac{\pi_{i,11}}{\pi_{i3}+\pi_{i,11}}\right) &=& \eta_{i,11}\ ;\\
\text{logit}(P(Z_{i4}=1\mid Z_{i1}=0, Z_{i2}=0, Z_{i3}=1))=\text{logit}\left(\frac{\pi_{i,12}}{\pi_{i4}+\pi_{i,12}}\right) &=& \eta_{i,12}\ ;\\
\text{logit}(P(Z_{i4}=1\mid Z_{i1}=1, Z_{i2}=0, Z_{i3}=1))=\text{logit}\left(\frac{\pi_{i,13}}{\pi_{i5}+\pi_{i,13}}\right) &=& \eta_{i,13}\ ;\\
\text{logit}(P(Z_{i4}=1\mid Z_{i1}=0, Z_{i2}=1, Z_{i3}=1))=\text{logit}\left(\frac{\pi_{i,14}}{\pi_{i6}+\pi_{i,14}}\right) &=& \eta_{i,14}\ ;\\
\text{logit}(P(Z_{i4}=1\mid Z_{i1}=1, Z_{i2}=1, Z_{i3}=1))=\text{logit}\left(\frac{\pi_{i,15}}{\pi_{i7}+\pi_{i,15}}\right) &=& \eta_{i,15}\ .
\end{eqnarray*}

The corresponding  $\mathbf{L}$, $\mathbf{R}$, and $\mathbf{b}$ as described in a general multinomial link model~\eqref{eq:mlm_in_matrix} are 

\setcounter{MaxMatrixCols}{20}

\begin{equation*}
{\mathbf L}= \begin{pmatrix}
1 & 0 & 1 & 0 & 1 & 0 & 1 & 0 & 1 & 0 & 1 & 0 & 1 & 0 & 1\\
0 & 1 & 0 & 0 & 0 & 1 & 0 & 0 & 0 & 1 & 0 & 0 & 0 & 1 & 0\\
0 & 0 & 1 & 0 & 0 & 0 & 1 & 0 & 0 & 0 & 1 & 0 & 0 & 0 & 1\\
0 & 0 & 0 & 1 & 0 & 0 & 0 & 0 & 0 & 0 & 0 & 1 & 0 & 0 & 0\\
0 & 0 & 0 & 0 & 1 & 0 & 0 & 0 & 0 & 0 & 0 & 0 & 1 & 0 & 0\\
0 & 0 & 0 & 0 & 0 & 1 & 0 & 0 & 0 & 0 & 0 & 0 & 0 & 1 & 0\\
0 & 0 & 0 & 0 & 0 & 0 & 1 & 0 & 0 & 0 & 0 & 0 & 0 & 0 & 1\\
0 & 0 & 0 & 0 & 0 & 0 & 0 & 1 & 0 & 0 & 0 & 0 & 0 & 0 & 0\\
0 & 0 & 0 & 0 & 0 & 0 & 0 & 0 & 1 & 0 & 0 & 0 & 0 & 0 & 0\\
0 & 0 & 0 & 0 & 0 & 0 & 0 & 0 & 0 & 1 & 0 & 0 & 0 & 0 & 0\\
0 & 0 & 0 & 0 & 0 & 0 & 0 & 0 & 0 & 0 & 1 & 0 & 0 & 0 & 0\\
0 & 0 & 0 & 0 & 0 & 0 & 0 & 0 & 0 & 0 & 0 & 1 & 0 & 0 & 0\\
0 & 0 & 0 & 0 & 0 & 0 & 0 & 0 & 0 & 0 & 0 & 0 & 1 & 0 & 0\\
0 & 0 & 0 & 0 & 0 & 0 & 0 & 0 & 0 & 0 & 0 & 0 & 0 & 1 & 0\\
0 & 0 & 0 & 0 & 0 & 0 & 0 & 0 & 0 & 0 & 0 & 0 & 0 & 0 & 1\\
\end{pmatrix}
\end{equation*}

\begin{equation*}
{\mathbf R}= \begin{pmatrix}
1 & 1 & 1 & 1 & 1 & 1 & 1 & 1 & 1 & 1 & 1 & 1 & 1 & 1 & 1\\
0 & 1 & 0 & 1 & 0 & 1 & 0 & 1 & 0 & 1 & 0 & 1 & 0 & 1 & 0\\
1 & 0 & 1 & 0 & 1 & 0 & 1 & 0 & 1 & 0 & 1 & 0 & 1 & 0 & 1\\
0 & 0 & 0 & 1 & 0 & 0 & 0 & 1 & 0 & 0 & 0 & 1 & 0 & 0 & 0\\
1 & 0 & 0 & 0 & 1 & 0 & 0 & 0 & 1 & 0 & 0 & 0 & 1 & 0 & 0\\
0 & 1 & 0 & 0 & 0 & 1 & 0 & 0 & 0 & 1 & 0 & 0 & 0 & 1 & 0\\
0 & 0 & 1 & 0 & 0 & 0 & 1 & 0 & 0 & 0 & 1 & 0 & 0 & 0 & 1\\
0 & 0 & 0 & 0 & 0 & 0 & 0 & 1 & 0 & 0 & 0 & 0 & 0 & 0 & 0\\
1 & 0 & 0 & 0 & 0 & 0 & 0 & 0 & 1 & 0 & 0 & 0 & 0 & 0 & 0\\
0 & 1 & 0 & 0 & 0 & 0 & 0 & 0 & 0 & 1 & 0 & 0 & 0 & 0 & 0\\
0 & 0 & 1 & 0 & 0 & 0 & 0 & 0 & 0 & 0 & 1 & 0 & 0 & 0 & 0\\
0 & 0 & 0 & 1 & 0 & 0 & 0 & 0 & 0 & 0 & 0 & 1 & 0 & 0 & 0\\
0 & 0 & 0 & 0 & 1 & 0 & 0 & 0 & 0 & 0 & 0 & 0 & 1 & 0 & 0\\
0 & 0 & 0 & 0 & 0 & 1 & 0 & 0 & 0 & 0 & 0 & 0 & 0 & 1 & 0\\
0 & 0 & 0 & 0 & 0 & 0 & 1 & 0 & 0 & 0 & 0 & 0 & 0 & 0 & 1\\
\end{pmatrix}, \quad
{\mathbf b} = \left(\begin{array}{c}
1 \\ 1 \\ 0 \\ 1 \\ 0 \\ 0 \\ 0 \\ 1 \\ 0 \\ 0 \\ 0 \\ 0 \\ 0 \\ 0 \\ 0
\end{array}\right)\ .
\end{equation*}

The model matrix ${\mathbf X}_i \in \mathbb{R}^{(J-1)\times p}$ as described in \eqref{eq:mlm_in_matrix} first takes the form of 
\[
	{\mathbf X}_i= \begin{pmatrix}
		1\ x_i &   &  \\
		&   \ddots &   \\
		&   & 1\ x_i
	\end{pmatrix}\> \in \mathbb{R}^{(J-1) \times (d+1)(J-1)}
	\]
for a main-effects npo model (in this case, $J=16$, $d=1$, and $p=(d+1)(J-1) = 30$). 
After applying Algorithm~\ref{algo:mixture_model} with intercepts on the initial npo model with $\eta_{ij} = \beta_{j1} + \beta_{j2} x_i$~, the most appropriate po-npo mixture model takes
\[
{\mathbf X}_i= \begin{pmatrix}
 1 &   &   &   &   &     &     \\
   & 1 &   &   &   &     & x_i \\
   &   & 1 &   &   &     & x_i \\
   & 1 &   &   &   &     &     \\
 1 &   &   &   &   &     & x_i \\
   &   & 1 &   &   &     & x_i \\
   &   &   & 1 &   &     &     \\
   &   &   &   & 1 &     &     \\
   & 1 &   &   &   & x_i &     \\
   & 1 &   &   &   &     &     \\
 1 &   &   &   &   &     & x_i \\
 1 &   &   &   &   &     &     \\
   &   & 1 &   &   &     &     \\
   &   & 1 &   &   &     & x_i \\
   &   &   & 1 &   &     &     \\
\end{pmatrix}
\]
with seven distinct parameters $\beta_{11} = \beta_{51} = \beta_{11,1} = \beta_{12,1}$~, $\beta_{21} = \beta_{41} = \beta_{91} = \beta_{10,1}$~, $\beta_{31} = \beta_{61} = \beta_{13,1} = \beta_{14,1}$~, $\beta_{71} = \beta_{15,1}$~, $\beta_{81}$~, $\beta_{92}$~, $\beta_{22} = \beta_{32} = \beta_{52} = \beta_{62} = \beta_{11,2} = \beta_{14,2}$~.
The corresponding 95\% confidence intervals for the seven distinct parameters in the final model are listed in Table~\ref{tab:six_cities_ci}. All parameters are significantly nonzero at $5\%$ level.

\begin{table}[ht]
    \centering
    \caption{95\% Confidence Intervals for  Parameters in po-npo Mixture Model for Six Cities Data}
    \label{tab:six_cities_ci}
    \begin{tabular}{rrr}
    \hline
     Lower Bound & Estimated Parameter & Upper Bound\\
     \hline
     -1.808  & $\hat{\beta}_{11} = \hat{\beta}_{51} = \hat{\beta}_{11,1} = \hat{\beta}_{12,1} = -1.611$  &  -1.414\\
     -2.613 & $\hat{\beta}_{21} = \hat{\beta}_{41} = \hat{\beta}_{91} = \hat{\beta}_{10,1} = -2.383$  &  -2.154\\
     -0.833  & $\hat{\beta}_{31} = \hat{\beta}_{61} = \hat{\beta}_{13,1} = \hat{\beta}_{14,1} = -0.506$  &  -0.180\\
     0.179 & $\hat{\beta}_{71} = \hat{\beta}_{15,1} = 0.671$ & 1.164\\
     -3.576  & $\hat{\beta}_{81} = -3.100$  &  -2.624\\
      0.340 & $\hat{\beta}_{92} = 1.539$ & 2.738\\
      0.206 & $\hat{\beta}_{22} = \hat{\beta}_{32} = \hat{\beta}_{52} = \hat{\beta}_{62} = \hat{\beta}_{11,2} = \hat{\beta}_{14,2} = 0.555$ & 0.903\\

    \hline
    \end{tabular}
\end{table}

\section{Proofs}\label{sec:proofs}

\medskip\noindent
{\bf Proof of Lemma~\ref{lem:pi_from_rho}}

\noindent
Since $\boldsymbol\rho_i = \left({\mathbf L}{\boldsymbol\pi}_i\right)/\left({\mathbf R}\boldsymbol\pi_i + \pi_{iJ} {\mathbf b}\right)$ and $\pi_{iJ} = 1 - \pi_{i1} - \cdots -\pi_{i,J-1} = 1 - {\mathbf 1}_{J-1}^T \boldsymbol\pi_i$~, we have
\begin{eqnarray*}
& & {\mathbf L}{\boldsymbol\pi}_i = {\rm diag} \left(\boldsymbol\rho_i\right) \left({\mathbf R}\boldsymbol\pi_i + \pi_{iJ} {\mathbf b}\right)\\
&\Longleftrightarrow & {\rm diag} \left(\boldsymbol\rho_i^{-1}\right) {\mathbf L}{\boldsymbol\pi}_i =  {\mathbf R}\boldsymbol\pi_i + \pi_{iJ} {\mathbf b} = {\mathbf R}\boldsymbol\pi_i - {\mathbf b} {\mathbf 1}_{J-1}^T \boldsymbol\pi_i +  {\mathbf b}\\
&\Longleftrightarrow & \left[{\rm diag} \left(\boldsymbol\rho_i^{-1}\right) {\mathbf L} - {\mathbf R} + {\mathbf b} {\mathbf 1}_{J-1}^T\right]{\boldsymbol\pi}_i =  {\mathbf b} \\
&\Longleftrightarrow & \left({\mathbf D}_i + {\mathbf b} {\mathbf 1}_{J-1}^T\right){\boldsymbol\pi}_i =  {\mathbf b}\ .
\end{eqnarray*}
According to the Sherman-Morrison-Woodbury formula (see, for example, Section~2.1.4 in \cite{golub2013}), $\left({\mathbf D}_i + {\mathbf b} {\mathbf 1}_{J-1}^T\right)^{-1}$ exists if ${\mathbf D}_i^{-1}$ exists and $1 + {\mathbf 1}_{J-1}^T {\mathbf D}_i^{-1} {\mathbf b} \neq 0$, which is guaranteed since all components of ${\mathbf D}_i^{-1} {\mathbf b}$ are positive, and thus \begin{eqnarray*}
\boldsymbol\pi_i &=& \left({\mathbf D}_i + {\mathbf b} {\mathbf 1}_{J-1}^T\right)^{-1} {\mathbf b}\\
&=& \left[{\mathbf D}_i^{-1} - {\mathbf D}_i^{-1} {\mathbf b} \left(1 + {\mathbf 1}_{J-1}^T {\mathbf D}_i^{-1} {\mathbf b}\right)^{-1} {\mathbf 1}_{J-1}^T {\mathbf D}_i^{-1}\right] {\mathbf b}\\
&=& {\mathbf D}_i^{-1} {\mathbf b} - {\mathbf D}_i^{-1} {\mathbf b} \cdot \left(1 + {\mathbf 1}_{J-1}^T {\mathbf D}_i^{-1} {\mathbf b}\right)^{-1} {\mathbf 1}_{J-1}^T {\mathbf D}_i^{-1}{\mathbf b} \\
&=& \frac{{\mathbf D}_i^{-1} {\mathbf b}}{1 + {\mathbf 1}_{J-1}^T {\mathbf D}_i^{-1} {\mathbf b}}\ .
\end{eqnarray*}
The rest conclusions are straightforward.
\hfill{$\Box$}

\bigskip\noindent
{\bf Proof of Lemma~\ref{thm:rho_ij_in_0_1}}

\noindent
Suppose $\rho_{ij}\in (0,1)$ and ${\mathbf L}_j \boldsymbol{\pi}_i > 0$ for all $\boldsymbol{\pi}_i \in \boldsymbol{\Pi}_0$ and $j=1, \ldots, J-1$. We show that Assumption~\ref{as:A1} is satisfied.

First of all, we must have $0\preceq {\mathbf L}_j$ for each $j=1, \ldots, J-1$. Otherwise, if any coordinate of ${\mathbf L}_j$ is strictly less than zero, we can always find a $\boldsymbol{\pi}_i \in \boldsymbol{\Pi}_0$~, such that ${\mathbf L}_j \boldsymbol{\pi}_i < 0$. Furthermore, we must have ${\mathbf 1}_{J-1}^T {\mathbf L}_j >0$. Otherwise, ${\mathbf 1}_{J-1}^T {\mathbf L}_j = 0$ leads to ${\mathbf L}_j = {\mathbf 0}_{J-1}$~, thus  ${\mathbf L}_j \boldsymbol{\pi}_i = 0$ and contradiction.

If $b_j < 0$ for any $j=1, \ldots, J-1$, we can always find a $\boldsymbol{\pi}_i \in \boldsymbol{\Pi}_0$~, such that ${\mathbf R}^T_j \boldsymbol\pi_i + \pi_{iJ} b_j < 0$, which violates $\rho_{ij} > 0$. Therefore $b_j \geq 0$ for each $j=1, \ldots, J-1$.

We let ${\mathbf L}_j = (L_{j1}, \ldots, L_{j,J-1})^T$ and ${\mathbf R}_j = (R_{j1}, \ldots, R_{j,J-1})^T$. If $L_{jl} > R_{jl}$ for any $j=1, \ldots, J-1$, we can always find a $\boldsymbol{\pi}_i \in \boldsymbol{\Pi}_0$, such that $\rho_{ij} \notin (0,1)$. The contradiction implies $L_{jl} \leq R_{jl}$ for all $j=1, \ldots, J-1$. That is, ${\mathbf L}_j \preceq {\mathbf R}_j$~.

If $b_j=0$, then we must have ${\mathbf 1}_{J-1}^T ({\mathbf R}_j - {\mathbf L}_j) > 0$. Otherwise, if ${\mathbf 1}_{J-1}^T ({\mathbf R}_j - {\mathbf L}_j) = 0$, then ${\mathbf L}_j = {\mathbf R}_j$ and thus $\rho_{ij}=1$.

Now we assume that Assumption~\ref{as:A1} is satisfied. It can be verified that we always have $0 < {\mathbf L}^T_j \boldsymbol\pi_i  < {\mathbf R}^T_j \boldsymbol\pi_i + \pi_{iJ} b_j$ and thus $\rho_{ij}\in (0,1)$, for all $\boldsymbol{\pi}_i \in \boldsymbol{\Pi}_0$~.
\hfill{$\Box$}

\bigskip\noindent
{\bf Proof of Lemma~\ref{thm:unique_pi}}

\noindent
If ${\mathbf 1}_{J-1}^T {\mathbf b} = 0$, then $b_j=0$ for each $j=1, \ldots, J-1$. If there exists a $\boldsymbol{\pi}_i \in \boldsymbol{\Pi}_0$~, such that
\[
\rho_{ij} = \frac{{\mathbf L}^T_j \boldsymbol\pi_i}{{\mathbf R}^T_j \boldsymbol\pi_i + \pi_{iJ} b_j} = \frac{{\mathbf L}^T_j \boldsymbol\pi_i}{{\mathbf R}^T_j \boldsymbol\pi_i}\ ,
\]
then 
\[
\frac{{\mathbf 1}_{J-1}^T \boldsymbol{\pi}_i + 1}{2\cdot {\mathbf 1}_{J-1}^T \boldsymbol{\pi}_i} \cdot \boldsymbol{\pi}_i \ \in \ \boldsymbol{\Pi}_0
\]
satisfies \eqref{eq:rho_ij_general} as well.
\hfill{$\Box$}

\bigskip\noindent
{\bf Proof of Theorem~\ref{thm:feasibility_with_assumptions}}

\noindent
Assumptions~\ref{as:A1} and \ref{as:1_b>0} are necessary conditions to make the multinomial link model~\eqref{eq:mlm_in_matrix} or \eqref{eq:mlm_j} well-defined. That is, $\rho_{ij}\in (0,1)$ for all $\boldsymbol{\pi}_i \in \boldsymbol{\Pi}_0$ (according to Lemma~\ref{thm:rho_ij_in_0_1}), and the solution for $\boldsymbol{\pi}_i$ as functions of $\boldsymbol{\rho}_i$ is unique when existing (according to Lemma~\ref{thm:unique_pi}).

Suppose ${\mathbf L}, {\mathbf R}, {\mathbf b}$ also satisfy Assumption~\ref{as:Di_inverse}. According to the proof of Lemma~\ref{lem:pi_from_rho},  for any $\rho_{ij}\in (0,1)$, $j=1, \ldots, J-1$, $({\mathbf D}_i + {\mathbf b} {\mathbf 1}_{J-1}^T)^{-1}$ exists and $\boldsymbol{\pi}_i$ can be solved uniquely via \eqref{eq:pi_from_rho}.

If ${\mathbf L}, {\mathbf R}, {\mathbf b}$ further satisfy Assumption~\ref{as:Di_b>0}, then $\boldsymbol{\pi}_i$ obtained  via \eqref{eq:pi_from_rho} belongs to $\boldsymbol{\Pi}_0$~.
\hfill{$\Box$}

\bigskip\noindent
{\bf Proof of Theorem~\ref{thm:models_assumption_1234}}

\noindent
For baseline-category, adjacent-categories, and  continuation-ratio mixed-link models, no matter with ppo (see Example~\ref{ex:mixed_link_ppo}) or po-npo mixture (see Example~\ref{ex:po_ppo_mixture}), ${\mathbf L}, {\mathbf R}, {\mathbf b}$ are provided in Appendices~\ref{sec:baseline-category}, \ref{sec:adjacent-categories}, and \ref{sec:continuation-ratio}, respectively. It can be verified that Assumptions~\ref{as:A1}, \ref{as:1_b>0}, \ref{as:Di_inverse} and \ref{as:Di_b>0} are all satisfied.

For baseline-adjacent (two-group) mixed-link models, if the two groups of indices share the same baseline category $J$, ${\mathbf L}, {\mathbf R}, {\mathbf b}$ are provided in Appendix~\ref{sec:b-a-c_link_model} and it can be verified that Assumptions~\ref{as:A1}, \ref{as:1_b>0}, \ref{as:Di_inverse} and \ref{as:Di_b>0} are all satisfied. If the first group of response categories is with baseline $s\neq J$, ${\mathbf L}, {\mathbf R}, {\mathbf b}$, ${\mathbf D}_i$~, ${\mathbf D}_i^{-1}$, and ${\mathbf D}_i^{-1} {\mathbf b}$ are provided in \eqref{eq:L_R_b_two_group_s}, \eqref{eq:D_i_two_group_s}, \eqref{eq:D_i_inverse_two_group_s}, and \eqref{eq:D_i_b_two_group_s}, respectively, with 
\begin{equation}\label{eq:L_R_b_(2)_baseline_adjacent_two_group}
{\mathbf L}_{(2)} = {\mathbf I}_{J-1-k}~, \> {\mathbf R}_{(2)} = \left[\begin{array}{ccccc}
1 & 1 &    &  & \\
  & 1 & 1  &  & \\
  &   & \ddots & \ddots & \\
  &   &        & 1 & 1\\
  &   &        &   & 1
\end{array}\right] \in \mathbb{R}^{(J-1-k)\times (J-1-k)},\> 
{\mathbf b}_{(2)} = \left[\begin{array}{c}
0\\ \vdots \\ 0 \\ 1
\end{array}\right] \in \mathbb{R}^{J-1-k} .
\end{equation}
It can be verified that Assumptions~\ref{as:A1}, \ref{as:1_b>0}, \ref{as:Di_inverse} and \ref{as:Di_b>0} are all satisfied.

For baseline-continuation (two-group) mixed-link models, if the two groups of indices share the same baseline category $J$, ${\mathbf L}, {\mathbf R}, {\mathbf b}$ are provided in Appendix~\ref{sec:b-c_link_model} and it can be verified that Assumptions~\ref{as:A1}, \ref{as:1_b>0}, \ref{as:Di_inverse} and \ref{as:Di_b>0} are all satisfied. If the first group of response categories is with baseline $s\neq J$, ${\mathbf L}, {\mathbf R}, {\mathbf b}$, ${\mathbf D}_i$~, ${\mathbf D}_i^{-1}$, and ${\mathbf D}_i^{-1} {\mathbf b}$ are provided in \eqref{eq:L_R_b_two_group_s}, \eqref{eq:D_i_two_group_s}, \eqref{eq:D_i_inverse_two_group_s}, and \eqref{eq:D_i_b_two_group_s}, respectively, with 
\begin{equation}\label{eq:L_R_b_(2)_baseline_continuation_two_group}
{\mathbf L}_{(2)} = {\mathbf I}_{J-1-k}~, \> {\mathbf R}_{(2)} = \left[\begin{array}{cccc}
1 & 1 & \cdots   & 1 \\
  & 1 & \cdots   & 1 \\
  &   & \ddots & \vdots \\
  &   &        & 1 
\end{array}\right] \in \mathbb{R}^{(J-1-k)\times (J-1-k)},\> 
{\mathbf b}_{(2)} = {\mathbf 1}_{J-1-k}\ .
\end{equation}
It can be verified that Assumptions~\ref{as:A1}, \ref{as:1_b>0}, \ref{as:Di_inverse} and \ref{as:Di_b>0} are all satisfied. 

For dichotomous conditional link models, Assumptions~\ref{as:A1} and \ref{as:1_b>0} are satisfied as a direct conclusion of Lemma~\ref{lem:L_R_b_binary_clm} in Appendix~\ref{sec:more_on_dichotomous_condi_link_model}, and Assumptions~\ref{as:Di_inverse} and \ref{as:Di_b>0} are satisfied according to Corollary~\ref{cor:|Di|_binary_clm}.

According to Theorem~\ref{thm:feasibility_with_assumptions}, $\boldsymbol{\Theta} = \mathbb{R}^p$ for the above models.
\hfill{$\Box$}

\bigskip\noindent
{\bf Proof of Theorem~\ref{thm:cumulative_models_assumption_1234}}

\noindent
For cumulative mixed-link models, according to Appendix~\ref{sec:cumulative}, it can be verified that Assumptions~\ref{as:A1} and \ref{as:1_b>0} are satisfied. Furthermore, ${\mathbf D}_i$ is invertible and $1 + {\mathbf 1}_{J-1}^T {\mathbf D}_i^{-1} {\mathbf b} = (1-\rho_{i,J-1})^{-1} > 0$, that is, Assumption~\ref{as:Di_inverse} is satisfied as well. However, since ${\mathbf D}_i^{-1} {\mathbf b} = (1-\rho_{i,J-1})^{-1} (\rho_{i1}, \rho_{i2} - \rho_{i1}, \ldots, \rho_{i,J-2} - \rho_{i,J-3}, \rho_{i,J-1} - \rho_{i,J-2})^T$, whose coordinates are not necessarily strictly positive, Assumption~\ref{as:Di_b>0} is not satisfied. Therefore, its feasible parameter space is $\boldsymbol\Theta = \{ \boldsymbol\theta \in \mathbb{R}^p \mid \rho_{i1} < \cdots < \rho_{i,J-1}, i=1, \ldots, m\}$.

For baseline-cumulative (two-group) mixed-link models with $s=J$, that is, the two groups of response categories share the same baseline $J$, according to Appendix~\ref{sec:b-c_link_model}, it can be verified that Assumptions~\ref{as:A1} and \ref{as:1_b>0} are satisfied. Furthermore, ${\mathbf D}_i$ is invertible and $1 + {\mathbf 1}_{J-1}^T {\mathbf D}_i^{-1} {\mathbf b} = 1 + \sum_{l=1}^k \rho_{il}/(1-\rho_{il}) + \rho_{i,J-1}/(1-\rho_{i,J-1}) > 0$, that is, Assumption~\ref{as:Di_inverse} is satisfied as well. However, since
\[
{\mathbf D}_i^{-1} {\mathbf b} = \left(\frac{\rho_{i1}}{1-\rho_{i1}}, \ldots, 
\frac{\rho_{ik}}{1-\rho_{ik}}, \frac{\rho_{i,k+1}}{1-\rho_{i,J-1}}, \frac{\rho_{i,k+2}-\rho_{i,k+1}}{1-\rho_{i,J-1}}, \ldots, \frac{\rho_{i,J-1} - \rho_{i,J-2}}{1-\rho_{i,J-1}}\right)^T
\]
whose coordinates are not necessarily strictly positive, Assumption~\ref{as:Di_b>0} is not satisfied. Therefore, its feasible parameter space is $\boldsymbol\Theta = \{ \boldsymbol\theta \in \mathbb{R}^p \mid \rho_{i, k+1} < \cdots < \rho_{i,J-1}, i=1, \ldots, m\}$.

For baseline-cumulative (two-group) mixed-link models with $s\neq J$, that is, $s=k+1, \ldots, J-1$, ${\mathbf L}, {\mathbf R}, {\mathbf b}$, ${\mathbf D}_i$~, ${\mathbf D}_i^{-1}$, and ${\mathbf D}_i^{-1} {\mathbf b}$ are provided in \eqref{eq:L_R_b_two_group_s}, \eqref{eq:D_i_two_group_s}, \eqref{eq:D_i_inverse_two_group_s}, and \eqref{eq:D_i_b_two_group_s}, respectively, with 
\begin{equation}\label{eq:L_R_b_(2)_baseline_cumulative_two_group}
{\mathbf L}_{(2)} = \left[\begin{array}{cccc}
1 &  &    &   \\
1 & 1 &   &   \\
\vdots  & \vdots  & \ddots &   \\
1 & 1 &  \cdots & 1 \\
\end{array}\right], \> {\mathbf R}_{(2)} = \left[\begin{array}{cccc}
1 & 1 & \cdots  & 1\\
1 & 1 & \cdots  & 1\\
\vdots  & \vdots  & \ddots & \vdots  \\
1 & 1 & \cdots  & 1
\end{array}\right] \in \mathbb{R}^{(J-1-k)\times (J-1-k)},\> 
{\mathbf b}_{(2)} = {\mathbf 1}_{J-1-k}\ .
\end{equation}
It can be verified that Assumptions~\ref{as:A1}, \ref{as:1_b>0} are satisfied, and ${\mathbf D}_i$ is invertible. However, $1 + {\mathbf 1}_{J-1}^T {\mathbf D}_i^{-1} {\mathbf b}$ is not necessarily positive unless $s=k+1$, and not all coordinates of ${\mathbf D}_i^{-1} {\mathbf b}$ are strictly positive. That is, Assumption~\ref{as:Di_inverse} is true only if $s=k+1$ and Assumption~\ref{as:Di_b>0} is not true. It can be verified that $\boldsymbol\Theta = \{ \boldsymbol\theta \in \mathbb{R}^p \mid \rho_{i, k+1} < \cdots < \rho_{i,J-1}, i=1, \ldots, m\}$ in this case.
\hfill{$\Box$}

\bigskip\noindent
{\bf Proof of Lemma~\ref{lem:cumulative_assumption_5}}

\noindent
For cumulative mixed-link models, for any $\boldsymbol{\rho}_i \in {\mathbf P}_0$~, there exists a $\boldsymbol{\pi}_i = (\pi_{i1}, \ldots, \pi_{i,J-1})^T \in \boldsymbol{\Pi}_0$, such that, 
\begin{equation}\label{eq:rho_ij_assumption5}
\rho_{ij} = \frac{{\mathbf L}_j^T \boldsymbol{\pi}_i}{{\mathbf R}_j^T \boldsymbol{\pi}_i + (1-{\mathbf 1}_{J-1}^T \boldsymbol{\pi}_i) b_j}\ , j=1, \ldots, J-1.
\end{equation}
According to Theorem~\ref{thm:cumulative_models_assumption_1234}, ${\mathbf D}_i^{-1}$ exists and $\boldsymbol{\pi}_i$ is also the unique solution solved from $\boldsymbol{\rho}_i = (\rho_{i1}, \ldots, \rho_{i,J-1})^T$ via \eqref{eq:pi_from_rho}. According to \eqref{eq:rho_ij}, $\rho_{ij} = \pi_{i1} + \cdots + \pi_{ij}$~. Therefore, we always have $\rho_{i1} < \cdots < \rho_{i,J-1}$~, which implies all the $J-1$ coordinates of ${\mathbf D}_i^{-1} {\mathbf b}$ are strictly positive.

For baseline-cumulative (two-group) mixed-link models, for any $\boldsymbol{\rho}_i \in {\mathbf P}_0$~, there exists a $\boldsymbol{\pi}_i = (\pi_{i1}, \ldots, \pi_{i,J-1})^T \in \boldsymbol{\Pi}_0$~, such that $\rho_{ij}$~, $j=1, \ldots, J-1$ are derived via \eqref{eq:rho_ij_assumption5}.
According to Theorem~\ref{thm:cumulative_models_assumption_1234}, ${\mathbf D}_i^{-1}$ exists and $\boldsymbol{\pi}_i$ is also the unique solution solved from $\boldsymbol{\rho}_i = (\rho_{i1}, \ldots, \rho_{i,J-1})^T$ via \eqref{eq:pi_from_rho}. According to \eqref{eq:rho_ij_two_group_s}, 
\[
\rho_{ij} = \left\{\begin{array}{cl}
\frac{\pi_{ij}}{\pi_{ij} + \pi_{is}}\ , & \mbox{ for }j=1, \ldots, k;\\
\frac{\pi_{i,k+1} + \cdots + \pi_{ij}}{\pi_{i,k+1} + \cdots + \pi_{iJ}}\ , & \mbox{ for }j=k+1, \ldots, J-1,
\end{array}\right.
\]
where $s=k+1, \ldots, J$.
Therefore, we always have $\rho_{i,k+1} < \cdots < \rho_{i,J-1}$~, which implies all the $J-1$ coordinates of ${\mathbf D}_i^{-1} {\mathbf b}$ are strictly positive.
\hfill{$\Box$}

\bigskip\noindent
{\bf Proof of Theorem~\ref{thm:fisher_information}}

\noindent
As described in Appendix~\ref{subsec:supp_fisher_mat}, the score vector
\[
\frac{\partial l}{\partial \boldsymbol\theta^T} = \sum_{i=1}^{m}{\mathbf Y}_i^T{\rm diag}(\bar{\boldsymbol\pi}_i)^{-1}\frac{\partial \bar{\boldsymbol\pi}_i}{\partial \boldsymbol\theta^T}
\]
with
\[
	\frac{\partial \bar{\boldsymbol\pi}_i}{\partial \boldsymbol\theta^T}
	= \frac{\partial \bar{\boldsymbol\pi}_i}{\partial \boldsymbol\rho_i^T} \cdot \frac{\partial \boldsymbol\rho_i}{\partial \boldsymbol\eta_i^T}  \cdot \frac{\partial \boldsymbol\eta_i}{\partial \boldsymbol\theta^T}\\
	= \frac{\partial \bar{\boldsymbol\pi}_i}{\partial \boldsymbol\rho_i^T} \cdot {\rm diag}\left(\left({\mathbf g}^{-1}\right)'\left(\boldsymbol\eta_i\right)\right) \cdot {\mathbf X}_i\ ,
\]
and (see Lemma~\ref{lem:part_pi_part_rho} in Appendix~\ref{subsec:supp_fisher_mat})
\[
\frac{\partial \bar{\boldsymbol{\pi}}_i}{\partial \boldsymbol\rho_i^T} = {\mathbf E}_i {\mathbf D}_i^{-1} \cdot {\rm diag}\left({\mathbf L}\boldsymbol\pi_i\right) \cdot {\rm diag} \left(\boldsymbol\rho_i^{-2} \right) \ .
\]
Then 
\[
		\frac{\partial \bar{\boldsymbol\pi}_i}{\partial \boldsymbol\theta^T}
	= {\mathbf E}_i {\mathbf D}_i^{-1} \cdot {\rm diag}\left({\mathbf L}\boldsymbol\pi_i\right) \cdot {\rm diag} \left(\boldsymbol\rho_i^{-2} \right) \cdot {\rm diag}\left(\left({\mathbf g}^{-1}\right)'\left(\boldsymbol\eta_i\right)\right) \cdot {\mathbf X}_i\ .
\]
As another direct conclusion of Lemma~\ref{lemma:for_partial_l} in Appendix~\ref{subsec:supp_fisher_mat},
\[
	E\left(\frac{\partial l}{\partial \boldsymbol\theta^T}\right) = \sum_{i=1}^{m} n_i \bar{\boldsymbol{\pi}}_i^T{\rm diag}(\bar{\boldsymbol\pi}_i)^{-1}\frac{\partial \bar{\boldsymbol\pi}_i}{\partial \boldsymbol\theta^T} = {\mathbf 0}\ .
\]

Since $E\left(\partial l/\partial \boldsymbol\theta^T\right) = {\mathbf 0}$, the Fisher information matrix (see, for example, Section~2.3.1 in \cite{schervish1995}) can be defined as
\begin{eqnarray*}
	{\mathbf F} &=& E\left(\frac{\partial l}{\partial \boldsymbol\theta}\cdot\frac{\partial l}{\partial \boldsymbol\theta^T}\right)\\
	&=& E\left(\sum_{i=1}^{m} \left(\frac{\partial \bar{\boldsymbol\pi}_i}{\partial \boldsymbol\theta^T}\right)^T {\rm diag}(\bar{\boldsymbol\pi}_i)^{-1}{\mathbf Y}_i\cdot\sum_{j=1}^{m}{\mathbf Y}_j^T{\rm diag}(\bar{\boldsymbol\pi}_j)^{-1}\frac{\partial \bar{\boldsymbol\pi}_j}{\partial \boldsymbol\theta^T}\right)\\
	&=& E\left(\sum_{i=1}^{m}\sum_{j=1}^{m} \left(\frac{\partial \bar{\boldsymbol\pi}_i}{\partial \boldsymbol\theta^T}\right)^T {\rm diag}(\bar{\boldsymbol\pi}_i)^{-1}{\mathbf Y}_i{\mathbf Y}_j^T{\rm diag}(\bar{\boldsymbol\pi}_j)^{-1}\frac{\partial \bar{\boldsymbol\pi}_j}{\partial \boldsymbol\theta^T}\right) \\
	&=& \sum_{i=1}^{m}\sum_{j=1}^{m} \left(\frac{\partial \bar{\boldsymbol\pi}_i}{\partial \boldsymbol\theta^T}\right)^T {\rm diag}(\bar{\boldsymbol\pi}_i)^{-1} E\left({\mathbf Y}_i{\mathbf Y}_j^T\right) {\rm diag}(\bar{\boldsymbol\pi}_j)^{-1}\frac{\partial \bar{\boldsymbol\pi}_j}{\partial \boldsymbol\theta^T}\ ,
\end{eqnarray*}
where
\[
E\left({\mathbf Y}_i{\mathbf Y}_j^T\right) = \left\{
\begin{array}{cl}
	n_i(n_i-1)\bar{\boldsymbol\pi}_i\bar{\boldsymbol\pi}_i^T+n_i{\rm diag}(\bar{\boldsymbol\pi}_i) & \mbox{ if } i = j\ ;\\
	n_in_j\bar{\boldsymbol\pi}_i\bar{\boldsymbol\pi}_j^T & \mbox{ if } i\neq j\ .
\end{array}
\right.
\]
According to Lemma~\ref{lemma:for_partial_l}, $\bar{\boldsymbol{\pi}}_i^T {\rm diag}(\bar{\boldsymbol\pi}_i)^{-1} \cdot \partial \bar{\boldsymbol\pi}_i/\partial \boldsymbol\theta^T = {\mathbf 0}$, the Fisher information matrix
\begin{eqnarray*}
    {\mathbf F} &=& 	\sum_{i=1}^{m} \left(\frac{\partial \bar{\boldsymbol\pi}_i}{\partial \boldsymbol\theta^T}\right)^T {\rm diag}(\bar{\boldsymbol\pi}_i)^{-1} \cdot n_i {\rm diag}(\bar{\boldsymbol\pi}_i) \cdot  {\rm diag}(\bar{\boldsymbol\pi}_i)^{-1}\frac{\partial \bar{\boldsymbol\pi}_i}{\partial \boldsymbol\theta^T} \\
    &=& \sum_{i=1}^{m} n_i \left(\frac{\partial \bar{\boldsymbol\pi}_i}{\partial \boldsymbol\theta^T}\right)^T {\rm diag}(\bar{\boldsymbol\pi}_i)^{-1}\frac{\partial \bar{\boldsymbol\pi}_i}{\partial \boldsymbol\theta^T}\ .
\end{eqnarray*}
\hfill{$\Box$}

\medskip\noindent
{\bf Proof of Lemma~\ref{lem:Fi}}

\noindent
Since $\boldsymbol{\theta} \in \boldsymbol{\Theta}$, ${\mathbf D}_i^{-1}$ exists and is nonsingular. Due to $\rho_{ij} \in (0,1), j=1, \ldots, J-1$, all coordinates of ${\mathbf L}\boldsymbol\pi_i$ are nonzero and both ${\rm diag}\left({\mathbf L}\boldsymbol\pi_i\right)$ and ${\rm diag} \left(\boldsymbol\rho_i^{-2} \right)$ are nonsingular. Since $\left(g_j^{-1}\right)'(\eta_{ij}) \neq 0$, $j=1, \ldots, J-1$, ${\rm diag}\left(\left({\mathbf g}^{-1}\right)'\left(\boldsymbol\eta_i\right)\right)$ is nonsingular as well. According to Theorem~\ref{thm:fisher_information}, the only thing left is to verify that ${\mathbf E}_i^T {\rm diag}\left(\bar{\boldsymbol\pi}_i\right)^{-1} {\mathbf E}_i$ is nonsingular. Actually, it can be verified that
\[
\left| {\mathbf E}_i^T {\rm diag}\left(\bar{\boldsymbol\pi}_i\right)^{-1} {\mathbf E}_i\right| = \left| {\rm diag} \left(\boldsymbol\pi_i\right)^{-1} - {\mathbf 1}_{J-1} {\mathbf 1}_{J-1}^T \right| = \pi_{iJ} \prod_{l=1}^{J-1} \pi_{il}^{-1} \neq 0\ .
\]
\hfill{$\Box$}

\medskip\noindent
{\bf A lemma needed for the proof of Theorem~\ref{thm:F=HUH^T}:}

\begin{lemma}\label{lem:U_mat}
Suppose $\boldsymbol{\theta} \in \boldsymbol{\Theta}$, $(g_j^{-1})'(\eta_{ij}) \neq 0$ and $n_i > 0$ for all $i=1, \ldots, m$ and $j=1, \ldots, J-1$. Then ${\mathbf U}$ is positive definite and $|{\mathbf U}| = \left(\prod_{i=1}^m n_i\right)^{J-1} \cdot \prod_{i=1}^m |{\mathbf U}_i|$, where
\[
|{\mathbf U}_i| = \left[\prod_{l=1}^{J-1} \left(g_l^{-1}\right)'(\eta_{il})\right]^2 \cdot \prod_{l=1}^{J-1} \rho_{il}^{-4}\cdot |{\mathbf D}_i^{-1}|^2 \cdot |{\rm diag} ({\mathbf L} \boldsymbol\pi_i)|^2 \cdot \pi_{iJ} \prod_{l=1}^{J-1} \pi_{il}^{-1}\ .
\]
\hfill{$\Box$}
\end{lemma}

\medskip\noindent
{\bf Proof of Lemma~\ref{lem:U_mat}}

\noindent
We denote an $mJ\times m(J-1)$ matrix $\tilde{\mathbf C} = ({\rm diag}\{{\mathbf c}_{11}, \ldots, {\mathbf c}_{m1}\},$ $\ldots,$ ${\rm diag}\{{\mathbf c}_{1,J-1}, \ldots, {\mathbf c}_{m,J-1}\})$ and an $mJ\times mJ$ matrix $\tilde{\mathbf W} = {\rm diag}\{ n_1 {\rm diag}(\bar{\boldsymbol\pi}_1)^{-1},$ $\ldots,$ $n_m {\rm diag}(\bar{\boldsymbol\pi}_m)^{-1}\}$. Then ${\mathbf U} = \tilde{\mathbf C}^T \tilde{\mathbf W} \tilde{\mathbf C}$.

If $\boldsymbol{\theta} \in \boldsymbol{\Theta}$ and $n_i > 0$ for all $i=1, \ldots, m$, then $\tilde{\mathbf W}$ is positive definite. 
By rearranging columns, we can verify that ${\rm rank}(\tilde{\mathbf C}) = {\rm rank} \left( {\rm diag} \{{\mathbf C}_1, \ldots, {\mathbf C}_m\}\right) = m(J-1)$ given that $\left(g_j^{-1}\right)'(\eta_{ij}) \neq 0$ for all $i=1, \ldots, m$ and $j=1, \ldots, J-1$. That is, $\tilde{\mathbf C}$ is of full rank, and thus ${\mathbf U}$ is positive definite.

As a direct conclusion of Theorem~S.4 in the Supplementary Material of \cite{bu2020}, $|{\mathbf U}| = \left(\prod_{i=1}^m n_i\right)^{J-1} \cdot \prod_{i=1}^m |{\mathbf U}_i|$, where ${\mathbf U}_i = {\mathbf C}_i^T {\rm diag} (\bar{\boldsymbol\pi}_i)^{-1} {\mathbf C}_i$ in our case. Then
\begin{eqnarray*}
|{\mathbf U}_i| &=& \left|{\rm diag}\left(\left({\mathbf g}^{-1}\right)'\left(\boldsymbol\eta_i\right)\right) \right|^2 \cdot \left|{\rm diag} \left(\boldsymbol\rho_i^{-2} \right) \right|^2 \cdot \left|{\mathbf D}_i^{-1}\right|^2 \cdot \left|{\rm diag}\left({\mathbf L}\boldsymbol\pi_i\right)\right|^2 \cdot \left|{\mathbf E}_i^T {\rm diag} (\bar{\boldsymbol\pi}_i)^{-1} {\mathbf E}_i\right|   \\
&=& \left[\prod_{l=1}^{J-1} \left(g_l^{-1}\right)'(\eta_{il})\right]^2 \cdot \prod_{l=1}^{J-1} \rho_{il}^{-4}\cdot |{\mathbf D}_i^{-1}|^2 \cdot |{\rm diag} ({\mathbf L} \boldsymbol\pi_i)|^2 \cdot \pi_{iJ} \prod_{l=1}^{J-1} \pi_{il}^{-1}\ .
\end{eqnarray*}
\hfill{$\Box$}

\medskip\noindent
{\bf Proof of Theorem~\ref{thm:F=HUH^T}}

\noindent
According to Theorem~\ref{thm:fisher_information}, ${\mathbf F}=\sum_{i=1}^{m}n_i{\mathbf F}_i$ with 
\[
{\mathbf F}_i = \left(\frac{\partial \bar{\boldsymbol\pi}_i}{\partial \boldsymbol\theta^T}\right)^T {\rm diag}(\bar{\boldsymbol\pi}_i)^{-1}\frac{\partial \bar{\boldsymbol\pi}_i}{\partial \boldsymbol\theta^T} = \left({\mathbf C}_i {\mathbf X}_i\right)^T {\rm diag}\left(\bar{\boldsymbol\pi}_i\right)^{-1} {\mathbf C}_i {\mathbf X}_i
= {\mathbf X}_i^T {\mathbf U}_i {\mathbf X}_i\ ,
\]
where ${\mathbf X}_i = (f_{jl}({\mathbf x}_i)) \in \mathbb{R}^{(J-1)\times p}$ and ${\mathbf U}_i = (u_{st}(\boldsymbol\pi_i))_{s,t=1, \ldots, J-1}$~. Then ${\mathbf F}_i$ can be rewritten as
\[
{\mathbf F}_i = \left[\begin{array}{ccc}
\sum_{l=1}^{J-1}\sum_{j=1}^{J-1} f_{j1}({\mathbf x}_i) u_{jl}(\boldsymbol\pi_i) f_{l1}({\mathbf x}_i) & \cdots & \sum_{l=1}^{J-1}\sum_{j=1}^{J-1} f_{j1}({\mathbf x}_i) u_{jl}(\boldsymbol\pi_i) f_{lp}({\mathbf x}_i)\\
\vdots & \cdots & \vdots\\
\sum_{l=1}^{J-1}\sum_{j=1}^{J-1} f_{jp}({\mathbf x}_i) u_{jl}(\boldsymbol\pi_i) f_{l1}({\mathbf x}_i) & \cdots & \sum_{l=1}^{J-1}\sum_{j=1}^{J-1} f_{jp}({\mathbf x}_i) u_{jl}(\boldsymbol\pi_i) f_{lp}({\mathbf x}_i) \end{array}\right] \ .
\]

Using \eqref{eq:H_general} and ${\mathbf U} = ({\mathbf U}_{st})_{s,t=1, \ldots, J-1}$ with ${\mathbf U}_{st} = {\rm diag}\{n_1 u_{st}(\boldsymbol\pi_1), \ldots, n_m u_{st}(\boldsymbol\pi_m)\}$, it can be verified that
\footnotesize
\begin{eqnarray*}
& & {\mathbf H} {\mathbf U} {\mathbf H}^T\\
&=& \left[\begin{array}{ccc}
\sum_{l=1}^{J-1} \sum_{j=1}^{J-1} {\mathbf F}_{j1}^T {\mathbf U}_{jl} {\mathbf F}_{l1} & \cdots & \sum_{l=1}^{J-1} \sum_{j=1}^{J-1} {\mathbf F}_{j1}^T {\mathbf U}_{jl} {\mathbf F}_{lp}\\
\vdots & \cdots & \vdots\\
\sum_{l=1}^{J-1} \sum_{j=1}^{J-1} {\mathbf F}_{jp}^T {\mathbf U}_{jl} {\mathbf F}_{l1} & \cdots & \sum_{l=1}^{J-1} \sum_{j=1}^{J-1} {\mathbf F}_{jp}^T {\mathbf U}_{jl} {\mathbf F}_{lp}
\end{array}\right]\\
&=& \left[\begin{array}{ccc}
\sum_{i=1}^m \sum_{l=1}^{J-1}\sum_{j=1}^{J-1} n_i f_{j1}({\mathbf x}_i) u_{jl}(\boldsymbol\pi_i) f_{l1}({\mathbf x}_i) & \cdots & \sum_{i=1}^m \sum_{l=1}^{J-1}\sum_{j=1}^{J-1} n_i f_{j1}({\mathbf x}_i) u_{jl}(\boldsymbol\pi_i) f_{lp}({\mathbf x}_i)\\
\vdots & \cdots & \vdots\\
\sum_{i=1}^m \sum_{l=1}^{J-1}\sum_{j=1}^{J-1} n_i f_{jp}({\mathbf x}_i) u_{jl}(\boldsymbol\pi_i) f_{l1}({\mathbf x}_i) & \cdots & \sum_{i=1}^m \sum_{l=1}^{J-1}\sum_{j=1}^{J-1} n_i f_{jp}({\mathbf x}_i) u_{jl}(\boldsymbol\pi_i) f_{lp}({\mathbf x}_i) \end{array}\right]\\
&=& \sum_{i=1}^m n_i {\mathbf F}_i\\
&=& {\mathbf F}\ .
\end{eqnarray*}
\normalsize
The rest statement is a direct conclusion of Lemma~\ref{lem:U_mat}.
\hfill{$\Box$}

\bigskip\noindent
{\bf Proof of Theorem~\ref{thm:validation_algorithm_3}}

\noindent
We only need to show that $\boldsymbol{\theta}^{(00)} \in \boldsymbol{\Theta}$.

According to Algorithm~\ref{algo:initial_theta_supplement}, $\boldsymbol{\pi}^{(0)} \in \boldsymbol{\Pi}_0$ and thus $\boldsymbol{\rho}^{(0)} = (\rho_1^{(0)}, \ldots, \rho_{J-1}^{(0)})^T \in {\mathbf P}_0$~. Since the model satisfies Assumptions~\ref{as:A1} and \ref{as:1_b>0}, we must have $\rho_j^{(0)} \in (0,1)$, $j=1, \ldots, J-1$.  Since
\[
\theta_l^{(00)} = \left\{\begin{array}{cl}
\eta_j^{(0)} = g_j(\rho_j^{(0)})\ , & \mbox{ if }l=l_j\ ,\ j=1, \ldots, J-1;\\
0\ , & \mbox{ otherwise}\ ,
\end{array}\right.
\]
then $\eta_{ij} = {\mathbf f}_j^T({\mathbf x}_i) \boldsymbol{\theta}^{(00)} = g_j(\rho_j^{(0)})$, for all ${\mathbf x}_i \in \mathbb{R}^d$, and $\rho_{ij} = g_j^{-1}(\eta_{ij}) = \rho_j^{(0)} \in (0,1)$.

Since the model satisfies Assumption~\ref{as:P_0_to_Pi_0_assumption} and $\boldsymbol{\rho}_i = (\rho_{i1}, \ldots, \rho_{i,J-1})^T = \boldsymbol{\rho}^{(0)} \in {\mathbf P}_0$~, then ${\mathbf D}_i^{-1}$ exists and all the $J-1$ coordinates of ${\mathbf D}_i^{-1} {\mathbf b}$ are strictly positive. That is, $\boldsymbol{\theta}^{(00)} \in \boldsymbol{\Theta}$.
\hfill{$\Box$}

\end{appendices}

\section*{Acknowledgement}
This work was supported in part by the U.S.~National Science Foundation grant DMS-1924859.

%\bibliographystyle{plain}
%\bibliographystyle{rss}
%\bibliography{reference}

\end{document}